\documentclass[a4paper]{article}
\usepackage{graphicx}
\usepackage{epstopdf, epsfig}
\usepackage{amsmath}
\usepackage{dashrule,xcolor}
\usepackage{natbib}
\usepackage{array}
\usepackage{makecell}
\usepackage{latexsym}
\usepackage{lscape}
\usepackage{authblk}
\usepackage{amssymb}

\usepackage[left=2.5cm,right=2.5cm,top=2cm,bottom=2cm]{geometry}

\usepackage{booktabs}  
\usepackage{multirow}

\title{Machine-learning wall model of large-eddy simulation for low- and high-speed flows over rough surfaces}

\author{Rong Ma$^{1*}$}
\author{Adri{\'a}n Lozano-Dur{\'a}n$^{1,2}$}
\affil{$^1$Department of Aeronautics and Astronautics, Massachusetts Institute of Technology, Cambridge, MA 02139, USA \\
$^2$Graduate Aerospace Laboratories, California Institute of Technology, \\
Pasadena, CA 91125, USA}
\affil{$^*$rongma@mit.edu}
\date{}

\begin{document}
\maketitle

\begin{abstract}
We present a wall model for large-eddy simulation (LES) that
incorporates surface-roughness effects and is applicable across low-
and high-speed flow regimes, for both transitional and fully rough
conditions. The model, implemented using an artificial neural network,
is trained on a database of direct numerical simulations (DNS) of
compressible turbulent channel flows over rough walls. The dataset
contains 372 cases spanning a wide range of irregular roughness
topographies, including Gaussian and Weibull distributions, Mach
numbers from 0 to 3.3, and friction Reynolds numbers between 180 and
2,000.
We employ an information-theoretic, dimensionless learning method to
identify the inputs with the highest predictive power for the
dimensionless wall friction and wall heat flux. Predictions are also
accompanied by a confidence score derived from a spectrally normalized
neural Gaussian process, which quantifies uncertainty in regions that
deviate from the training data distribution.
The performance of the model is first evaluated \textit{a-priori} on
110 turbulent channel flow cases, yielding prediction errors below
4\%. The model is then assessed \textit{a-posteriori} in wall-modeled
large-eddy simulations (WMLES) across a diverse set of test cases.
These include over 160 subsonic and supersonic turbulent channel flows
with rough walls, a transonic high-pressure turbine (HPT) blade with
Gaussian roughness, a high-speed compression ramp with sandpaper
roughness, and three hypersonic blunt bodies with sand-grain
roughness.  Results show that the proposed wall model typically
achieves \textit{a-posteriori} predictive accuracy within 10\% for
wall shear stress and within 15\% for wall heat flux, with high
confidence in the channel flows and HPT blade cases. In the rough-wall
compression ramp and hypersonic blunt bodies, the model captures the
heating augmentation with errors ranging from 0\% to 20\%.  In
the cases with the highest errors, the reduced performance is
correctly detected by a drop in the confidence score.
\end{abstract}



\section{Introduction}
\label{sec:headings}

Turbulent flows over rough surfaces in high-speed flow regimes present
major challenges for both physical understanding and modeling. The
compressibility effects on smooth walls, such as strong temperature
gradients, aero-thermal coupling, and shock--boundary-layer
interactions, have been extensively investigated and are known to
fundamentally alter near-wall turbulence structures and transport
mechanisms~\citep{bradshaw1977compressible,papamoschou1988compressible,zhang2014generalized,nichols2017stability,yu2019genuine,gaitonde2023dynamics}. When
surface roughness is present, these effects are further compounded,
leading to enhanced drag, wall heating, and complex flow--surface
interactions that are difficult to predict~\citep{ekoto2008supersonic,tyson2013numerical,modesti2022direct,aghaei2023supersonic}. In the context of LES, existing wall models,
largely developed for smooth-wall or low-speed flows, usually fail to
capture the coupled aero--thermal dynamics induced by roughness in
high-speed flow regimes. Addressing these challenges requires improved
physical insight and modeling strategies that can account for
roughness effects in high-speed flows. In the present work, we develop
a wall model for LES that accounts for subgrid-scale surface roughness
and is applicable across a wide range of Mach numbers, from low- to
high-speed regimes.

\subsection{Physics of high-speed turbulence over rough surfaces}

The flow physics of rough-wall turbulence have been extensively
investigated in low-speed
regimes~\citep[e.g.,][]{orlandi2006dns,schultz2007rough,
  yuan2014roughness, ma2021direct, kadivar2021review} and, separately,
for smooth-wall high-speed
flows~\citep{spina1994physics,guarini2000direct,
  pirozzoli2008characterization, modesti2016reynolds, zhang2018direct,
  cheng2024reynolds}. In low-speed rough-wall turbulent boundary
layers, surface roughness disrupts the near-wall cycle, leading to
enhanced drag and a larger momentum
deficit~\citep{raupach1991rough,jimenez2004turbulent}. The fully-rough
regime is relatively well understood: form drag dominates and the wall
friction becomes largely independent of Reynolds number. By contrast,
the transitionally-rough regime remains challenging even at low speed.
In this regime, viscous and form drag contribute simultaneously,
rendering wall friction highly sensitive to Reynolds number and to the
details of the surface topography. This sensitivity complicates the
development of universal scaling laws and predictive drag models.

The challenges are further amplified at high speeds, where
comparatively little attention has been devoted to rough-wall
supersonic and hypersonic flows. The interplay among roughness,
compressibility, and thermal effects introduces additional mechanisms
that remain largely unresolved. In particular, roughness modifies
near-wall temperature gradients and can weaken the Reynolds analogy
between momentum and heat transfer~\citep{kadivar2021review}. Recent
work by \citet{kadivar2025turbulent} further suggests that a universal
rough-wall heat-transfer theory, even in low-speed flows, must account
for multiple factors, including roughness topology, roughness thermal
conductivity, roughness Reynolds number, and the Prandtl number.

To advance the understanding of compressible turbulent flows over
rough surfaces, several experimental and DNS studies have been
conducted. On the experimental front, the work of \cite{latin2000flow,
  latin2002temporal} examined the influence of various surface
roughness configurations, including two uniformly distributed and
three sand-grain roughness types, on the mean and turbulent properties
of supersonic turbulent boundary layers. Their findings revealed that
while kinematic turbulence quantities scaled with the local mean flow,
thermodynamically dependent turbulence quantities exhibited a strong
linear dependence on roughness height.  Additionally, roughness was
observed to extend the region where inner scaling holds for several
turbulence intensity metrics.  In a separate study,
\cite{ekoto2008supersonic} investigated the effects of large-scale
periodic surface roughness on a supersonic turbulent boundary layer by
comparing square and diamond roughness topologies against a smooth
wall. They found that square roughness exhibited canonical rough-wall
behavior, whereas the diamond pattern induced strong shock–expansion
interactions that significantly distorted the flow and amplified
turbulence production across the boundary layer.

From the computational side, several DNS of compressible flows over
rough surfaces have been conducted over the past decade to investigate
roughness effects. \cite{muppidi2012direct} used DNS to examine
laminar-to-turbulent transition in a Mach 2.9 supersonic flat-plate
boundary layer induced by distributed surface roughness. Similarly,
\cite{bernardini2012compressibility} performed DNS to analyze
compressibility effects on boundary layer transition caused by an
isolated three-dimensional cubic element in both low- and high-speed
regimes. Both studies highlight the critical role of counter-rotating
vortices in triggering transition, independent of Mach number.
\cite{tyson2013numerical} conducted DNS of compressible turbulent
channel flow over two-dimensional wavy rough surfaces at Mach numbers
0.3, 1.5, and 3. Their results indicate that at Mach 3, roughness
peaks generate strong shock waves that significantly impact the mean
flow across the entire channel. \cite{modesti2022direct} carried out
DNS of supersonic turbulent channel flows over distributed cubical
roughness elements for Mach numbers ranging from 0.3 to 4. They found
that while velocity statistics in the outer layer remain similar to
those of smooth-wall flows, the thermal field is substantially
modified by roughness throughout the channel core.
\cite{aghaei2023supersonic} performed DNS of supersonic turbulent
channel flows over both two-dimensional and three-dimensional
sinusoidal rough walls. Their findings demonstrate that flow behavior
strongly depends on roughness topology: two-dimensional surfaces
produce strong oblique shocks and elevated entropy generation, whereas
three-dimensional surfaces generate only weaker shocklets.

The DNS studies of compressible flows discussed above have primarily
focused on uniformly distributed regular roughness, with limited
attention to more realistic, multiscale, and irregular roughness. In
this work, we address this limitation by creating a DNS database of
compressible turbulent channel flows over irregular Gaussian rough
surfaces, spanning flow regimes from subsonic to supersonic, with bulk
Mach numbers from 0.4 to 3.3 and bulk Reynolds numbers from 7,500 to
15,500. This database not only complements existing canonical datasets
but also serves as a critical foundation for the training of our wall
model.

\subsection{Wall models for  turbulence over rough surfaces}

Modeling and prediction of roughness effects in rough-wall flows are
essential tasks in engineering, as surface roughness can substantially
amplify hydrodynamic drag and wall heat transfer, reducing the overall
efficiency and durability of engineering systems
\citep{bons2010review,chung2021predicting}. In practical applications
such as turbine blades in gas-turbine engines, and external surfaces
of entry, landing, descent (EDL) vehicles, surface roughness arises
from erosion, deposition, or thermal protection system (TPS) ablation
\citep{bons2010review,hollis2014distributed}. Accurately capturing the
impact of such roughness is crucial for predicting aerodynamic
performance, heat loading, and material lifetime under realistic
operating conditions. However, performing computational fluid dynamics
(CFD) analysis with full-vehicle grids that explicitly resolve surface
roughness is computationally prohibitive due to the extremely large
number of grid cells required. While roughness-resolved simulations
can provide valuable physical insight into the local flow structures
and mechanisms, their applicability to high-Reynolds-number flows and
realistic engineering configurations remains severely limited by the
associated computational cost.

In the past decades, a range of models with differing levels of
fidelity have been developed to capture wall-roughness effects,
especially the drag induced by the roughness, spanning from empirical
correlations derived from Moody charts to wall functions for
Reynolds-averaged Navier--Stokes (RANS) simulations, as well as WMLES
approaches~\citep{flack2010review, forooghi2017toward,
  garcia2024challenges, volino2024effects,zhou2024data}. An extended overview of
these modeling approaches for low-speed flows can be found in
\cite{ma2025machine}. Many of these methods are developed based on the
equivalent sand-grain roughness height $k_s$, which quantifies the
hydrodynamic drag through a logarithmic relationship with the
roughness function in the fully rough regime. However, $k_s$ is not a
purely geometrical parameter that can be directly inferred from the
surface topology. Rather, it is a hydrodynamic parameter that
characterizes the roughness effect on the flow and must be determined
empirically through experiments or calibrated using high-fidelity
simulations. This has limited the generalizability and applicability
of the models.

Compared to low-speed flows, modeling for high-speed flows over rough
surfaces poses additional challenges due to the complex coupling
between aerodynamic and thermal processes~\citep{bose2018wall}. In
the context of EDL vehicles, roughness effects on aeroheating are
typically modeled using RANS together with empirical correlations
based on the roughness Reynolds number. For instance, in the Earth
Entry System design, the Dahm roughness model has been applied in
post-processing to estimate heating augmentation from surface
roughness~\citep{cheatwood2024mars}. However, this approach is
inherently limited: correlations that rely solely on the roughness
Reynolds number lack generality and may lose predictive accuracy
across different roughness types, spatial distributions, and flow
regimes. Moreover, RANS modeling for EDL flows faces additional
challenges due to the complex aerodynamic and thermodynamic phenomena
involved, including roughness-induced heat transfer and skin-friction
enhancement in ablated TPS, separation behind blunt bodies,
shock--boundary-layer interactions, high-enthalpy effects, and
laminar-to-turbulent transition~\citep{dirling1973method,
  finson1980effect, bowersox2007survey, gaitonde2023dynamics}. As
noted by \citet{hollis2014distributed}, existing closure models often
fail to capture these effects, particularly under conditions of
strong compressibility and roughness-enhanced aero-thermal loads.

WMLES has demonstrated improved fidelity for high-speed flows
\citep{kawai2010dynamic, yang2018aerodynamic, iyer2019analysis,
  mettu2022wall, chen2022wall, griffin2023near}. By explicitly
resolving the majority of length scales away from the wall while
modeling the subgrid scales and the near-wall region, WMLES can
improve predictions of wall shear stress, heat flux, and complex
aero-thermal dynamics in supersonic and hypersonic regimes. In WMLES,
the classical wall model for compressible flow consists of two coupled
ordinary differential equations (ODEs) for velocity and temperature,
solved with boundary conditions at the wall and the matching
location---the interface between the wall-model layer and the LES
outer layer~\citep{cabot2000approximate, larsson2016large,
  bose2018wall}. However, \citet{fu2021shock} demonstrated that the
iterative solution of this coupled boundary-value problem introduces
greater nonlinearity than in the incompressible case and shows poor
convergence in flows with large temperature gradients. The accuracy of
the wall model also deteriorates under strong heat-transfer
conditions, as observed by \citet{griffin2023near}. Moreover, the
mixing-length eddy viscosity model with two empirical constants (the
von K{\'a}rm{\'a}n constant $\kappa_v$ and the damping function
constant $A^+$) commonly used for incompressible flow may not be
appropriate for high-speed flows. \citet{yang2018semi} suggested that
enhanced accuracy for high-speed flows over cold walls can be achieved
through semi-local scaling in the damping function, while
\citet{iyer2019analysis} reported that the prediction of skin friction
and wall heat flux in high-speed flows is sensitive to the values of
$\kappa_v$ and $A^+$ used in the eddy-viscosity model and damping
function. Recently, \citet{griffin2023near} coupled a compressible
velocity transformation that accurately represents both diabatic and
adiabatic turbulent boundary layers~\citep{griffin2021velocity}, along
with the temperature--velocity
relationship~\citep{zhang2014generalized}, as the smooth-wall model
for WMLES, and obtained more accurate predictions of wall shear stress
and heat transfer in cases with strong heat transfer. Nevertheless,
existing wall models often neglect the critical role of surface
roughness in shaping aero-thermal dynamics and still exhibit reduced
accuracy when applied to flow over rough surfaces.

In our recent work~\citep{ma2025machine}, we developed a rough-wall
model for WMLES that directly leverages both geometrical roughness
parameters and local flow information. This model, referred to as the
\emph{Building-Block-Flow Wall Model for Rough Walls} (BFWM-rough),
builds upon our previous efforts on wall-modeling for low-speed and
high-speed flows over smooth and rough walls~\citep{lozano2023machine,
  arranz2024building, Ma_CTR_2024, ma2025machine,
  ling2025general}. The approach is grounded in the idea that a
limited set of canonical flow configurations, termed
\emph{building-block flows}, can encapsulate the essential physics
required to accurately predict wall shear stress and heat flux in
complex flow environments.  The first version of BFWM-rough,
introduced in~\cite{ma2025machine}, was based on an artificial neural
network trained on DNS of incompressible turbulent channel flows over
rough walls. Validation results demonstrated superior accuracy
compared to existing models in both canonical configurations and
high-pressure turbine (HPT) blades, achieving prediction errors for
wall shear stress within the range of 1\% to 15\% across both
transitionally and fully rough regimes.  In the present work, we
re-design BFWM-rough to explicitly incorporate both compressibility
and surface roughness effects. This new version is denoted as
\emph{BFWM-rough-v2}.

The manuscript is structured as follows: \S\ref{database} introduces
the roughness generation procedure and the DNS training database,
comprising both incompressible and compressible
flows. \S\ref{formulation} provides an overview of the proposed wall
model, the formulation of dimensionless inputs and outputs, details
about the wall model training, and model uncertainty
quantification. \S\ref{validation} presents \textit{a-posteriori}
results from WMLES applied to turbulent channel flows, HPT blade, a
compression ramp, and three hypersonic blunt bodies with surface
roughness. Concluding remarks are offered in \S\ref{conclusion}.

\section{DNS database}\label{database}

We conducted DNS of low- and high-speed turbulent channel flows over
irregular rough surfaces. This DNS database forms the training set for
the development of BFWM-rough-v2.  The simulations were carried out
using the minimal-span channel approach,
following~\citep{jimenez1991minimal, chung2015fast,
  macdonald2017minimal}. As will be shown, these simulations can
accurately reproduce near-wall flow dynamics, provided that the
associated domain constraints are properly satisfied.  Hereafter, the
streamwise, wall-normal, and spanwise directions are denoted by $x$,
$y$, and $z$, respectively. The channel half-height is denoted by
$\delta$. The details of surface roughness generation and DNS setups
are described below.

\subsection{Surface roughness generation}
\label{sec:rough_gen}

The roughness repository comprises a set of rough surface profiles
designed to support the development of the DNS database used for
training and validating the wall model. These surfaces are intended to
reflect the variety of roughness types encountered in real-world
engineering applications and are characterized by distinct probability
density functions (PDFs) and power spectra (PS). Using a surface
generator developed by \citet{perez2019generating}, irregular
isotropic rough surfaces are synthesized based on prescribed PDFs and
PS. Two primary statistical families are included: Gaussian and
Weibull. The Gaussian distribution is selected due to its widespread
occurrence in both natural and engineered systems
\citep{williamson1969shape, whitehouse2023handbook}, while Weibull
roughness is included to allow the repository to span a wider range of
non-Gaussian and asymmetric roughness patterns, improving the
generalization of the model across diverse surface conditions.

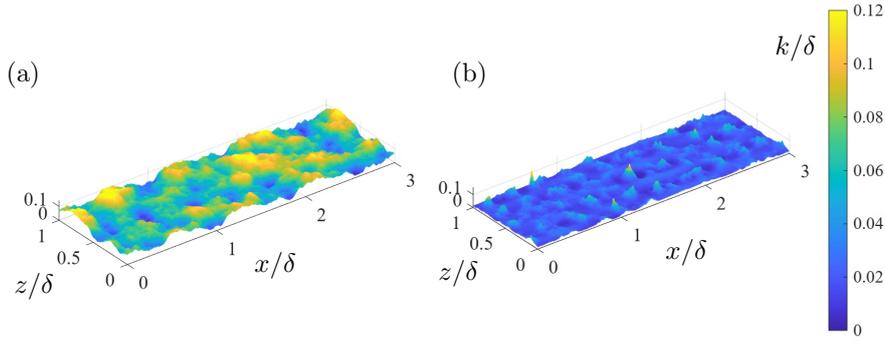
\begin{figure}
\centering
\includegraphics[width=53mm,trim={0.5cm 0.5cm 0.5cm 0.5cm},clip]{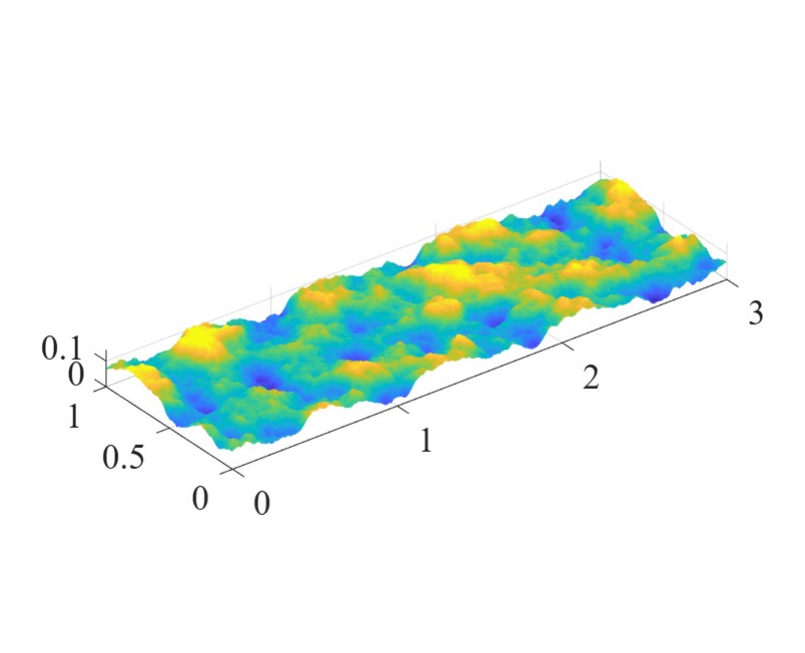}
\put(-62,28){$x/\delta$}
\put(-152,20){$z/\delta$}
\put(-155,100){(a)}
\hspace{1mm}
\includegraphics[width=60mm,trim={0.5cm 0.8cm 0.5cm 0.5cm},clip]{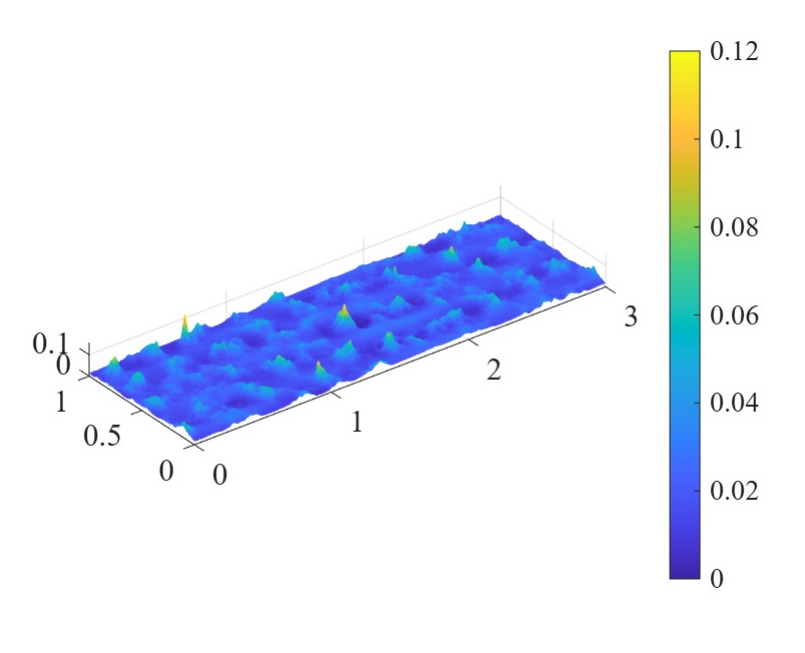}
\put(-85,33){$x/\delta$}
\put(-170,25){$z/\delta$}
\put(-165,100){(b)}
\put(-44,110){$k/\delta$}
\caption{Roughness height $(k)$ contour map of selected rough surface
  samples: (a) Gaussian roughness GS06; (b) Weibull roughness
  WB10. The roughness map is applied to turbulent channel walls, where
  $\delta$ is the channel half-height.}
\label{fig:rough_samples}
\end{figure}

Following our previous work~\citep{ma2025machine}, 19 Gaussian rough
surfaces and 13 Weibull rough surfaces are used to create the
rough-wall DNS database. A visualization of two roughness samples is
shown in figure~\ref{fig:rough_samples}. The Gaussian roughness cases
are labeled GS1 through GS19, while the Weibull roughness cases are
labeled WB1 through WB13. Here, we summarize the range of key
geometric roughness parameters represented in the roughness repository,
including the root-mean-square roughness height ($k_{rms}$), mean
roughness height ($R_a$), skewness ($S_k$), kurtosis ($K_u$), and
effective slope ($ES$). The definitions of all roughness parameters
are provided in Appendix~\ref{appA}. The corresponding ranges,
$1.1 < k_{rms}/R_a < 1.5$, $-0.7 < S_k < 2.2$, $2.2 < K_u < 12.0$, and
$0.1 < ES < 0.7$, encompass a broad spectrum of surface characteristics
representative of realistic roughness conditions encountered in
engineering applications~\citep{bons2010review, chung2021predicting}.

\begin{figure}
\centering
\includegraphics[width=130mm,trim={2cm 0.0cm 0.0cm 0.0cm},clip]{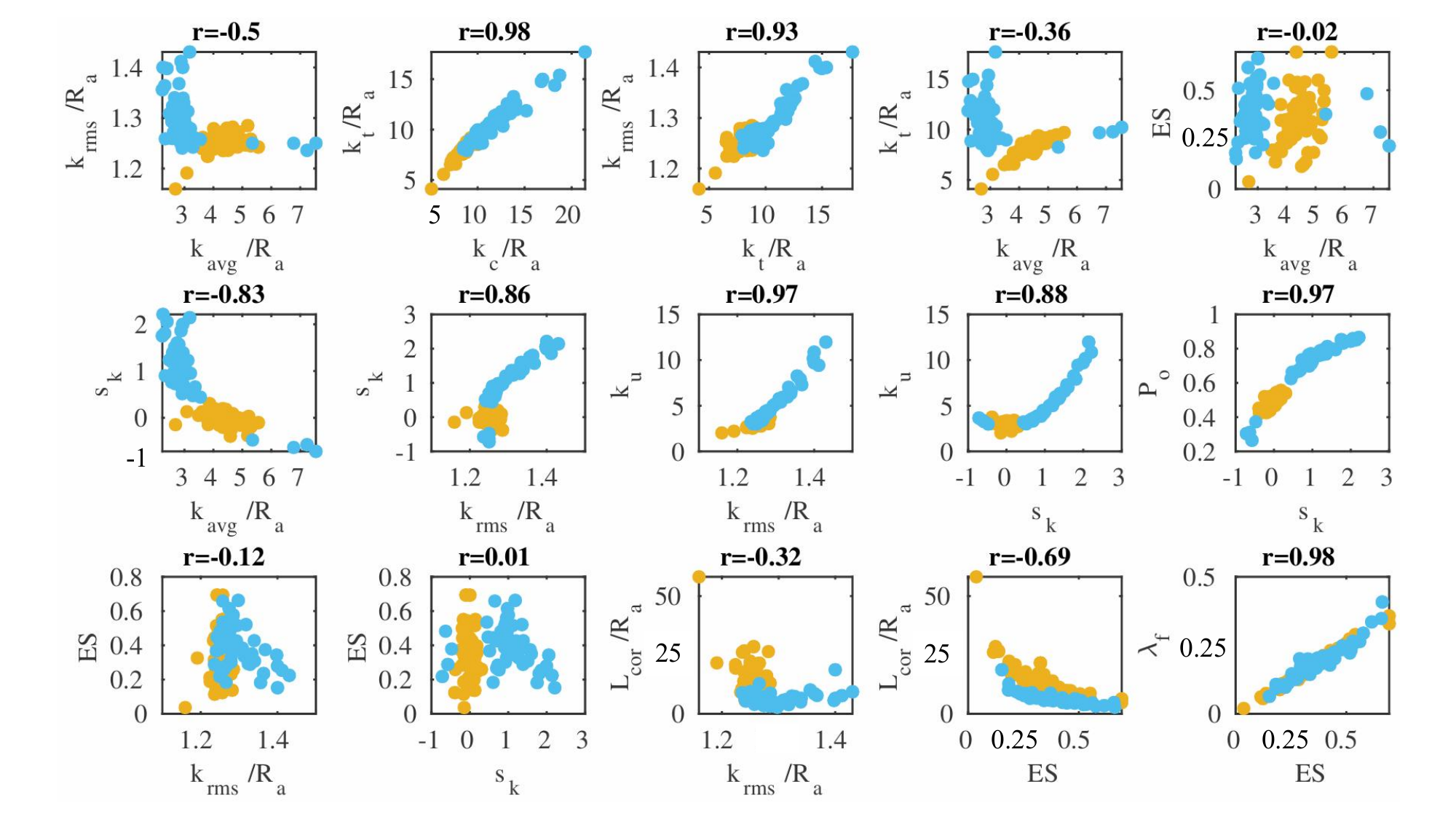}
\put(-343,151){\colorbox{white}{\scriptsize{$k_{avg}/R_a$}}}
\put(-378,170){\rotatebox{90}{{\colorbox{white}{\scriptsize{$k_{rms}/R_a$}}}}}
\put(-270,151){\colorbox{white}{\scriptsize{$k_{c}/R_a$}}}
\put(-303,176){\rotatebox{90}{{\colorbox{white}{\scriptsize{$k_{t}/R_a$}}}}}
\put(-200,151){\colorbox{white}{\scriptsize{$k_{t}/R_a$}}}
\put(-233,170){\rotatebox{90}{{\colorbox{white}{\scriptsize{$k_{rms}/R_a$}}}}}
\put(-129,151){\colorbox{white}{\scriptsize{$k_{avg}/R_a$}}}
\put(-157,176){\rotatebox{90}{{\colorbox{white}{\scriptsize{$k_{t}/R_a$}}}}}
\put(-57,151){\colorbox{white}{\scriptsize{$k_{avg}/R_a$}}}
\put(-85,178){\rotatebox{90}{{\colorbox{white}{\scriptsize{$ES$}}}}}
\put(-343,81){\colorbox{white}{\scriptsize{$k_{avg}/R_a$}}}
\put(-372,110){\rotatebox{90}{{\colorbox{white}{\scriptsize{$S_k$}}}}}
\put(-272,81){\colorbox{white}{\scriptsize{$k_{rms}/R_a$}}}
\put(-300,110){\rotatebox{90}{{\colorbox{white}{\scriptsize{$S_k$}}}}}
\put(-200,81){\colorbox{white}{\scriptsize{$k_{rms}/R_a$}}}
\put(-230,110){\rotatebox{90}{{\colorbox{white}{\scriptsize{$K_u$}}}}}
\put(-118,81){\colorbox{white}{\scriptsize{$S_k$}}}
\put(-157,110){\rotatebox{90}{{\colorbox{white}{\scriptsize{$K_u$}}}}}
\put(-46,81){\colorbox{white}{\scriptsize{$S_k$}}}
\put(-87,110){\rotatebox{90}{{\colorbox{white}{\scriptsize{$P_o$}}}}}
\put(-345,11){\colorbox{white}{\scriptsize{$k_{rms}/R_a$}}}
\put(-376,40){\rotatebox{90}{{\colorbox{white}{\scriptsize{$ES$}}}}}
\put(-265,11){\colorbox{white}{\scriptsize{$S_k$}}}
\put(-303,40){\rotatebox{90}{{\colorbox{white}{\scriptsize{$ES$}}}}}
\put(-200,11){\colorbox{white}{\scriptsize{$k_{rms}/R_a$}}}
\put(-231,30){\rotatebox{90}{{\colorbox{white}{\scriptsize{$L_{cor}/R_a$}}}}}
\put(-120,11){\colorbox{white}{\scriptsize{$ES$}}}
\put(-157,30){\rotatebox{90}{{\colorbox{white}{\scriptsize{$L_{cor}/R_a$}}}}}
\put(-46,11){\colorbox{white}{\scriptsize{$ES$}}}
\put(-87,38){\rotatebox{90}{{\colorbox{white}{\scriptsize{$\lambda_f$}}}}}
\caption{Scatter plot of pairs of roughness parameters and their
  correlation for Gaussian (orange) and Weibull (blue) rough
  surfaces.}
\label{fig:correlation}
\end{figure}
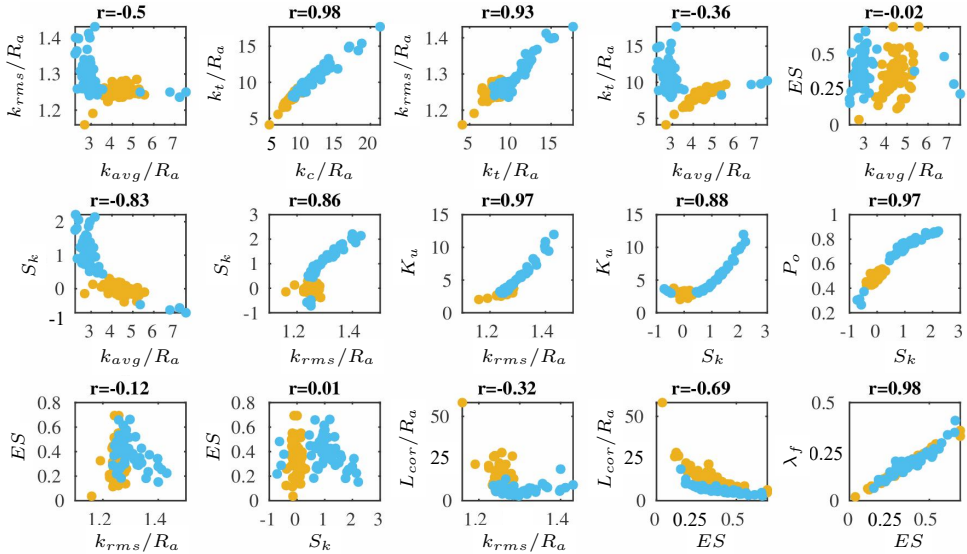

We examined the correlation between pairwise roughness parameters, as
shown in figure~\ref{fig:correlation}. The goal is to filter out
parameters containing redundant information, enabling the
identification of the subset of roughness parameters that most
effectively inform the wall model. The candidate roughness parameters
considered include height-based measures such as mean roughness height
$k_{avg}$, crest height $k_c$, mean peak-to-valley height $k_t$,
root-mean-square height $k_{rms}$, and the first-order moment of
height fluctuations $R_a$; higher-order statistical parameters
including skewness $S_k$ and kurtosis $K_u$; as well as spatial and
geometric descriptors such as effective slope $ES$, surface porosity
$P_o$, frontal solidity $\lambda_f$, and correlation length
$L_{cor}$. The results show that $k_{rms}/R_a$ is strongly correlated
with $k_t/R_a$, $k_c/R_a$, $S_k$, $K_u$, and $P_o$, while only weakly
correlated with $k_{avg}/R_a$, $ES$, and $L_{cor}$. In addition, $ES$
shows a strong correlation with $L_{cor}$ and $\lambda_f$, indicating
that they contain redundant information. Therefore, the most
representative parameters for input construction are $k_{rms}$,
$k_{avg}$, $R_a$, and $ES$.

\subsection{Incompressible channel flow database}

The incompressible channel flow database was generated in our previous
work~\citep{ma2025machine}. The roughness tiles described in
\S\ref{sec:rough_gen} are applied to the walls of open-channel
flows. Although these simulations were performed in open channels, we
use $\delta$ to denote the simulation height, for consistency with
full-height channel notation.  Turbulent open-channel flows over 32
rough surfaces were simulated using DNS at six distinct friction
Reynolds numbers: $Re_{\tau} = u_{\tau} \delta / \nu = 180, 360, 540,
720, 900$, and $1,000$, where $u_{\tau}$ is the friction velocity and
$\nu$ is the kinematic viscosity. As a result, the database contains a
total of 192 incompressible flow cases. The range of $Re_{\tau}$
values was determined based on the roughness Reynolds number $k_s^+ =
k_s u_{\tau} / \nu$, which spans from 0 to 300 and represents
conditions commonly observed in practical rough-wall flow scenarios,
encompassing both transitionally rough and fully rough regimes.  These
simulations provide only the training data for the wall model that
predicts the wall shear stress. To construct training data consistent
with the compressible flow database, the flow variables such as
density, temperature, dynamic viscosity, and thermal conductivity, are
set to constant values, and the bulk Mach number, defined as $M_b =
U_b / \sqrt{\gamma R T_w}$, is set to 0.1, where $U_b$ is the bulk
velocity, $\gamma$ is the ratio of specific heats, $R$ is the specific
gas constant, and $T_w$ is the wall temperature.  Here, ``bulk''
refers to averaging across the entire cross-section of the channel, as
well as over time.

The flow solver integrates the incompressible Navier--Stokes equations
and resolves the roughness geometry using an immersed boundary
approach based on the volume-of-fluid method~\citep{keating2004large,
  yuan2014roughness}. The simulation domain has dimensions $(L_x, L_y,
L_z) = (3\delta, \delta, \delta)$, following established guidelines
for minimal-span channel simulations~\citep{chung2015fast,
  macdonald2017minimal}. Periodic boundary conditions are applied in
the streamwise and spanwise directions, while no-slip and symmetry
boundary conditions are enforced at the bottom and top boundaries,
respectively. More details of the simulations and validation can be
found in \citet{ma2025machine}.


\subsection{Compressible channel flow database}

The training dataset also includes DNS of compressible channel flows
over irregular rough surfaces \citep{Ma_CTR_2024}. The rough-wall
cases comprise 15 Gaussian rough surfaces created by specifying a
Gaussian PDF and the power spectrum (PS) of isotropic self-affine
fractals, following the method of \citet{perez2019generating}. The
channel flows are driven by a uniform volumetric source term in the
momentum and energy equations to achieve bulk Mach numbers $M_b =
0.4,\ 0.9,\ 1.7,\ 3.3$, and bulk Reynolds numbers $Re_b = \rho_b U_b
\delta / \mu_w = 7,500,\ 10,000,\ 15,500$, where $\rho_b$ is the bulk
density and $\mu_w$ is the wall dynamic viscosity. In all simulations,
the Prandtl number is $Pr = 0.72$, and the dynamic viscosity increases
with temperature following a three-quarter power-law relationship,
i.e., $\mu \propto T^{3/4}$.

The values $M_b = 0.4,\ 0.9,\ 1.7,\ 3.3$ are selected to cover
subsonic and supersonic regimes relevant to practical conditions such
as turbine blades~\citep{nardini2024direct} and high-speed
vehicles~\citep{hollis2014distributed}. In these cases, typical flow
conditions yield $k_s^+$ values spanning nearly two orders of
magnitude, from approximately $k_s^+ \sim 1$ to
200~\citep{bons2010review, hollis2014distributed}. Accordingly, $Re_b
= 7,500,\ 10,000,\ 15,500$ are selected to achieve roughness Reynolds
numbers in the range $0 < k_s^+ < 200$, where the flow spans
transitionally rough and fully rough regimes. In total, the DNS
database includes 180 rough-wall cases and 12 reference smooth-wall
cases.

A summary of the reference simulation parameters is provided in
Table~\ref{table2}. For the sake of brevity, only smooth-wall cases
are reported. Each smooth-wall case is labeled using the convention
M$\langle M_b \rangle$--R$\langle Re_b \rangle$; for example,
M0.4--R7500 corresponds to a case with $M_b = 0.4$ and $Re_b = 7,500$.
For each smooth-wall case, there are 15 rough-wall cases at similar
$M_b$ and $Re_b$, denoted by GS\#--M$\langle M_b \rangle$--R$\langle
Re_b \rangle$, where ``\#'' is the roughness index number. These are
not shown in the table.
\begin{table}
\begin{center}
\begin{tabular}{l c c c c c c c c l}
  Case & $M_b$ & $Re_b$ & $Re_{\tau}$ & $Re_{\tau}^*$ & $-B_q$  & $\Delta x^+$ & $\Delta y_{min}^+$ & $\Delta y_{max}^+$ & $\Delta z^+$  \\
\hline
M0.4-R7500 & 0.4 & 7500 & 416 & 403 & 0.004 & 3.12 & 0.06 & 3.74 & 2.60 \\
M0.4-R10000 & 0.4 & 10000 & 550 & 532 & 0.004 & 4.13 & 0.08 & 4.95 & 3.44 \\
M0.4-R15500 & 0.4 & 15500 & 818 & 795 & 0.004 & 4.09 & 0.12 & 7.36 & 3.41 \\
M0.9-R7500 & 0.9 & 7500 & 432 & 373 & 0.016 & 3.24 & 0.06 & 3.89 & 2.70 \\
M0.9-R10000 & 0.9 & 10000 & 570 & 493 & 0.015 & 4.28 & 0.09 & 5.13 & 3.56 \\
M0.9-R10000 & 0.9 & 15500 & 862 & 744 & 0.014 & 4.31 & 0.13 & 7.76 & 3.59 \\
M1.7-R7500 & 1.7 & 7500 & 490 & 309 & 0.06 & 3.68 & 0.07 & 4.14 & 3.06 \\
M1.7-R10000 & 1.7 & 10000 & 660 & 400 & 0.05 & 4.95 & 0.10 & 5.94 & 4.13 \\
M1.7-R15500 & 1.7 & 15500 & 970 & 600 & 0.05 & 4.85 & 0.15 & 8.73 & 4.04 \\
M3.3-R7500 & 3.3 & 7500 & 671 & 194 & 0.19 & 5.03 & 0.10 & 6.04 & 4.19 \\
M3.3-R10000 & 3.3 & 10000 & 869 & 251 & 0.16 & 4.35 & 0.13 & 7.82 & 3.62 \\
M3.3-R15500 & 3.3 & 15500 & 1344 & 384 & 0.16 & 5.04 & 0.20 & 7.53 & 4.20 \\
\hline
\end{tabular}
\caption{Simulation parameters for the reference smooth-wall cases in
  the DNS database. Although not shown, for each reference case, 15
  Gaussian-roughness simulations are performed, labeled as
  \mbox{GS\#--M$\langle M_b\rangle$--R$\langle Re_b\rangle$}. The
  friction Reynolds number is $Re_{\tau}=\rho_w u_{\tau}\delta/\mu_w$,
  with $u_{\tau}=\sqrt{\tau_w/\rho_w}$. The transformed friction
  Reynolds number is $Re_{\tau}^{*}=\rho_c u_{\tau}^{*}\delta/\mu_c$
  \citep{coleman1995numerical}, where
  $u_{\tau}^{*}=\sqrt{\tau_w/\rho_c}$ and the subscript $c$ denotes
  mean quantities at the channel centerline. The wall heat-transfer
  rate is $B_q=q_w/(\rho_w c_p u_{\tau} T_w)$. Superscript $+$ denotes
  viscous units based on $u_{\tau}$ and $\nu_w$, where $\nu_w$ is the
  kinematic viscosity at the wall. \label{table2}}
\end{center}
\end{table}

\begin{figure}
\begin{center}
    \includegraphics[width=1\textwidth,trim={0.0cm 4.0cm 0.0cm
        2.0cm},clip]{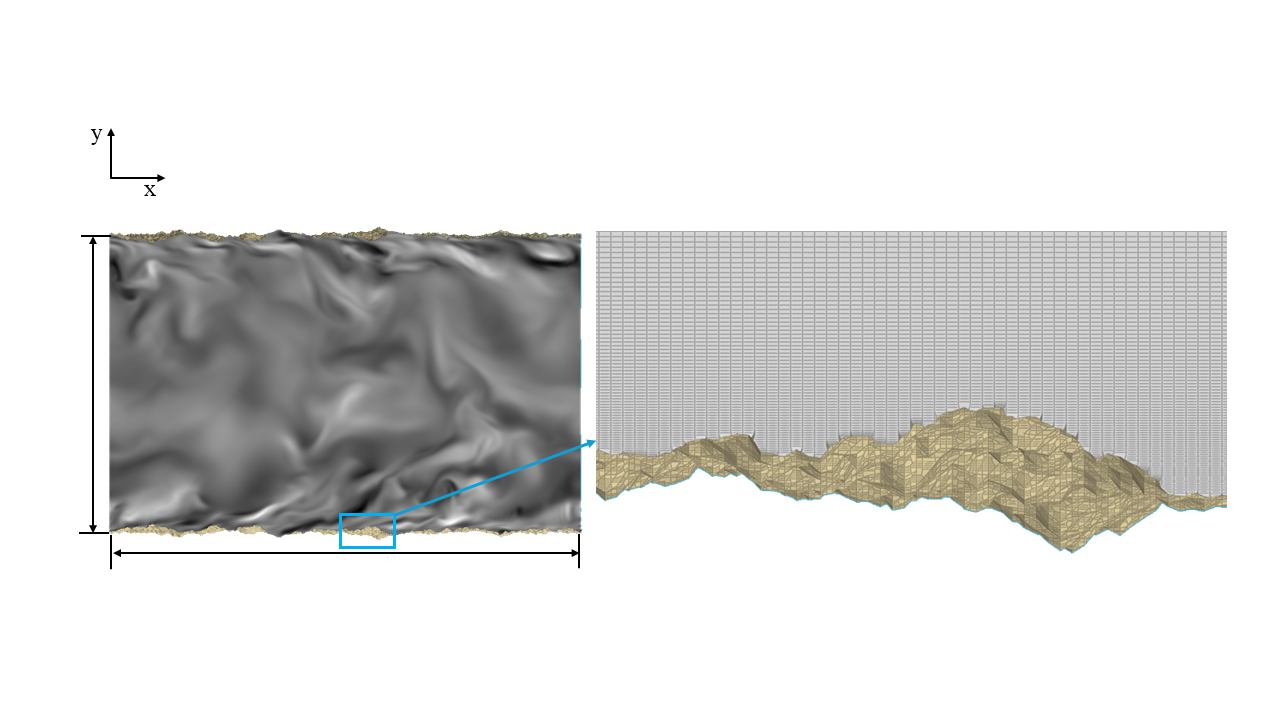}
        \put(-460,62){$L_y=2\delta$}
        \put(-350,-7){$L_x=3\delta$}
    \caption{Instantaneous spanwise velocity normalized by the maximum
      spanwise velocity $w/w_\text{max}$ for DNS of a minimal-span
      channel flow over a rough wall. The contour legend is from -1
      (black) to 1 (white). The case shown is GS1-M1.7-R10000. The
      zoom-in view shows the grid in the near-wall region. The wetted
      rough surfaces are colored in dark yellow. \label{fig:DNS}}
\end{center}
\end{figure}

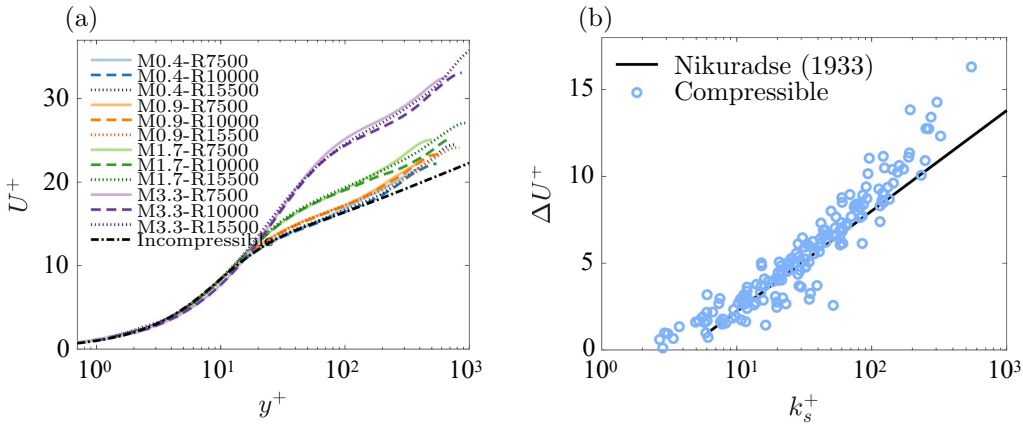
\begin{figure}
\centering
\includegraphics[width=70mm,trim={5cm 0cm 3cm 0cm},clip]{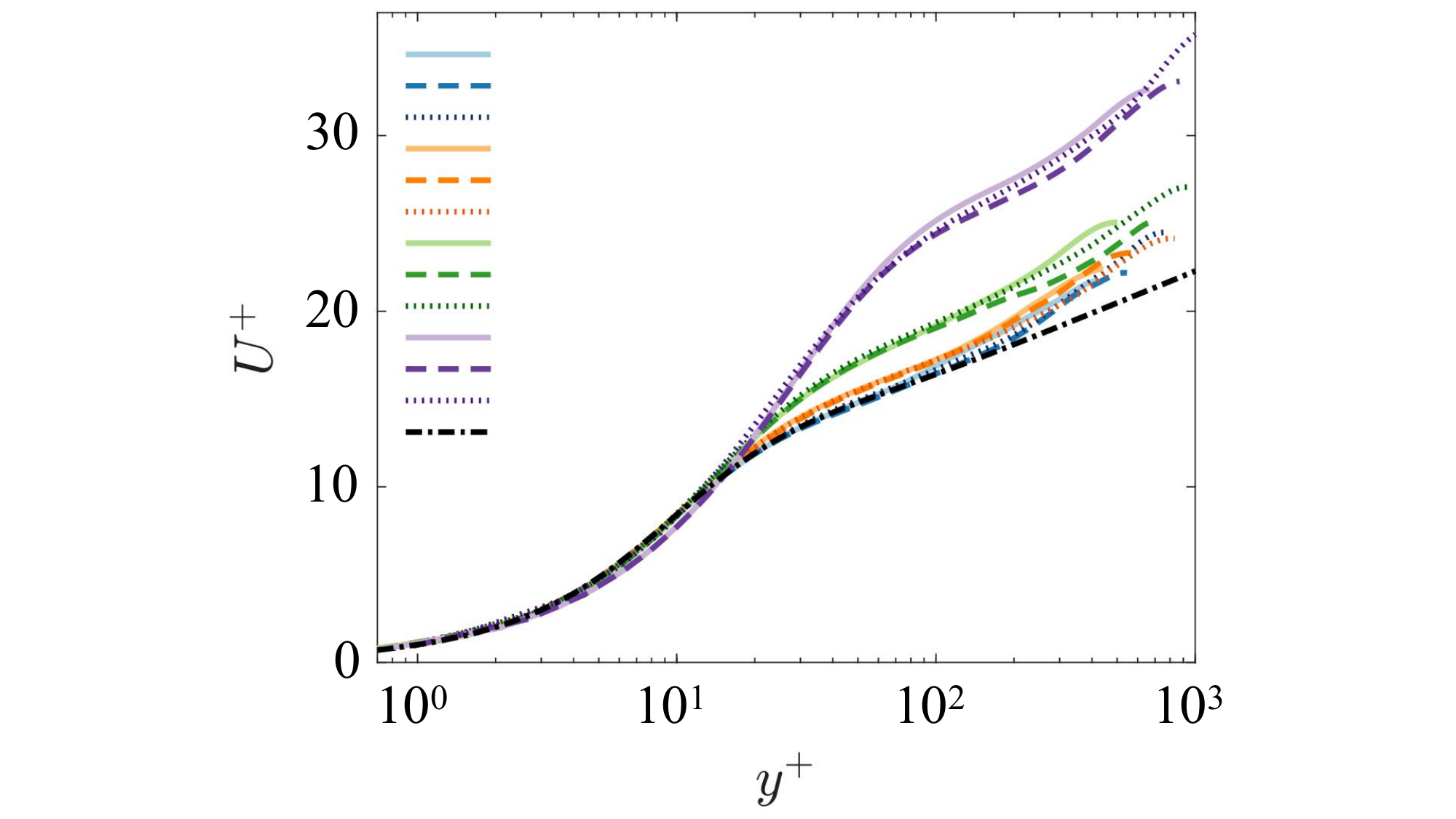}
\put(-175,150){(a)}
\put(-148,134.5){\scriptsize{M0.4-R7500}}
\put(-148,129){\scriptsize{M0.4-R10000}}
\put(-148,123.5){\scriptsize{M0.4-R15500}}
\put(-148,117.5){\scriptsize{M0.9-R7500}}
\put(-148,112){\scriptsize{M0.9-R10000}}
\put(-148,106.5){\scriptsize{M0.9-R15500}}
\put(-148,101){\scriptsize{M1.7-R7500}}
\put(-148,95.5){\scriptsize{M1.7-R10000}}
\put(-148,90){\scriptsize{M1.7-R15500}}
\put(-148,84){\scriptsize{M3.3-R7500}}
\put(-148,78){\scriptsize{M3.3-R10000}}
\put(-148,72){\scriptsize{M3.3-R15500}}
\put(-148,67){\scriptsize{Incompressible}}
\includegraphics[width=70mm,trim={5cm 0cm 3cm 0cm},clip]{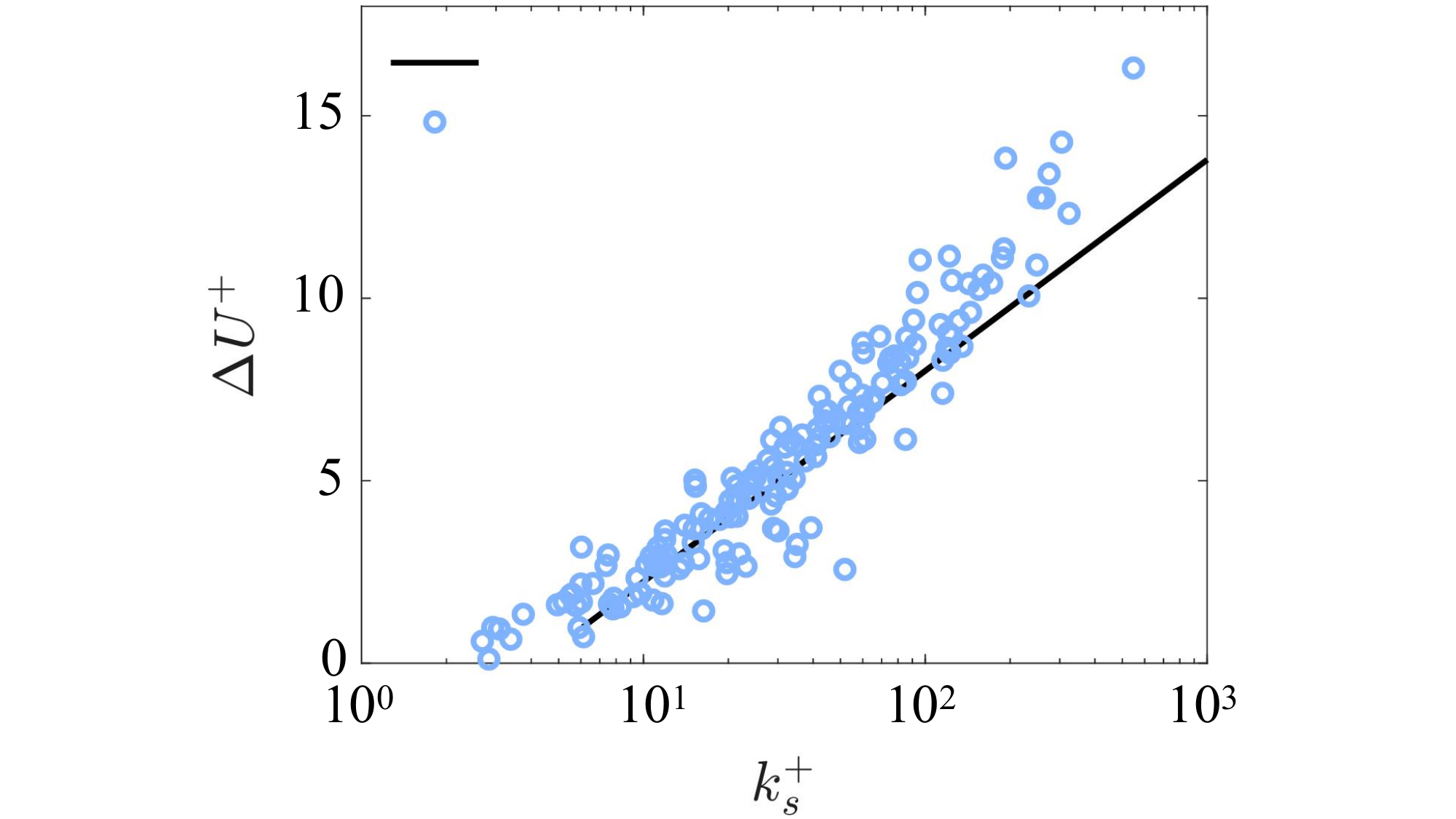}
\put(-182,150){(b)}
\put(-146,132){\cite{nikuradse1933laws}}
\put(-146,122){Compressible}
\caption{(a) Streamwise mean velocity profiles of smooth-wall
  reference cases, with a comparison to the incompressible channel
  flow DNS data from \cite{lee2015direct}. (b) Roughness function
  $\Delta U^+$ as a function of $k_s^+$ for current compressible
  channel flows with rough walls. The line denotes the Nikuradse's law
  for incompressible rough-wall turbulent flow
  \citep{nikuradse1933laws}. }
\label{fig:Umean}
\end{figure}

The DNS is conducted with the explicit, unstructured, finite-volume
solver \textit{charLES}, developed by Cascade Technology, which
integrates the compressible Navier--Stokes equations. The solver has
been previously used and validated in high-speed
flows~\citep{fu2021shock}. A sample visualization of the flow field
and near-wall mesh is presented in figure~\ref{fig:DNS}. The same
minimal-span domain size as in the incompressible cases is employed,
with $(L_x, L_y, L_z) = (3\delta, \delta, \delta)$. Uniform grid
spacing is applied in the streamwise and spanwise directions, while a
hyperbolic tangent stretching function is used to cluster points near
the wall in the wall-normal direction. The grid resolution for
smooth-wall cases is determined based on a grid convergence study. The
rough-wall cases at the same $M_b$ and $Re_b$ are refined to achieve a
grid resolution with $\Delta x^+ < 5$, $\Delta z^+ < 4.2$, and $\Delta
y_{\min}^+ < 0.2$, which is comparable to the DNS of supersonic flows
over roughness conducted by \citet{modesti2022direct} and
\citet{aghaei2023supersonic}. Figure \ref{fig:Umean} shows the
streamwise mean velocity profiles of smooth reference cases and the
roughness function against $k_s^+$ for each compressible rough case.
%

Validation of the minimal-span configuration against full-span
simulations for both supersonic smooth- and rough-wall channel flows
is presented in Appendix~\ref{appB}. For the smooth-wall case, our DNS
results show agreement with the reference DNS of
\citet{trettel2016mean}, confirming both the accuracy of the flow
solver and the validity of the minimal-span channel
approach. Moreover, comparisons between minimal- and full-span
rough-wall channel flows demonstrate agreement in the mean velocity
profiles, with deviations within 3\% for $y/\delta = 0.15$. Based on
these results, the DNS mean data for $y/\delta < 0.15$ are deemed
reliable and used to provide flow information for training the
wall-model development.

\section{Formulation}
\label{formulation}

\subsection{Model overview}

The wall model follows the building-block flow model principle
introduced by \citet{lozano2023machine}. The central assumption is
that a set of simple, canonical flows can capture the essential
physics needed to accurately predict wall shear stress and heat flux
in more complex flow scenarios. Accordingly, it is postulated that a
finite set of \emph{building-block flows} (used as training data)
encapsulates the relevant physics required to construct generalizable
wall models. The building-block-flow concept has been applied to both
subgrid-scale (SGS) models and wall models and the reader is referred
to previous work for additional details~\citep{lozano2020self,
  arranz2024building}.
\begin{figure}
\begin{center}
  \includegraphics[width=1\textwidth]{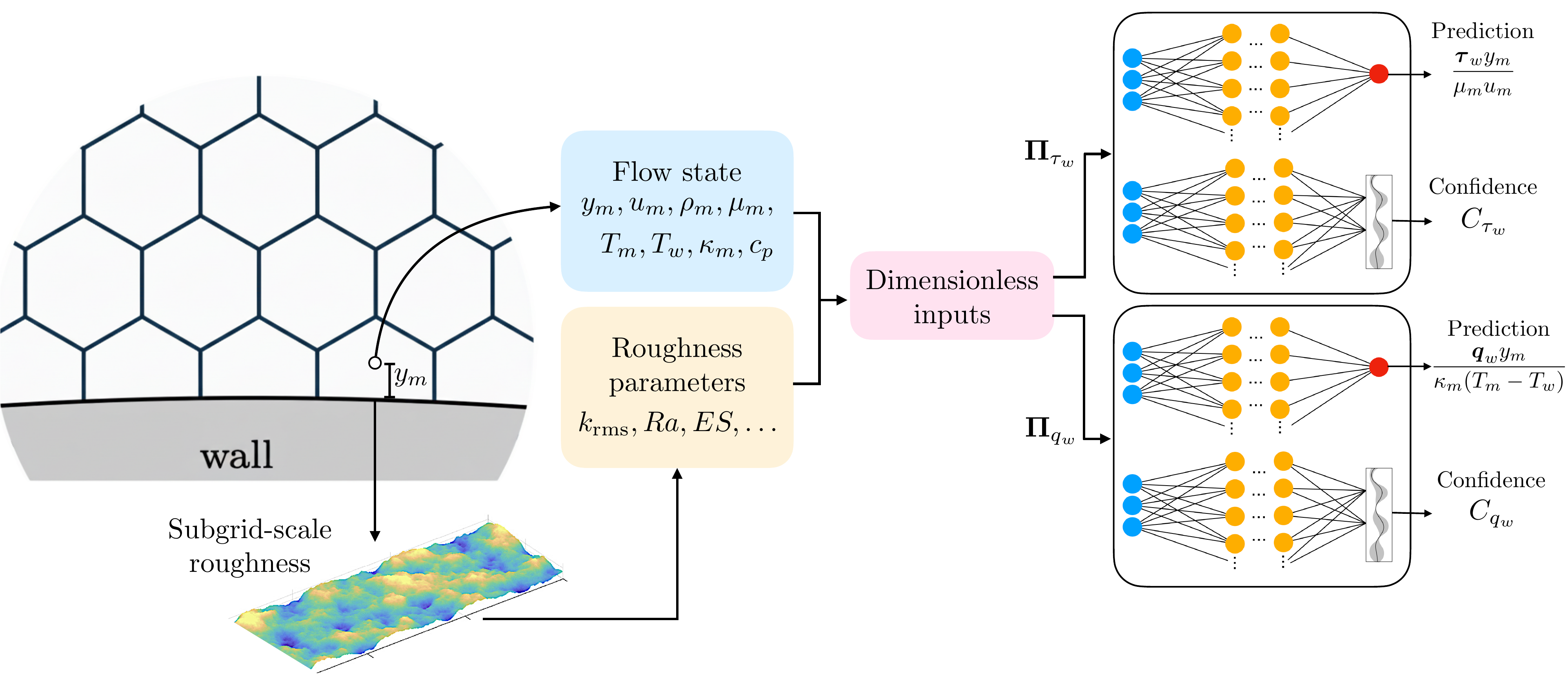}
  \caption{Overview of the proposed wall model (BFWM-rough-v2). The
    flow state at the wall and in the first off-wall control volume,
    together with statistical roughness parameters, are used to
    construct dimensionless inputs to the wall model. The model
    outputs the dimensionless wall-shear stress and wall heat flux,
    each accompanied by a confidence score. The wall-model mapping is
    represented with an FNN, while the confidence score in the
    prediction is quantified using an SNGP. \label{fig:overview}}
\end{center}
\end{figure}

An overview of the proposed wall model is shown in
figure~\ref{fig:overview}.  The approach is implemented using two
feedforward neural networks (FNNs) that predict the dimensionless wall
shear stress and wall heat flux, respectively:
\begin{equation}
  \frac{\boldsymbol{\tau}_w y_m}{\mu_m u_m} = \boldsymbol{\mathrm{ANN}}_{\tau_w}(\boldsymbol{\Pi}_{\tau_w}), 
  \quad
  \frac{\boldsymbol{q}_w y_m}{\kappa_m(T_m-T_w)} = \boldsymbol{\mathrm{ANN}}_{q_w}(\boldsymbol{\Pi}_{q_w}),
\end{equation}
where $\boldsymbol{\Pi}_{\tau_w}$ and $\boldsymbol{\Pi}_{q_w}$ are
dimensionless inputs constructed from dimensional variables from both
the wall and the first off-wall control volume located at height $y_m$
(i.e., the matching location hereafter denoted by the subscript
$m$). These variables include the density $\rho_m$, the magnitude of
the wall-parallel velocity $u_m$, temperature $T_m$, wall temperature
$T_w$, dynamic viscosity $\mu_m$, thermal conductivity $\kappa_m$, and
specific heat capacity $c_p$. The mean flow state, obtained through time and
spatial averaging in the wall normal range $0.02 < y/\delta < 0.15$,
is extracted from the DNS and used as input to the wall model. In addition, roughness parameters such
as the root-mean-square roughness height $k_{rms}$, arithmetic mean
roughness $R_a$, and effective slope $ES$, are also provided to the
wall model. An information-theoretic dimensionless learning
method~\citep{yuan2025dimensionless} is employed to identify the most
predictive inputs, as described in \S\ref{sec:input}. The final input
set used by the wall model is selected based on the minimum $L_2$ norm
error achieved during the training process detailed in
\S\ref{sec:training}.  To quantify model uncertainty and provide a
confidence estimate, a Spectral-normalized Neural Gaussian Process
(SNGP) is integrated into the model, as described in
\S\ref{sec:confidence}.

\subsection{Modeling assumptions}
\label{sec:assumption}

We summarize the primary modeling assumptions underlying the
BFWM-rough-v2 beyond those from WMLES.
\begin{itemize}
\setlength\itemsep{0em}
\item[(i)] \textit{Building-block flow assumption:} A finite
  collection of canonical turbulent channel flows over rough surfaces,
  covering a range of Mach numbers, Reynolds numbers, and roughness
  geometries, contains the essential physics required to construct a
  generalizable wall model for predicting both wall shear stress and
  wall heat flux.
\item[(ii)] \textit{Quasi-equilibrium assumption:} The near-wall
  region of rough-wall flows is assumed to be in a quasi-equilibrium
  state, i.e., the characteristic time scales of near-wall
  subgrid-scale motions affecting wall fluxes are much shorter than
  those of the grid-resolved large-scale motions.
\item[(iii)] \textit{Space/time locality assumption:} The
  local-in-space and local-in-time variables sampled at the first
  wall-normal control volume and at the wall (i.e., $y_m$, $u_m$,
  $\rho_m$, $T_m$, $T_w$, $\mu_m$, $\kappa_m$ and $c_p$) are assumed
  to contain sufficient information to predict the wall fluxes under
  assumption (ii).
\item[(iv)] \textit{Subgrid-scale roughness assumption:} Surface
  roughness is treated as a subgrid-scale feature whose primary effect
  on the resolved flow is represented via modified wall boundary
  conditions for momentum and energy (i.e., wall-shear stress and wall
  heat flux).
\item[(v)] \textit{Statistical roughness description assumption:} A
  finite set of statistical geometric parameters of the roughness (see
  Table~\ref{tab:roughness_para}) is assumed to be sufficient to
  characterize the influence of surface geometry on both wall momentum
  and heat transfer.
\item[(vi)] \textit{Velocity--shear and temperature--heat-flux
  alignment assumption:} The wall shear stress vector is assumed to be
  aligned with the local wall-parallel velocity at the first off-wall
  control volume in the WMLES grid. Similarly, the wall heat flux
  vector is assumed to be aligned with (and opposite to) the local
  temperature wall-normal gradient at the same location.
\item[(vii)] \textit{Dimesinonless-form generalizability assumption:}
  The dimensionless form of the inputs and outputs, identified by
  maximizing predictability on the training set, remains valid across
  arbitrary flow configurations.
\item[(viii)] \textit{Mean-flow training data assumption:}
  Input–output samples constructed from mean velocity and temperature
  profiles, along with mean wall shear stress and wall heat flux, are
  assumed to provide sufficient information to train an accurate and
  generalizable wall model for compressible rough-wall flows.
\end{itemize}

\subsection{Dimensionless learning for input variables}
\label{sec:input}

The model is formulated using dimensionless input and output variables
to ensure dimensional homogeneity. The objective is to identify the
minimal set of dimensionless input variables that retains strong
predictive capability for the target output. The dimensional variables
used to construct the dimensionless inputs include both flow and
roughness information, denoted as $\mathbf{q} =
[\mathbf{q}_\text{flow};\ \mathbf{q}_\text{roughness}]$, where
$\mathbf{q}_\text{flow} =
[y_m,\ u_m,\ \rho_m,\ T_m,\ T_w,\ \mu_m,\ \kappa_m,\ c_p]^T$.  The
vector of roughness variables, $\mathbf{q}_\text{roughness}$, is left
undefined at this stage. It may include parameters such as $k_{avg}$,
$k_{rms}$, $R_a$, and $ES$ (as discussed in
\S\ref{sec:rough_gen}). The optimal combination of these roughness
parameters will be determined later. The goal is to minimize the
number of geometric roughness features used by the model, as these
quantities may not be readily available or may be difficult to
quantify accurately in practical applications.

We used the dimensionless-learning method based on information
developed by \cite{yuan2025dimensionless} to discover the most
predictive dimensionless inputs for the wall model. We aim to find $p$
dimensionless inputs that provide the highest predictive power of the
dimensionless output $\Pi_o$.  First, consider the generic
dimensionless input $\mathbf{\Pi}= [\Pi_1, \ldots, \Pi_p]$. The
Buckingham-$\pi$ theorem is applied to derive dimensionless groups
$\boldsymbol{\Pi}$. Each dimensionless number $\Pi_i$ takes the form
$\Pi_i = \prod_{j=1}^n q_j^{a_{ij}} \equiv \mathbf{q}^{\mathbf{a}_i}$,
where $\mathbf{a}_i = [a_{i1}, a_{i2}, \ldots, a_{in}]^T$ is the
exponent vector defining $\Pi_i$. These vectors are determined by
solving the homogeneous linear system $\mathbf{D} \mathbf{a}_i =
\mathbf{0}$, where $\mathbf{D}$ is the dimension matrix containing the
fundamental unit powers of each physical quantity $q_j$ (i.e.,
$\mathbf{D} = [\mathbf{d}_1, \mathbf{d}_2, \ldots, \mathbf{d}_n]$ with
$\mathbf{d}_j$ being the dimension vector of $q_j$). For example, if
$q_1 = u$ (velocity) has dimensions $[\text{length}]^1 [\text{mass}]^0
[\text{time}]^{-1} [\text{temperature}]^0$, then $\mathbf{d}_1 = [1,
  0, -1, 0]^T$. The general solution to $\mathbf{D} \mathbf{a}_i =
\mathbf{0}$ can be written as a linear combination of basis vectors of
the null space of $\mathbf{D}$: $ \mathbf{a}_i= \sum_{j=1}^{n-k} c_{i
  j} \mathbf{w}_j = \mathbf{W} \mathbf{c}_i, $ where $\mathbf{W} =
[\mathbf{w}_1, \mathbf{w}_2, \ldots, \mathbf{w}_{n-k}]$ spans the null
space, $\mathbf{c}_i = [c_{i1}, c_{i2}, \ldots, c_{i(n-k)}]^T$ are the
coefficients and $k=4$ is the number of fundamental units (i.e.,
length, mass, time, and temperature). Therefore, the resulting
dimensionless input variables are given by
$\boldsymbol{\Pi(\mathbf{C})} \equiv \mathbf{q}^{\mathbf{W
    C}}=\left[\mathbf{q}^{\mathbf{W} \mathbf{c}_1},
  \mathbf{q}^{\mathbf{W} \mathbf{c}_2}, \ldots, \mathbf{q}^{\mathbf{W}
    \mathbf{c}_p}\right],$ where the matrix $\mathbf{C} =
[\mathbf{c}_1, \mathbf{c}_2, \ldots, \mathbf{c}_p]$ can be freely
chosen to define different candidate $\Pi$ groups.

Although infinitely many dimensionless input combinations can be
constructed, not all of them are optimal for predicting the
dimensionless output. Our goal is to identify the input combination
that offers the highest predictive power. Let the prediction error of
a model $\hat{f}$, which estimates the output $\Pi_o$ from input
features $\boldsymbol{\Pi}$, be denoted by $
\epsilon\left(\boldsymbol{\Pi},
\hat{f}\right)=\left\|\Pi_o-\hat{\Pi}_o\right\|_2$ where $\hat{\Pi}_o
= \hat{f}(\boldsymbol{\Pi})$. The minimum achievable prediction error
across all models $\hat{f}$ is lower bounded by the
information-theoretic irreducible error~\citep{yuan2025dimensionless}:
\begin{equation}
\epsilon_{\min}\left(\boldsymbol{\Pi}\right)=\min _{\hat{f}}
\epsilon\left(\boldsymbol{\Pi}, \hat{f}\right) \geq \frac{1}{\sqrt{2 e
    \pi}} e^{ h(\Pi_o) - I\left(\Pi_o ; \boldsymbol{\Pi}\right) },
\end{equation}
where $h(\Pi_o)$ is the differential entropy of the output and
$I(\Pi_o ; \boldsymbol{\Pi})$ denotes the mutual information between
$\Pi_o$ and the inputs $\boldsymbol{\Pi}$ \citep{cover1999elements}.
The differential entropy quantifies the expected uncertainty
associated with the output variable $\Pi_o$, and is defined as
\begin{equation}
h(\Pi_o) = - \int f_{\Pi_o}(\pi_o) \, \ln f_{\Pi_o}(\pi_o) \, \mathrm{d}\pi_o,
\end{equation}
where $f_{\Pi_o}$ denotes the probability density function of $\Pi_o$.
The mutual information measures the amount of shared information
between the output and the input variables, and is given by
\begin{equation}
I(\Pi_o; \boldsymbol{\Pi}) =
\int f_{\boldsymbol{\Pi}, \Pi_o}(\boldsymbol{\pi}, \pi_o)
\ln\!\left(
\frac{f_{\boldsymbol{\Pi}, \Pi_o}(\boldsymbol{\pi}, \pi_o)}
     {f_{\boldsymbol{\Pi}}(\boldsymbol{\pi}) \,
      f_{\Pi_o}(\pi_o)}
\right)
\mathrm{d}\boldsymbol{\pi} \, \mathrm{d}\pi_o,
\end{equation}
where $f_{\boldsymbol{\Pi}, \Pi_o}$ is the joint probability density
function, and $f_{\boldsymbol{\Pi}}$, $f_{\Pi_o}$ are the
corresponding marginal densities of the input and output variables,
respectively.  Since $h(\Pi_o)$ is fixed for a given output, the best
input candidate is the one that maximizes the mutual information:
\begin{equation}
 \mathbf{\Pi}(\mathbf{C}) = \arg \max _{\mathbf{\Pi}} I\left(\Pi_o;
\boldsymbol{\Pi}(\mathbf{C})\right) = \arg \max _{\mathbf{C}} I\left(\Pi_o;
\boldsymbol{\Pi}(\mathbf{C})\right).
\end{equation}
An advantage of using this method is that it provides an intrinsic
predictability limit given the input and output variables, regardless
of the modeling approach, allowing us to identify the most predictive
inputs and outputs without the need to construct an explicit model.

This optimization problem can be efficiently solved using the
Covariance Matrix Adaptation Evolution Strategy (CMA-ES) algorithm
proposed by \cite{hansen2003reducing}. The CMA-ES algorithm was used
to find optimal combinations of inputs, with a population size of 50,
coefficient bounds set to $[-2,2]$, and a maximum of 50,000
iterations. The initial standard deviation for the search was set to
0.5.



To determine the minimum number of roughness parameters required in
the wall-model inputs, we perform a dimensionless learning analysis by
systematically varying both the combinations of roughness parameters
and the number of dimensionless input variable. The results are
summarized in Tables~\ref{tab:RP} and~\ref{tab:qwall} for wall shear
stress and wall heat flux predictions, respectively. These tables
report $(1-\tilde{\epsilon}_{LB})$, where $\tilde{\epsilon}_{LB}$ is
the normalized irreducible error defined
as~\citep{yuan2025dimensionless} \[ \tilde{\epsilon}_{LB} =
e^{-I(\Pi_0;\boldsymbol{\Pi})} \in [0,1].\] The quantity
$(1-\tilde{\epsilon}_{LB})$ measures the \emph{degree of
predictability} (akin to a generalized coefficient of determination)
and represents the fraction of the target variable that can, in
principle, be recovered from the available input information: a value
of 0 indicates total unpredictability, whereas a value of 1
corresponds to perfect predictability.
\begin{table}
    \centering
    \begin{tabular}{c|l|cc|cc|cc}
        \toprule
        Flow quantity & Roughness parameter & \multicolumn{2}{c|}{2 inputs} & \multicolumn{2}{c|}{3 inputs} & \multicolumn{2}{c}{4 inputs} \\
        $\mathbf{q}_\text{flow}$ & $\mathbf{q}_\text{roughness}$ & $1-\tilde{\epsilon}_{LB}$ & $\epsilon$ (\%) & $1-\tilde{\epsilon}_{LB}$ & $\epsilon$ (\%) & $1-\tilde{\epsilon}_{LB}$ & $\epsilon (\%) $ \\ 
        \midrule   
        & $k_{rms}$, $R_a$                 & 0.6732 & 15.78   & 0.6758 & 15.65  & 0.6790 &  14.92  \\
        & $k_{avg}$, $R_a$                 & 0.6875 & 14.08 & 0.6882 & 11.87 & 0.6897 &  11.51 \\
        & $k_{avg}$, $k_{rms}$             & 0.6854 & 14.23 & 0.6878 & 12.95 & 0.7356  & 12.48  \\
         & $k_{avg}$, $k_{rms}$, $R_a$       & 0.6860 & 14.06 & 0.6882 & 11.66 & 0.6892 & 11.49   \\
        & $k_{avg}$, $ES$                  & 0.6925 & 10.74  & 0.7058 & 10.09 & 0.7102  & 9.30   \\
        \multicolumn{1}{l|}{$y_m, u_m, \rho_m, T_m, T_w, \mu_m, \kappa_m, c_p$} & $k_{rms}$, $ES$ & 0.7120 & 8.24  & 0.7275 & 5.37 & 0.7528 & 4.49 \\
        & $R_a$, $ES$                      & 0.7135 & 7.82 & 0.7302 & 5.30 & 0.7511 & 4.91  \\
        & $k_{rms}$, $k_{avg}$, $ES$       & 0.7204 & 7.47 & 0.7352 & 5.25 & 0.7519 &  4.85 \\
        & $k_{rms}$, $R_a$, $ES$           & 0.7187 & 7.59 & 0.7253 & 5.63 & 0.7521 & 4.81   \\
        & $k_{avg}$, $R_a$, $ES$           & 0.7192 & 7.57 & 0.7350 & 5.25 & {\color{green}{0.7536}} &  {\color{green}{4.07}} \\
        & $k_{avg}$, $k_{rms}$, $R_a$, $ES$ & 0.7199 & 7.49 & 0.7365 & 5.24  & 0.7539 & 4.04  \\
        & All parameters                   & 0.7238 & 7.43 & 0.7386 & 5.21  & 0.7545 & 4.02  \\
        \bottomrule
    \end{tabular}
    \caption{Dimensionless-learning analysis for selecting inputs to
      the wall-shear-stress model. Table entries report the degree of
      predictability, $(1-\tilde{\epsilon}_{LB})$, and the minimum
      $L_2$-norm error, $\epsilon$ (\%), achieved by an FNN for
      different combinations of roughness parameters included in the
      input. ``All parameters'' indicates that all candidate roughness
      parameters listed in Table~\ref{tab:roughness_para} are
      included. The configuration highlighted in green, yielding
      $\epsilon=4.07\%$, is selected as a compromise between
      predictability and the number of required roughness inputs.  }
    \label{tab:RP}
\end{table}
%
\begin{table}
    \centering
    \begin{tabular}{c|l|cc|cc|cc}
        \toprule
        Flow quantity & Roughness parameter & \multicolumn{2}{c|}{2 inputs} & \multicolumn{2}{c|}{3 inputs} & \multicolumn{2}{c}{4 inputs} \\
        $\mathbf{q}_\text{flow}$ & $\mathbf{q}_\text{roughness}$ & $1-\tilde{\epsilon}_{LB}$ & $\epsilon$ (\%) & $1-\tilde{\epsilon}_{LB}$ & $\epsilon$ (\%) & $1-\tilde{\epsilon}_{LB}$ & $\epsilon  (\%)$ \\ 
        \midrule
        & $k_{rms}$, $R_a$                  & 0.7262 & 6.37  & 0.7323 & 5.56 & 0.7440 & 5.17 \\
        & $k_{avg}$, $R_a$                  & 0.7281 & 6.14  & 0.7311 & 5.71 & 0.7471 & 5.57 \\
        & $k_{avg}$, $k_{rms}$              & 0.7272 & 6.17  & 0.7330 & 5.59 & 0.7433 & 5.40 \\
        & $k_{avg}$, $k_{rms}$, $R_a$        & 0.7292 & 6.08  & 0.7384 & 5.26 & 0.7525 & 5.12 \\
        & $k_{avg}$, $ES$                   & 0.7123 & 7.06  & 0.7261 & 6.91 & 0.7482 &  6.79\\
        \multicolumn{1}{l|}{$y_m, u_m, \rho_m, T_m, T_w, \mu_m, \kappa_m, c_p$} & $k_{rms}$, $ES$                   & 0.7344 & 5.44  & 0.7521 & 3.94 & {\color{green}0.7753} & {\color{green}2.89} \\
        & $R_a$, $ES$                       & 0.7330 & 5.54  & 0.7542 & 3.91 & 0.7721 & 3.12 \\
        & $k_{rms}$, $k_{avg}$, $ES$        & 0.7360 & 5.32 & 0.7562 & 3.90 & 0.7749 & 2.83 \\
        & $k_{rms}$, $R_a$, $ES$            & 0.7323 & 5.54  & 0.7553 & 3.88 & 0.7802 & 2.85 \\
        & $k_{avg}$, $R_a$, $ES$            & 0.7351 & 5.47  & 0.7570 & 3.85 & 0.7783 & 3.02\\
        & $k_{avg}$, $k_{rms}$, $R_a$, $ES$  & 0.7375 & 5.28  & 0.7610 & 3.79 & 0.7853 & 2.74 \\
        & All parameters                    & 0.7384 & 5.26  & 0.7641 & 3.77 & 0.7884 & 2.62 \\
        \bottomrule
    \end{tabular}
    \caption{Dimensionless-learning analysis for selecting inputs to
      the wall-heat-flux model. Table entries report the degree of
      predictability, $(1-\tilde{\epsilon}_{LB})$, and the minimum
      $L_2$-norm error, $\epsilon$ (\%), achieved by an FNN for
      different combinations of roughness parameters included in the
      input. ``All parameters'' indicates that all candidate roughness
      parameters listed in Table~\ref{tab:roughness_para} are
      included. The configuration highlighted in green, yielding
      $\epsilon=2.89\%$, is selected as a compromise between
      predictability and the number of required roughness inputs.}
    \label{tab:qwall}
\end{table}


Given a training database of size $O(10^4)$, the number of
dimensionless inputs that can be reliably tested is limited to four,
to ensure accurate estimation of the mutual information. Therefore,
models are evaluated for $p = 1, 2, 3$, and $4$. The results for $p =
1$ are omitted, as the analysis indicates that a single input yields
poor scaling; at least two input parameters are required to capture
meaningful relationships. An increase in $p$ generally improves the
degree of predictability across all parameter combinations, reflecting
higher mutual information $I(\Pi_0 ; \boldsymbol{\Pi})$, as shown in
Tables~\ref{tab:RP} and~\ref{tab:qwall}. We select the dimensionless
input sets with $p = 4$ for both wall shear stress and wall heat
flux. For wall shear stress, the chosen inputs consist of $[k_{avg},
  R_a, ES]^T$ combined with $\mathbf{q}_{\text{flow}}$, while for wall
heat flux, the selected model uses $[k_{rms}, ES]^T$ together with
$\mathbf{q}_{\text{flow}}$; these configurations are highlighted in
green in Tables~\ref{tab:RP} and~\ref{tab:qwall}. Although including
additional roughness descriptors can offer marginal improvements, the
gains are minor and do not justify the increased model complexity.
  
The optimal dimensionless input combinations for predicting wall-shear
stress are:
\[
\scalebox{0.9}{$
\begin{aligned}
\mathbf{\Pi}_{\tau_w} &= \arg \max_{\mathbf{\Pi}} I\left(\frac{\tau_w y_m}{\mu_m u_m}; \boldsymbol{\Pi}\right) = \left[ {\Pi}_{\tau_w,1}, {\Pi}_{\tau_w,2}, {\Pi}_{\tau_w,3}, {\Pi}_{\tau_w,4} \right],
\end{aligned}
$}
\]
where
\[
\scalebox{0.9}{$
\begin{aligned}
{\Pi}_{\tau_w,1} &= 
\frac{\left(\frac{T_m}{T_w}\right)^{0.2} Re_m^{0.7} Pr_m^{0.1} 
\left(\frac{y_m}{Ra}\right)^{0.1}}
{M_m^{1.9} ES^{0.1}}, 
&\quad
{\Pi}_{\tau_w,2} &= 
\frac{Re_m^{0.4} M_m^{0.2} Pr_m 
\left(\frac{y_m}{k_{avg}}\right)^{0.1}}
{\left(\frac{T_m}{T_w}\right)
 \left(\frac{y_m}{Ra}\right)^{0.5} ES^{0.4}}, \\[1pt]
{\Pi}_{\tau_w,3} &= 
\frac{\left(\frac{T_m}{T_w}\right) Re_m^{1.2} Pr_m^{0.9} ES^{0.2}}
{\left(\frac{y_m}{Ra}\right)^{0.1} M_m^{0.1}},
&\quad
{\Pi}_{\tau_w,4} &= 
\frac{\left(\frac{T_m}{T_w}\right)^{0.3} Re_m^{0.6} Pr_m^{0.5} 
\left(\frac{y_m}{k_{avg}}\right)^{0.2} ES^{0.5}}
{\left(\frac{y_m}{Ra}\right)}.
\end{aligned}
$}
\]
and for wall heat flux:
\[
\scalebox{0.9}{$
\begin{aligned}
\mathbf{\Pi}_{q_w} &= \arg \max_{\mathbf{\Pi}} I\left(\frac{q_w y_m}{\kappa_m(T_m-T_w)}; \boldsymbol{\Pi}\right) = \left[ {\Pi}_{q_w,1}, {\Pi}_{q_w,2}, {\Pi}_{q_w,3}, {\Pi}_{q_w,4} \right],
\end{aligned}
$}
\]
where
\[
\scalebox{0.9}{$
\begin{aligned}
{\Pi}_{q_w,1} &= 
\frac{\left(\frac{T_m}{T_w}\right)^{0.2} Re_m^{0.7} M_m^{0.2} Pr_m^{0.3} ES^{0.3}}{\left(\frac{y_m}{k_{rms}}\right)},
&\quad
{\Pi}_{q_w,2} &= 
\left(\frac{T_m}{T_w}\right)^{0.9} Re_m^{1.8} M_m^{0.3} Pr_m^{1.0} \left(\frac{y_m}{k_{rms}}\right)^{0.1} ES^{0.2}, \\[6pt]
{\Pi}_{q_w,3} &=
\frac{ Re_m^{0.7} \left(\frac{y_m}{k_{rms}}\right)^{0.1} ES^{0.1}}{\left(\frac{T_m}{T_w}\right) M_m^{0.2} Pr_m^{0.3} },
&\quad
{\Pi}_{q_w,4} &=
\frac{\left(\frac{T_m}{T_w}\right)^{0.3} Re_m^{0.5} Pr_m } {\left(\frac{y_m}{k_{rms}}\right)^{0.5} ES^{0.7}}.
\end{aligned}
$}
\]
Here, $M_m=u_m/\sqrt{\gamma R T_m}$, $Pr_m=c_p \mu_m/\kappa_m$, and
$Re_m=\rho_m u_m y_m/\mu_m$ are the local Mach, Prandtl, and Reynolds
numbers evaluated at the first grid point off the wall.  Note that the
temperature is assumed to be expressed on an absolute scale, where the
minimum possible temperature corresponds to zero (e.g., Kelvin or
Rankine). This ensures that ratios such as $T_m/T_w$ can be
interpreted as $(T_m - T_0)/(T_w - T_0)$ with $T_0 = 0$K. If the
temperature is provided in a relative scale (e.g., degrees Celsius or
Fahrenheit), it must first be converted to an absolute scale (e.g., by
adding 273.15 for Celsius) to maintain consistency.

\subsection{Wall model training}
\label{sec:training}

We train FNNs to construct wall models for $\tau_w$ and $q_w$ using
each input set listed in Tables~\ref{tab:RP} and~\ref{tab:qwall}. The
network layers employ hyperbolic tangent sigmoid functions as
activation functions, except for the output layer, which uses
rectified linear units (ReLUs). Training is performed using gradient
descent with momentum and an adaptive learning rate, following the
backpropagation method~\citep{yu2002backpropagation}. To mitigate
overfitting, an $L_2$ regularization term with a factor of 0.9 is
applied. To ensure robust model performance, 100 random splits of the
data into training (70\%), validation (15\%), and testing (15\%) sets
are conducted for each input configuration. For every split, a grid
search is used to identify the optimal network architecture, exploring
configurations with 3 to 6 hidden layers and 5 to 20 neurons per
layer. The minimum $L_2$ norm model error $\epsilon$ across different
splits and network architectures is reported in Table~\ref{tab:RP} for
wall shear stress and Table~\ref{tab:qwall} for wall heat flux.

As expected, the model error decreases as the degree of predictability
$1-\tilde{\epsilon}_{LB}$ rises. For the wall shear stress, models
incorporating both a roughness height parameter and $ES$ exhibit
significantly reduced errors, emphasizing the joint importance of
height and slope statistics in predicting $\tau_w$. For the wall heat
flux, the inclusion of $ES$ plays a less dominant role; combinations
using $k_{rms}$ (or $R_a$) and $ES$ yield lower errors than those
using $k_{avg}$ and $ES$.

When increasing the number of inputs from 2 to 4, the model error
typically decreases by approximately 3\%. However, with only two or
three inputs, some compressible cases still exhibit errors exceeding
40\%, confirming that four inputs are necessary to ensure consistent
scaling performance. As a result, the model for wall shear stress
achieves an error of $\epsilon = 4.07\%$, and the model for wall heat
flux attains an error of $\epsilon = 2.89\%$.

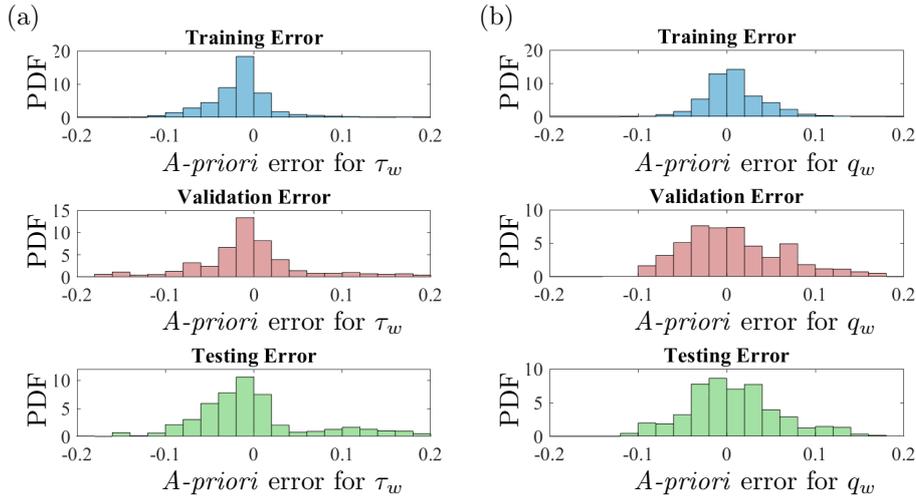
\begin{figure}
\centering
\includegraphics[width=60mm]{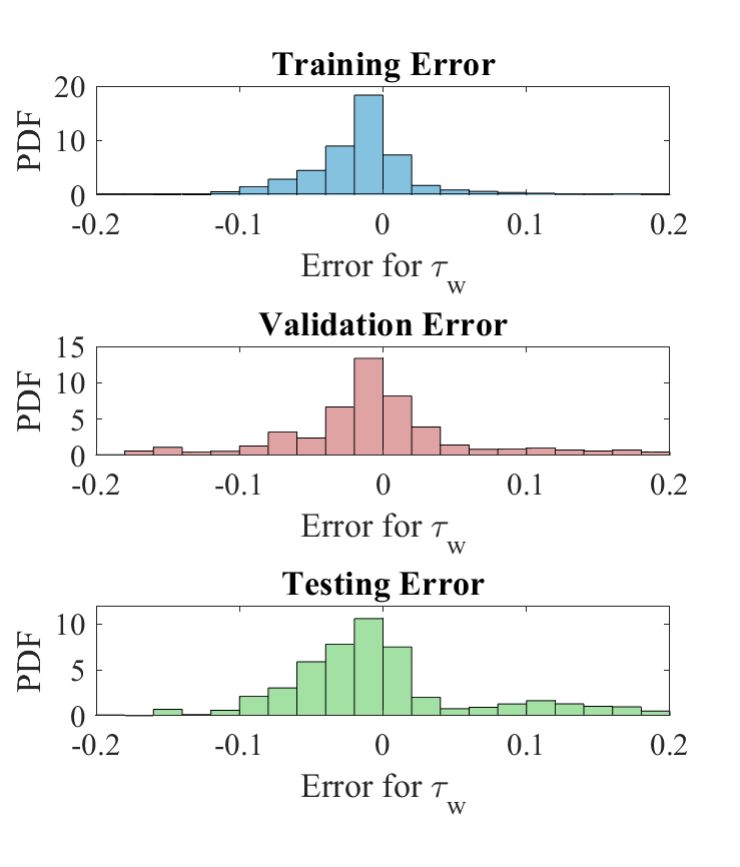}
\put(-175,190){(a)}
\put(-120,15){\colorbox{white}{\emph{A-priori} error for $\tau_w$}}
\put(-170,35){\rotatebox{90}{\colorbox{white}{PDF}}}
\put(-120,75){\colorbox{white}{\emph{A-priori} error for $\tau_w$}}
\put(-170,95){\rotatebox{90}{\colorbox{white}{PDF}}}
\put(-120,135){\colorbox{white}{\emph{A-priori} error for $\tau_w$}}
\put(-170,155){\rotatebox{90}{\colorbox{white}{PDF}}}
\hspace{1mm}
\includegraphics[width=60mm]{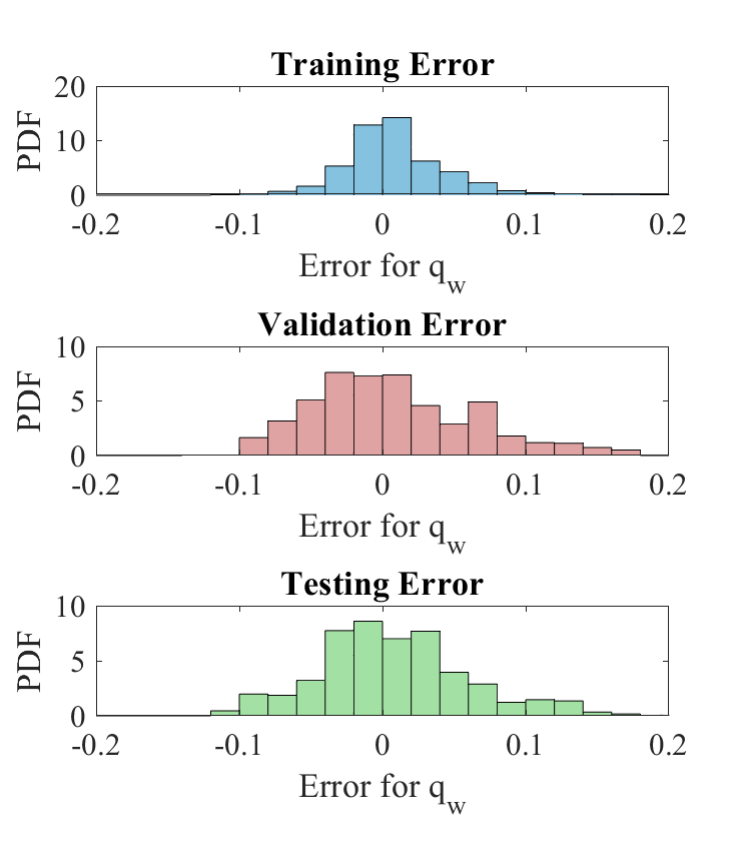}
\put(-175,190){(b)}
\put(-120,15){\colorbox{white}{\emph{A-priori} error for $q_w$}}
\put(-170,35){\rotatebox{90}{\colorbox{white}{PDF}}}
\put(-120,75){\colorbox{white}{\emph{A-priori} error for $q_w$}}
\put(-170,95){\rotatebox{90}{\colorbox{white}{PDF}}}
\put(-120,135){\colorbox{white}{\emph{A-priori} error for $q_w$}}
\put(-170,155){\rotatebox{90}{\colorbox{white}{PDF}}}
\caption{PDFs of \emph{a-priori} relative error of BFWM-rough-v2 for
  the training, validation, and testing datasets: (a) wall shear
  stress and (b) wall heat flux. The relative error is computed by
  $(\tau_{w,\text{predict}}-\tau_{w,\text{true}})/\tau_{w,\text{true}}$
  for $\tau_w$ and
  $(q_{w,\text{predict}}-q_{w,\text{true}})/q_{w,\text{true}}$ for
  $q_w$. }
\label{fig:pdf_error}
\end{figure}

The PDFs of the \emph{a-priori} error in figure~\ref{fig:pdf_error}
show the distribution of prediction errors for the training,
validation, and testing datasets, corresponding to the dataset split
used during the training of the optimal model. The resemblance of the
error distributions across these datasets indicates that the model for
both $\tau_w$ and $q_w$ performs consistently, regardless of whether
it is evaluated on seen (training) or unseen (validation and testing)
data. This similarity confirms that the model is not overfitting to
the training data and maintains predictive accuracy on new samples.
\subsection{Model confidence score}
\label{sec:confidence}

The model includes an additional module that outputs a
\emph{confidence score} alongside each prediction, providing separate
scores for the wall shear stress and wall heat flux estimates.  The
central idea is to quantify how far a previously `unseen' input lies
from the training data distribution: inputs that are close to the
training data are expected to yield high confidence, whereas inputs
that are far from it are flagged as potentially unreliable (low
confidence). We implement this distance-to-training principle using
the Spectral-normalized Neural Gaussian Process (SNGP) approach, which
enhances the distance-awareness and calibration of deep neural
networks. We adapt SNGP by introducing a Bayesian linear regression
output layer, and then map the resulting predictive uncertainty to a
scalar confidence score via an empirical cumulative distribution
function (ECDF).
%

\subsubsection{SNGP}
\label{sec:SNGP}

A Gaussian process (GP) with an appropriate kernel is a well-known
method that can achieve distance
awareness~\citep{rasmussen2006gaussian}.  This was used to obtain
confidence in our previous version of the model~\citep{ma2025machine}.
However, it often generalizes poorly to high-dimensional data, as
standard kernels may fail to capture the intrinsic structure, leading
to degraded performance due to the curse of dimensionality.  On the
other hand, a standard neural network is not distance-aware, since the
hidden mapping is not constrained to preserve geometric relationships
from the input space.

To address the limitations above, we use the SNGP introduced by
\cite{liu2023simple}. The approach produces uncertainty estimates
using a single deterministic model representation.  It is built on the
principle of distance awareness, which asserts that the predictions of
the model should account for the distance between new test inputs and
the training data. This property is essential for achieving
high-quality uncertainty quantification, as several prior studies have
highlighted the close connection between a neural network’s
uncertainty performance and its ability to discern distances in the
input space \citep{kristiadi2020being, van2021feature}.

The details of the implementation are discussed in
Appendix~\ref{appC}. Here, we provide a brief overview of the method.
For an input $\boldsymbol{\Pi}$ with feature representation
$\boldsymbol{\phi}=\boldsymbol{\phi}(\boldsymbol{h}(\boldsymbol{\Pi}))$,
where $\boldsymbol{h}$ denotes a spectrally-normalized deep neural
network feature extractor, the SNGP output layer models the scalar
response $\Pi_o$ using a Bayesian linear regression in feature space.
The posterior of the regression weights $\boldsymbol{\beta}$ yields
the predictive probability distribution of $\Pi_o$ conditioned on
$\boldsymbol{\Pi}$
\begin{equation}
f_{\Pi_o \mid \boldsymbol{\Pi}}
= \mathcal{N}\!\left(\mu_\beta^\top \boldsymbol{\phi},\,
\boldsymbol{\phi}^\top \Sigma_\beta \boldsymbol{\phi} + \sigma_n^2\right),
\end{equation}
where $\mathcal{N}$ is the normal distribution, $\mu_\beta$ and
$\Sigma_\beta$ are the posterior mean and covariance of
$\boldsymbol{\beta}$, and $\sigma_n^2$ is the observation-noise
variance.  The predictive variance decomposes into
$\boldsymbol{\phi}^\top \Sigma_\beta \boldsymbol{\phi} + \sigma_n^2$.
The first term, $\sigma_e^2=\boldsymbol{\phi}^\top \Sigma_\beta
\boldsymbol{\phi}$, captures epistemic uncertainty, which typically
increases for inputs whose feature representations are far from those
observed in training, reflecting lack of model knowledge. The second
term, $\sigma_n^2$, captures aleatoric uncertainty due to irreducible
observation noise. In this work, $\sigma_e^2$ is used to construct a
predictive confidence score.

The SNGP model is trained using the same DNS database employed for the
wall model, with identical input features and output targets. The
input features are normalized to the range $[0,\,1]$ based on their
componentwise extrema in the training dataset, while the output
targets are standardized to have zero mean and unit variance.
Training is performed using the Adam optimizer with a learning rate of
$10^{-3}$, a batch size of 128, and a total of 100 training epochs. A
validation split of 15\% of the training data is used to monitor
convergence and prevent overfitting.

\subsubsection{Confidence score}

To calibrate the confidence, we construct a reference distribution of
epistemic uncertainty across the normalized input space. This
procedure enables us to transform a raw value of $\sigma_e$ from any
test point into a percentile-based confidence score, reflecting how
typical or atypical the uncertainty of the model is relative to a
broad set of reference cases. Specifically, we generate a large pool
of reference inputs ($N_{\text{ref}}=20{,}000$ samples) in the
normalized input domain. This contains training points and random
draws.  For each reference input, the epistemic uncertainty
$\sigma_{\text{ref}}$ is computed using the SNGP model. The collection
of these values is then sorted and stored as a reference distribution.

During evaluation, for a given testing point $\boldsymbol{\Pi}$, a
test-point uncertainty $\sigma_e(\boldsymbol{\Pi})$ can be mapped to
its corresponding percentile in this distribution, yielding a
value for the confidence score.  The ECDF is defined as
\begin{equation}
F(\sigma_e) = \frac{1}{N_{\text{ref}}} \sum_{i=1}^{N_{\text{ref}}} \mathbf{1}\{\sigma_{\text{ref}, i} \leq \sigma_e \},
\end{equation}
where $\mathbf{1}\{\cdot\}$ is the indicator function. The confidence
score is then mapped as
\begin{equation}
C(\boldsymbol{\Pi}) = 1 - F\!\left(\sigma_e(\boldsymbol{\Pi})\right).
\end{equation}
Thus, low epistemic uncertainty corresponds to high confidence, while
inputs far from the training distribution yield higher epistemic
variance and lower confidence. 

\subsubsection{Demonstration of confidence score on a 1-D  problem}
%
\begin{figure}
\centering
\includegraphics[width=69mm,trim={4cm 0.2cm 3cm 0cm},clip]{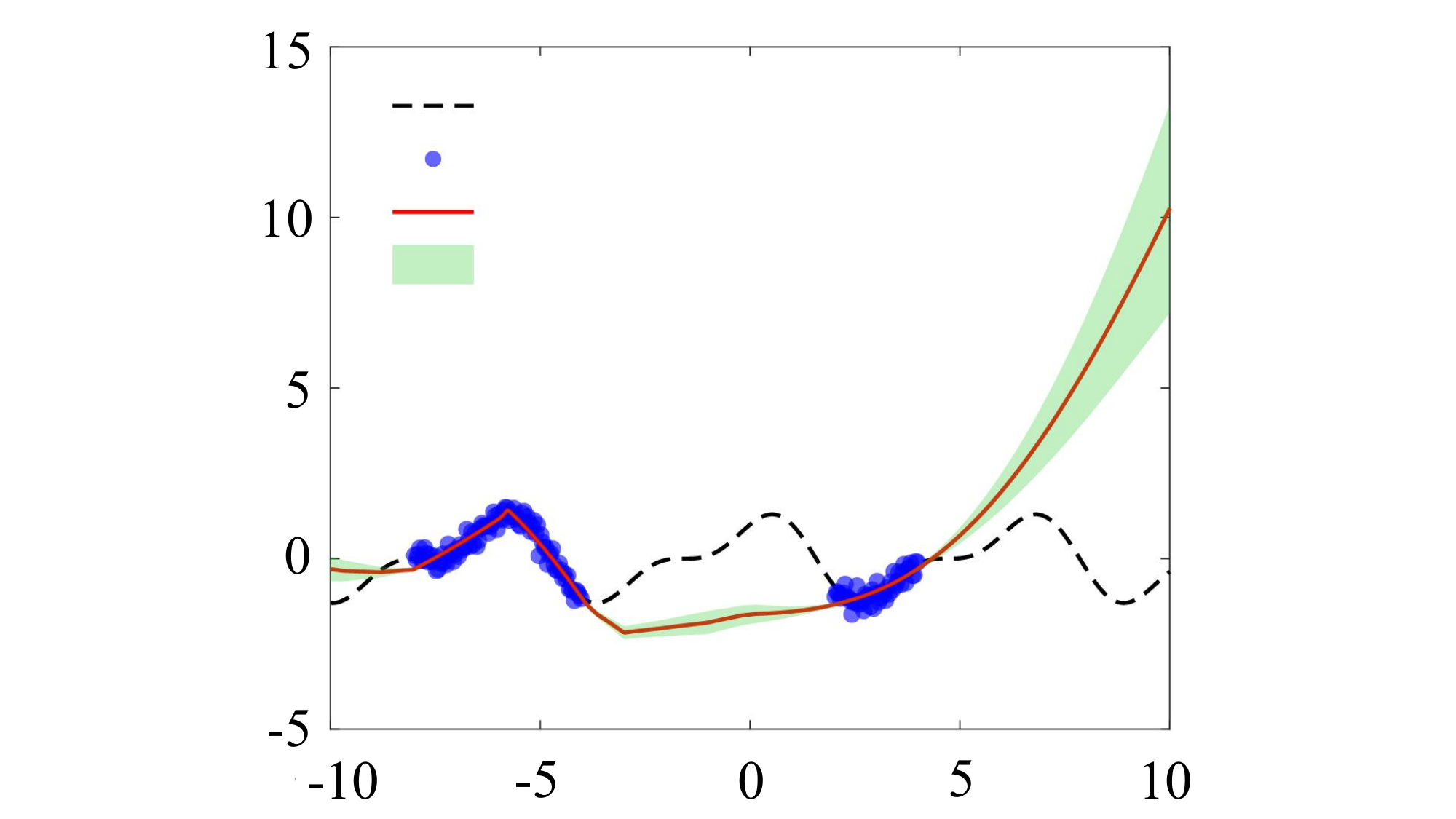}
\put(-180,140){$(a)$}
\put(-193,67){$\Pi_o$}
\put(-100,-10){$\Pi$}
\put(-136,118){True function}
\put(-136,108){Training data}
\put(-136,99){Prediction}
\put(-136,90){$\pm2\sigma_e$}
\includegraphics[width=66mm,trim={4cm 0.2cm 3cm 0cm},clip]{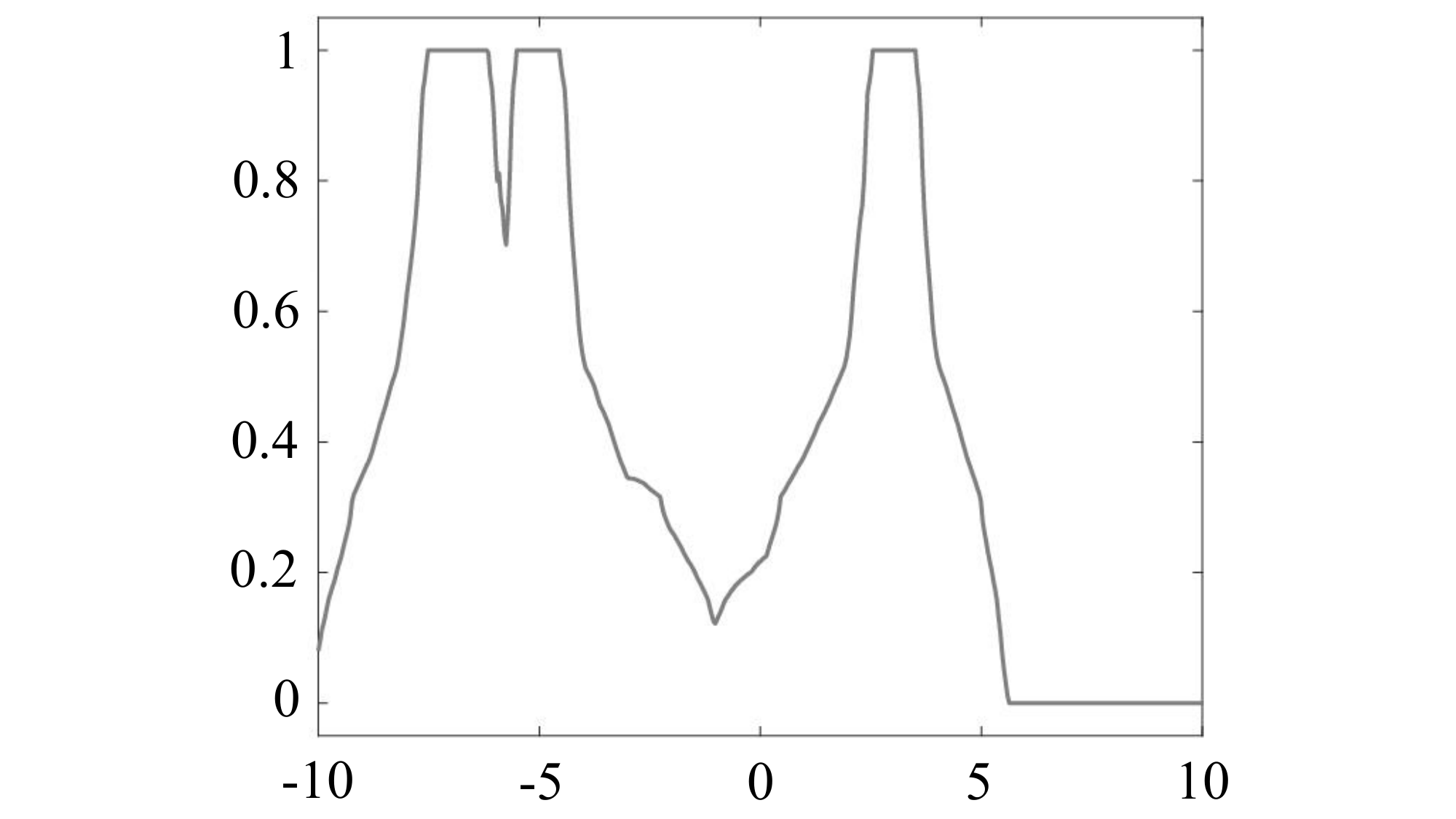}
\put(-180,140){$(b)$}
\put(-190,67){$C$}
\put(-95,-10){$\Pi$}
\caption{Demonstration of confidence score on a 1-D problem: $(a)$
  prediction and uncertainty results and $(b)$ confidence
  score.}
\label{fig:SNGP_toy}
\end{figure}
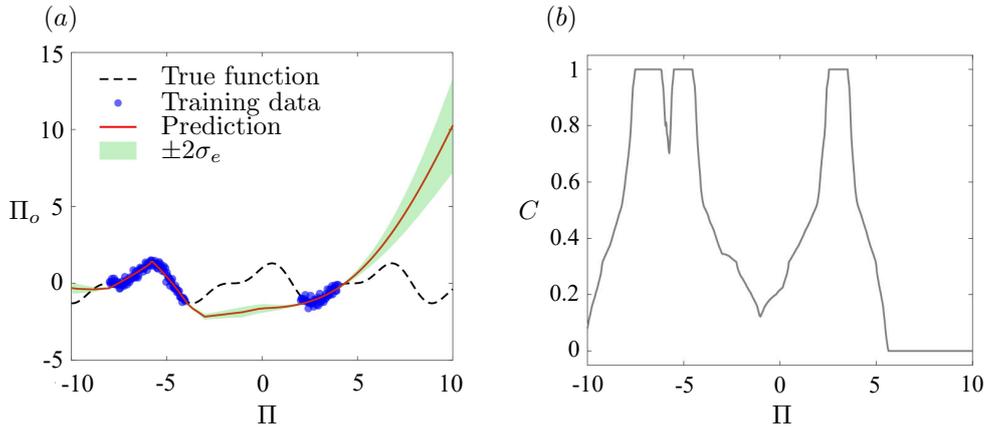

We illustrate the performance of the confidence score in a simple
one-dimensional problem. The ground-truth function is defined as
$\Pi_o = \cos(\Pi) + 0.5 \sin(2\Pi), \quad \Pi \in [-10, 10]$.
Training data are intentionally restricted to two narrow subregions,
\([-8, -4]\) and \([2, 4]\), leaving large intervals of the input
space unobserved. To mimic realistic measurement noise, Gaussian
perturbations are added to the training labels.

Figure~\ref{fig:SNGP_toy}$(a)$ shows the results. In the observed
training regions, the model accurately recovers the functional form,
with low uncertainty bands. In contrast, in unobserved intervals such
as $[-10, -8]$, $[-4, 2]$, and $[4, 10]$, the epistemic error
($\sigma_e^2$) grows systematically. Figure~\ref{fig:SNGP_toy}$(b)$
shows the confidence score as a function of the input location
$\Pi$. In regions covered by training data, the model maintains high
confidence ($C \approx 1$), consistent with its accurate predictions
and low predictive variance. Conversely, in regions far from the
training intervals, confidence decreases sharply, reflecting the
reduced reliability of the model.

\section{Validation}
\label{validation}

The performance of BFWM-rough-v2 is assessed \textit{a-posteriori} in
WMLES across a broad range of flow configurations. All validation
cases are unseen during model training. These include a total of 162
incompressible and compressible turbulent channel flows, a
high-pressure turbine (HPT) blade with Gaussian roughness, a
compression ramp with sandpaper roughness, and three hypersonic blunt
bodies with sandgrain roughness.

It is also useful to decompose the wall model errors into two primary
sources~\citep{lozano2023machine}: (\emph{i}) \emph{external errors},
which arise from inaccuracies in the outer LES input data at the
wall-model interface, and (\emph{ii}) \emph{internal errors}, which
are associated with the physical assumptions inherent to the wall
model itself. External errors are typically dominated by deficiencies
in the SGS model, while internal errors persist even when exact inputs
are provided, particularly if model assumptions, e.g.,
quasi-equilibrium, locality, or training on mean data (see
\S\ref{sec:assumption}) are violated.  The sum of these two
contributions constitutes the \emph{total error}.  The
\textit{internal errors} of BFWM-rough-v2 were assessed earlier in
Figure~\ref{fig:pdf_error} using DNS as input data. Here, we focus on
the \emph{total error} in actual WMLES to assess the overall
predictive performance of the model.

The BFWM-rough-v2 was implemented into the high-fidelity solver
charLES, developed by Cascade Technologies, Inc~\citep{bres2018large,
  fu2021shock}. For all the validation cases, the Vreman SGS
model~\citep{Vreman2004} is employed in the WMLES. The computational
mesh is generated by performing a Voronoi tessellation of a set of
seed points.

For comparison purposes, we employ an algebraic equilibrium wall model
with prescribed $k_s$ based on the rough-wall logarithmic law. The
wall heat flux is computed using an algebraic approximation derived
from the temperature law of the wall. We denote these two models for
wall shear stress and wall heat flux collectively as
``EQWM--$k_s$''. The detailed formulation of EQWM--$k_s$ is provided
in Appendix~\ref{appD}. Similar models have previously been applied to
low-speed flows, although their applicability to high-speed regimes
remains limited~\citep{li2022predictive, kadivar2025turbulent}.  It is
important to note, however, that EQWM--$k_s$ requires specification of
$k_s$, a flow-dependent parameter. In the cases below, $k_s$ is
obtained directly from DNS data or from a simple correlation-based
model calibrated for the case under consideration.  As such,
EQWM--$k_s$ is not a truly predictive model, but rather a reference
for assessing the performance of a wall model under the equilibrium
rough-wall logarithmic-law assumption.

\subsection{Turbulent channel flow}

\subsubsection{Incompressible flow}

We conduct WMLES with BFWM-rough-v2 for incompressible turbulent
channel flows over rough surfaces. The computational domain dimensions
are set to $L_x = \pi\delta$, $L_y = 2\delta$, and $L_z =
0.5\pi\delta$. The test cases are drawn from the testing dataset
introduced in \S\ref{sec:training}, and include 12 Gaussian and 9
Weibull rough surfaces spanning $Re_\tau = 180$ to $1000$ and $k_s^+ =
5$ to 230.  The model performance is evaluated at three different grid
resolutions: $\Delta/\delta = 1/20$, $1/10$, and $1/5$. In total, 78
unseen incompressible turbulent channel flow cases are tested.

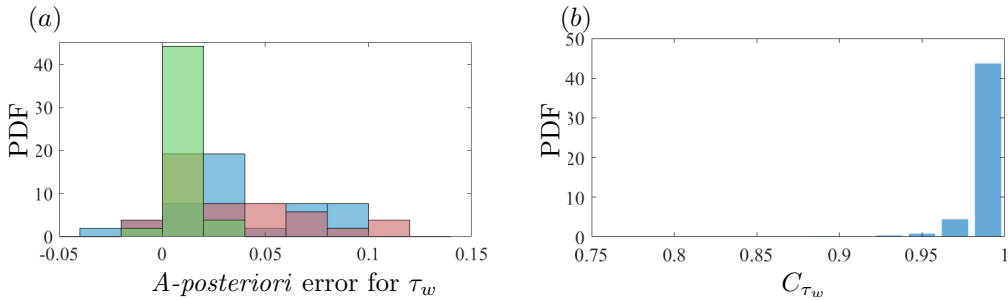
\begin{figure}
\centering
    \includegraphics[width=70mm]{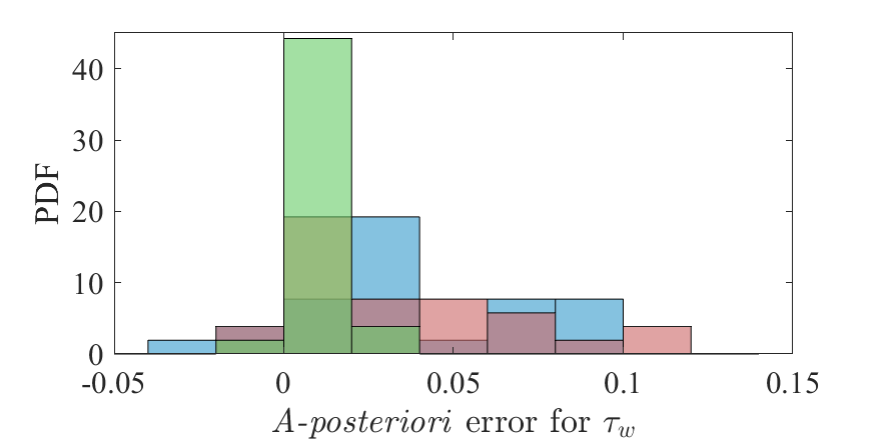}
    \put(-185,100){$(a)$}
    \put(-142,0){\colorbox{white}{\emph{A-posteriori} error for $\tau_w$}}
    \put(-195,47){\rotatebox{90}{{\colorbox{white}{{PDF}}}}}
    \includegraphics[width=70mm]{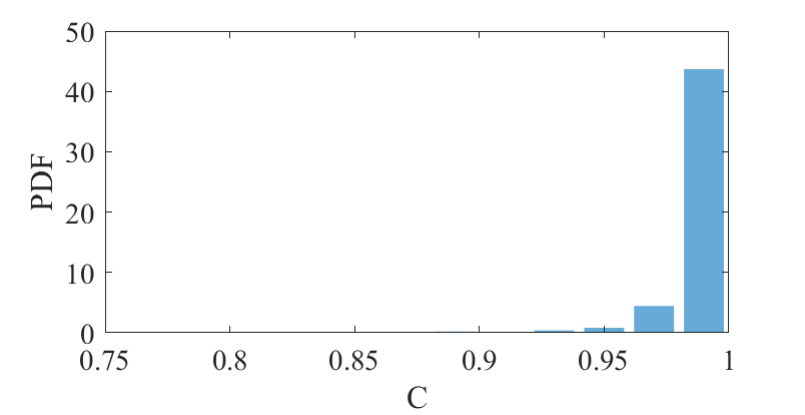}
    \put(-185,100){$(b)$}
    \put(-105,0){\colorbox{white}{$C_{\tau_w}$}}
    \put(-195,47){\rotatebox{90}{{\colorbox{white}{{PDF}}}}}
    \caption{ \emph{A-posteriori} evaluation of BFWM-rough-v2 in WMLES of
      incompressible turbulent channels. PDFs of $(a)$ relative error
      in wall shear stress and $(b)$ space-time averaged confidence
      score.  Results are shown at three different grid resolutions:
      $\Delta/\delta = 1/20$ (blue), $1/10$ (red), and $1/5$ (green).
\label{fig:post_incomp}}
\end{figure}

Figure~\ref{fig:post_incomp} presents the PDFs of the \emph{a
posteriori} error in the model predictions and the space-time averaged
confidence score. Results for the three different grid resolutions,
$\Delta/\delta = 1/20$, $1/10$, and $1/5$, are shown in blue, red, and
green, respectively. The error distributions remain within $15\%$
across all grid spacings, with the majority of cases exhibiting errors
below $5\%$.  Regarding model confidence, most cases cluster near $C =
1$, where the PDF exhibits a sharp peak, indicating high confidence in
the predictions for the majority of testing cases. Only a few cases
fall within the lower confidence range of $0.75 < C < 0.9$.

\begin{figure}
\centering
    \includegraphics[width=70mm]{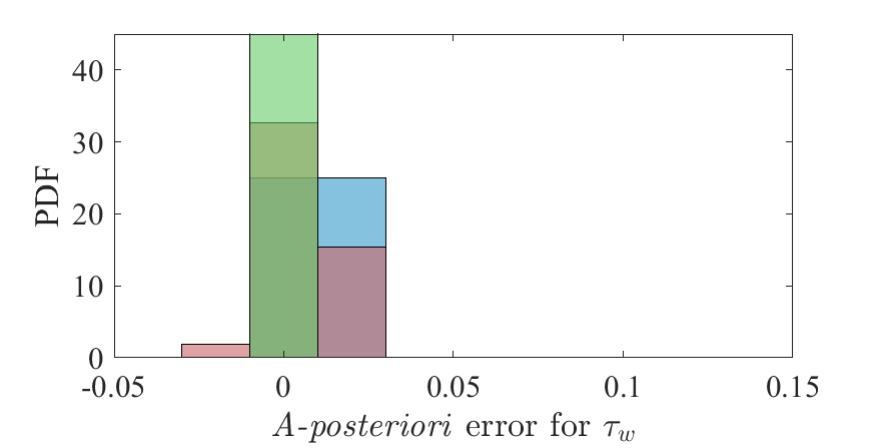}
    \put(-142,0){\colorbox{white}{\emph{A-posteriori} error for $\tau_w$}}
    \put(-195,47){\rotatebox{90}{{\colorbox{white}{{PDF}}}}}
    \caption{\emph{A-posteriori} evaluation of EQWM-$k_s$ in WMLES of
      incompressible turbulent channels.  PDFs of \emph{a-posteriori}
      relative error in wall shear stress. Results are shown at three
      different grid resolutions: $\Delta/\delta = 1/20$ (blue),
      $1/10$ (red), and $1/5$ (green). 
\label{fig:EQWMks_incomp}}
\end{figure}

The WMLES results using EQWM-$k_s$ for the same testing cases are
shown in figure~\ref{fig:EQWMks_incomp}, exhibiting errors within 5\%,
which are smaller than those obtained with BFWM-rough-v2. This outcome
is expected, as the true $k_s$ value extracted from DNS is prescribed
in EQWM-$k_s$ to enforce the correct momentum deficit in the
incompressible flow.

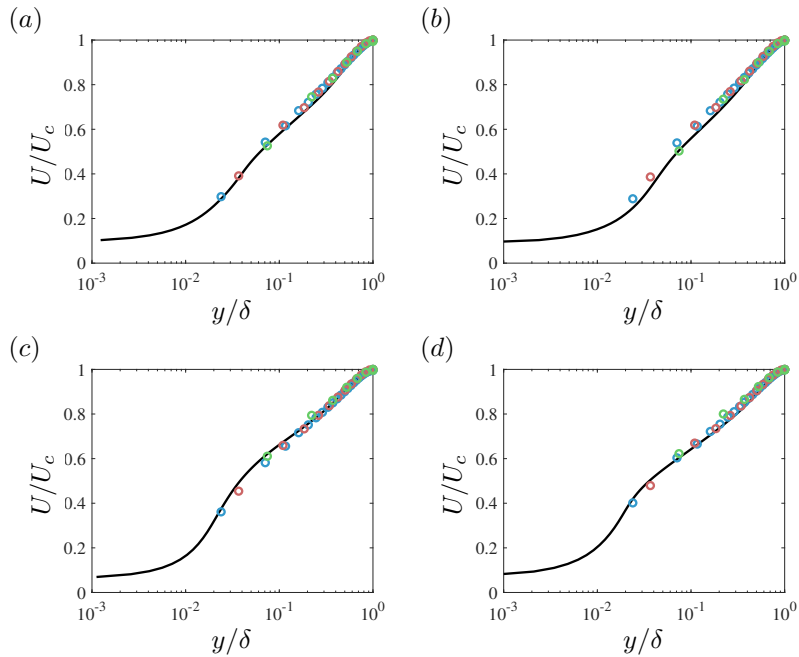
\begin{figure}
\begin{center}
    \includegraphics[width=50mm,trim={0cm 0.2cm 0cm 0.5cm},clip]{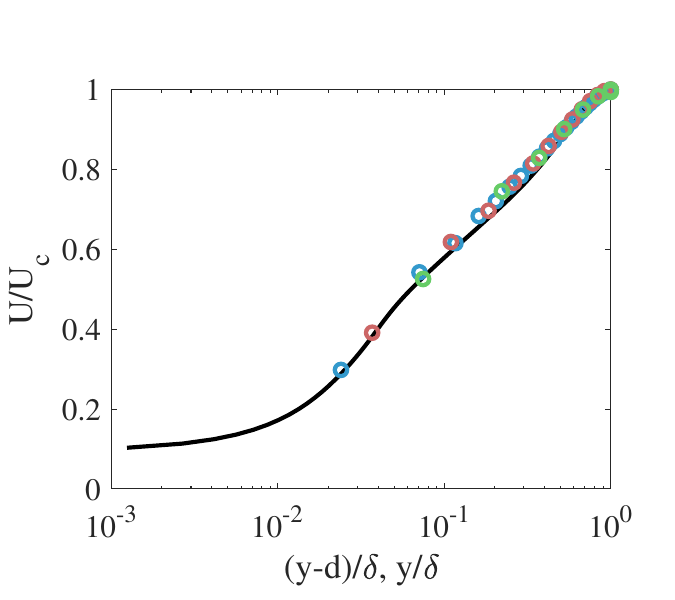}
    \put(-150,110){$(a)$}
    \put(-145,47){\rotatebox{90}{{\colorbox{white}{{$U/U_c$}}}}}
    \put(-83,0){\color{white}\rule{40pt}{10pt}}
    \put(-77,0){\colorbox{white}{$y/\delta$}}
    \hspace{3mm}
    \includegraphics[width=50mm,trim={0cm 0.2cm 0cm 0.5cm},clip]{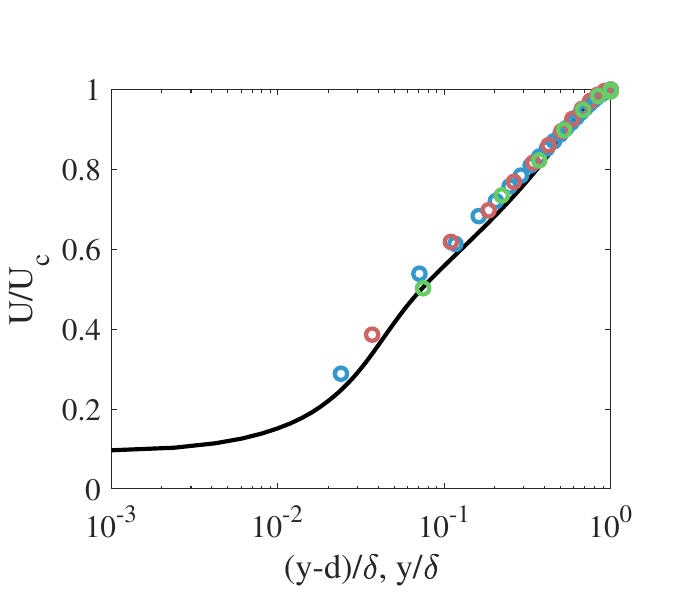}
    \put(-150,110){$(b)$}
    \put(-145,47){\rotatebox{90}{{\colorbox{white}{{$U/U_c$}}}}}
    \put(-83,0){\color{white}\rule{40pt}{10pt}}
    \put(-77,0){\colorbox{white}{$y/\delta$}}
    \hspace{7mm}
    \includegraphics[width=50mm,trim={0cm 0.2cm 0cm 0.5cm},clip]{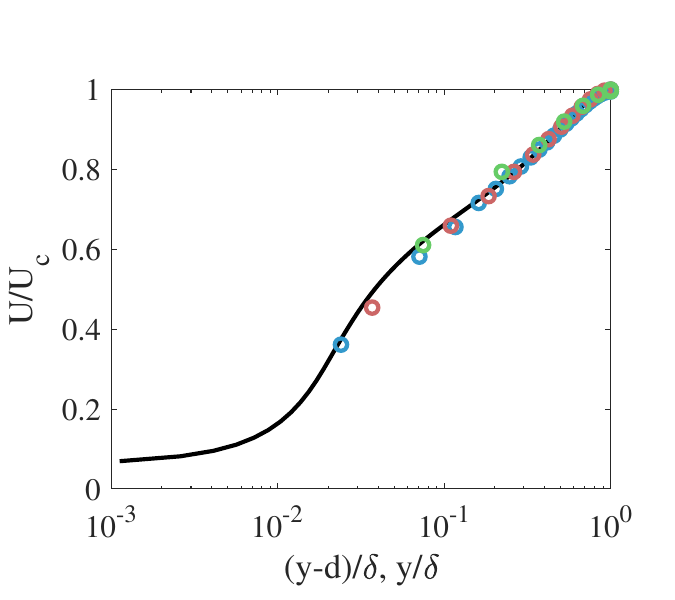}
    \put(-150,110){$(c)$}
    \put(-145,47){\rotatebox{90}{{\colorbox{white}{{$U/U_c$}}}}}
    \put(-83,0){\color{white}\rule{40pt}{10pt}}
    \put(-77,0){\colorbox{white}{$y/\delta$}}
    \hspace{3mm}
    \includegraphics[width=50mm,trim={0cm 0.2cm 0cm 0.5cm},clip]{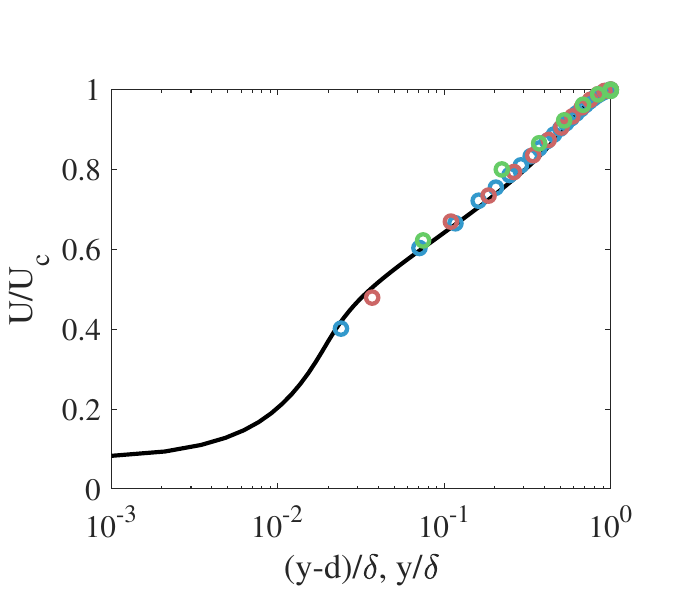}
    \put(-150,110){$(d)$}
    \put(-145,47){\rotatebox{90}{{\colorbox{white}{{$U/U_c$}}}}}
    \put(-83,0){\color{white}\rule{40pt}{10pt}}
    \put(-77,0){\colorbox{white}{$y/\delta$}}
    \caption{\emph{A-posteriori} evaluation of BFWM-rough-v2 in WMLES of
      incompressible turbulent channels. Mean velocity profiles for
      DNS (line) and WMLES (symbols) for $(a)$ GS3 at $Re_{\tau}$=720;
      $(b)$ GS11 at $Re_{\tau}$=900; $(c)$ WB8 at $Re_{\tau}$=540;
      $(d)$ WB11 at $Re_{\tau}$=1000. Results are shown at three
      different grid resolutions: $\Delta/\delta = 1/20$ (blue),
      $1/10$ (red), and $1/5$ (green).
\label{fig:umean_incomp}}
\end{center}
\end{figure}

Figure~\ref{fig:umean_incomp} compares the mean velocity profiles from
WMLES with DNS results for four selected cases spanning different
roughness types and $Re_{\tau}$. Overall, the WMLES results show
reasonable agreement with DNS across all grid resolutions for the
selected test cases.

\subsubsection{Compressible flow}

The BFWM-rough-v2 model is next validated in the WMLES of compressible
turbulent channel flow over rough walls. Simulations are conducted in
a domain of size $L_x = \pi\delta$, $L_y = 2\delta$, and $L_z =
0.5\pi\delta$. The test dataset, described in \S\ref{sec:training}, is
evaluated at three grid resolutions: $\Delta/\delta = 1/20$, $1/10$,
and $1/5$. These tests encompass 12 Gaussian rough surfaces, spanning
bulk Mach numbers from $M_b = 0.4$ to $3.3$, bulk Reynolds numbers
from $Re_b = 7,500$ to $155{,}000$, and roughness Reynolds numbers
from $k_s^+ = 5$ to $170$. In total, 84 cases are examined.

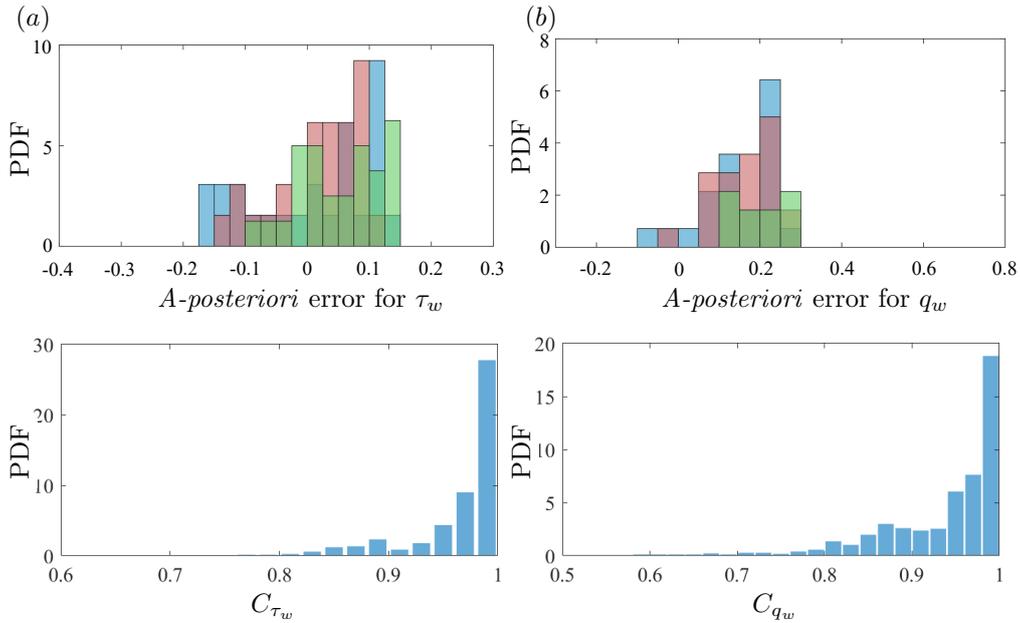
\begin{figure}
\begin{center}
    \includegraphics[width=67mm,trim={0.5cm 0.0cm 0.5cm 0.0cm},clip]{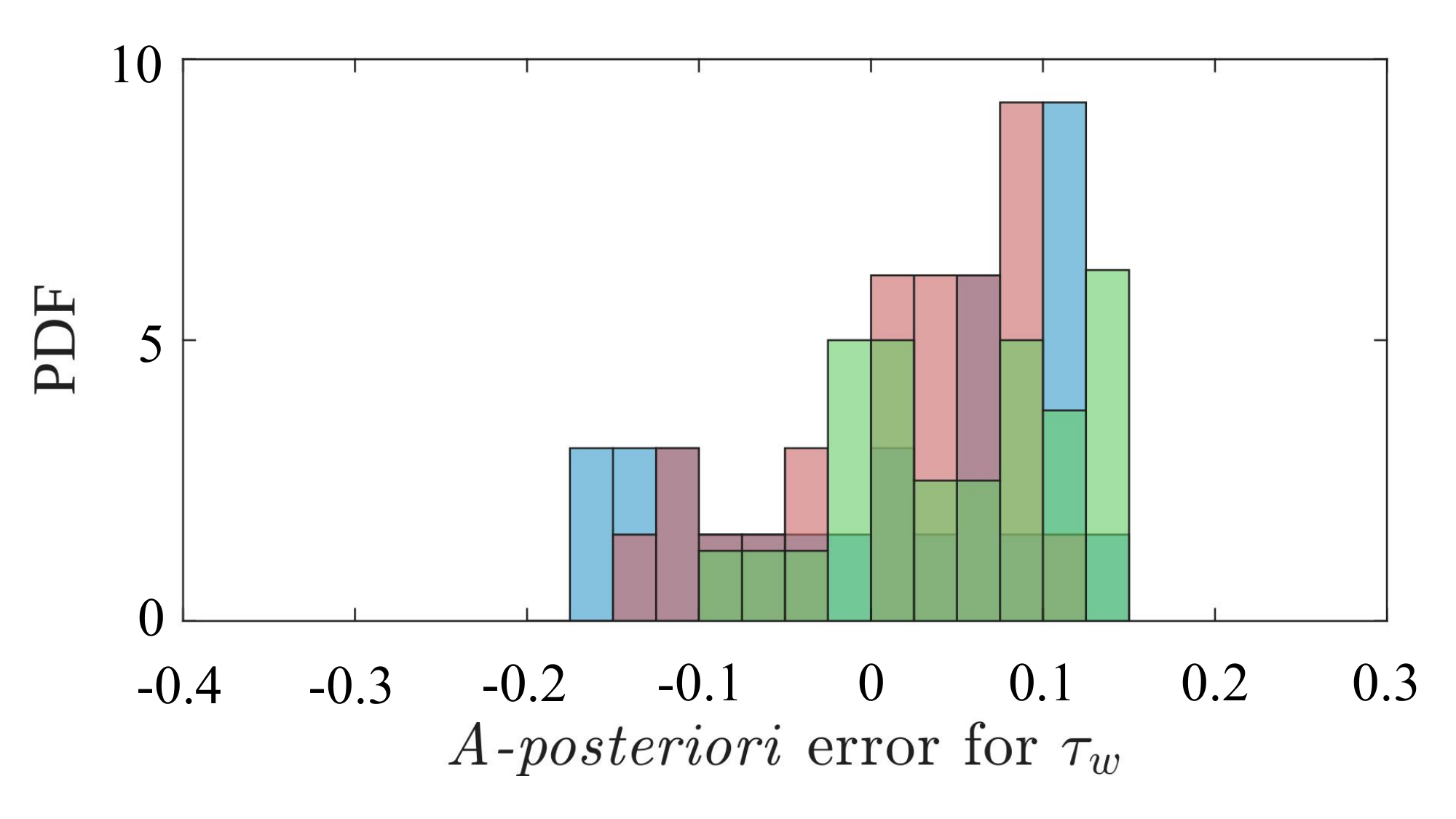}
    \put(-185,110){$(a)$}
    \put(-135,4){\colorbox{white}{\emph{A-posteriori} error for $\tau_w$}}
    \put(-190,50){\rotatebox{90}{{\colorbox{white}{{PDF}}}}}
    \includegraphics[width=67mm,trim={0.5cm 0.0cm 1.0cm 0.0cm},clip]{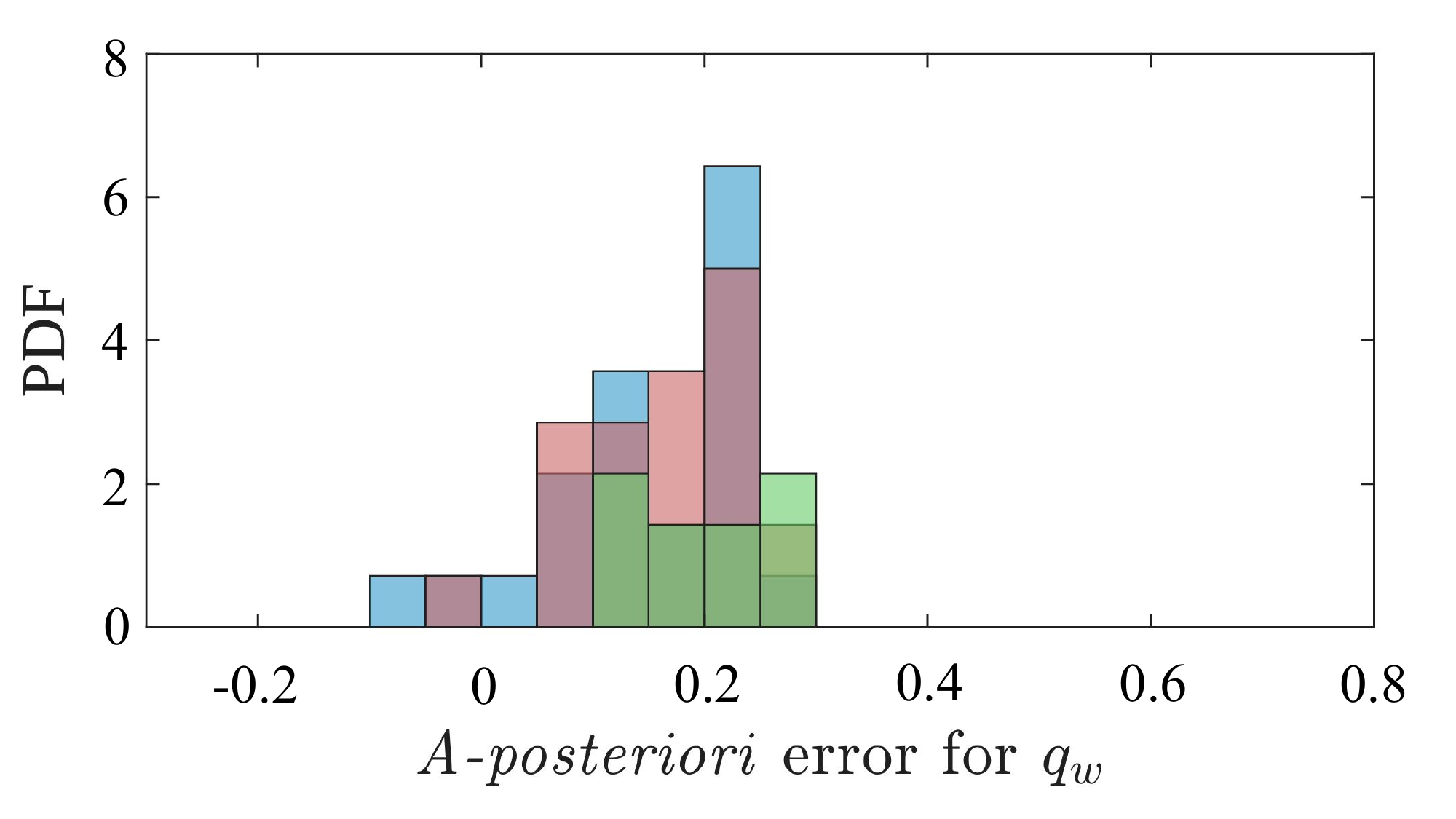}
    \put(-185,110){$(b)$}
    \put(-137,4){\colorbox{white}{\emph{A-posteriori} error for $q_w$}}
    \put(-193,51){\rotatebox{90}{{\colorbox{white}{{PDF}}}}}
    \hspace{3mm}
    \includegraphics[width=66mm,trim={0.5cm 0.0cm 1.0cm 0.0cm},clip]{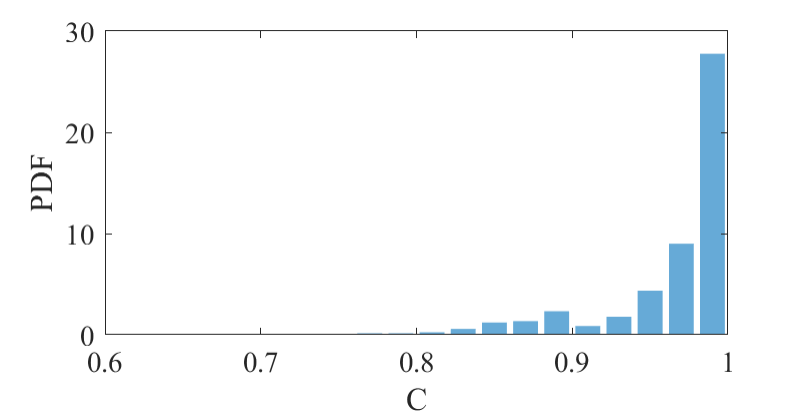}
    \put(-100,0){\colorbox{white}{$C_{\tau_w}$}}
    \put(-190,47){\rotatebox{90}{{\colorbox{white}{{PDF}}}}}
    \includegraphics[width=66mm,trim={0.5cm 0.0cm 1.0cm 0.0cm},clip]{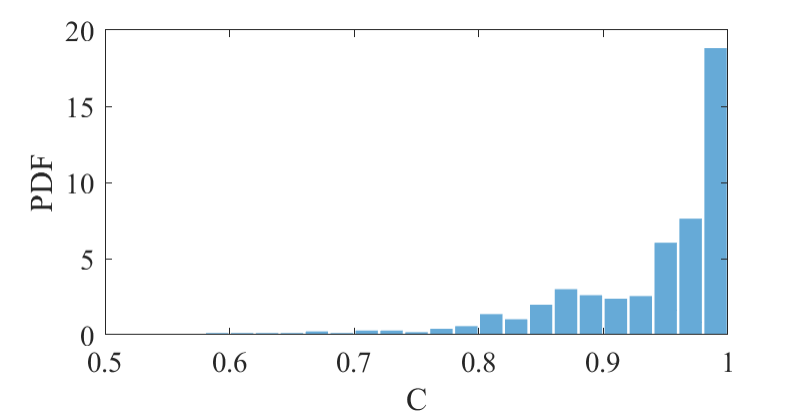}
    \put(-100,0){\colorbox{white}{$C_{q_w}$}}
    \put(-190,47){\rotatebox{90}{{\colorbox{white}{{PDF}}}}}
    \caption{\emph{A-posteriori} evaluation of BFWM-rough-v2 in WMLES of
      compressible turbulent channels. PDF of relative error for $(a)$
      wall shear stress and $(b)$ wall heat flux along with their
      respective confidence scores below.  Results are shown at three
      different grid resolutions: $\Delta/\delta = 1/20$ (blue),
      $1/10$ (red), and $1/5$ (green).  
\label{fig:post_comp}}
\end{center}
\end{figure}

Figure~\ref{fig:post_comp} shows the PDFs of the relative error in
wall shear stress and heat flux. The model performance is assessed at
the three grid resolutions mentioned above, represented by blue, red,
and green curves, respectively. The mean relative errors for $\tau_w$
and $q_w$ are 3.82\% and 7.85\%, respectively. Errors in $\tau_w$
remain within $15\%$, while larger errors of up to $30\%$ are observed
for $q_w$ in some testing cases.  Despite this, the majority of test
cases still exhibit high confidence scores ($0.9 \le C \le 1$) for
both wall shear stress and wall heat flux predictions. The confidence
score distribution for $\tau_w$ is sharply peaked around 1, indicating
strong model certainty. While the confidence score for $q_w$ also
peaks at 1, it displays a broader distribution with greater weight in
the lower-confidence range.

\begin{figure}
\begin{center}
    \includegraphics[width=67mm,trim={0.5cm 0.0cm 1.0cm 0.0cm},clip]{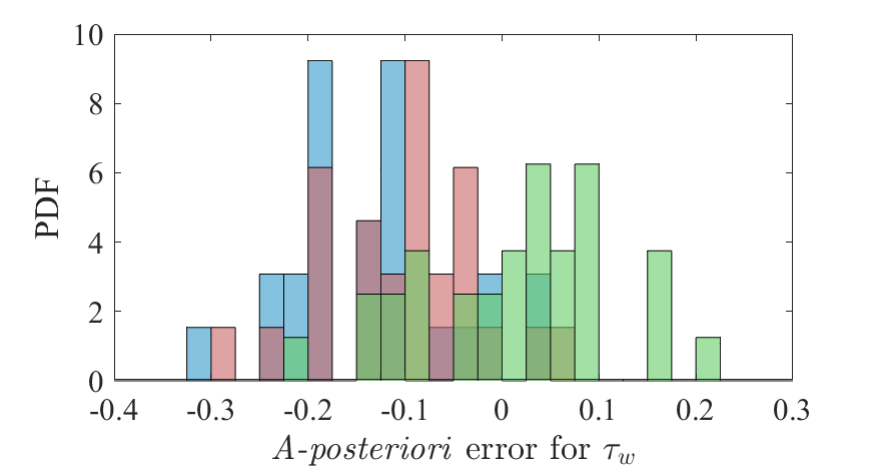}
    \put(-185,113){$(a)$}
    \put(-135,0){\colorbox{white}{\emph{A-posteriori} error for $\tau_w$}}
    \put(-190,50){\rotatebox{90}{{\colorbox{white}{{PDF}}}}}
    \includegraphics[width=67mm,trim={0.5cm 0.0cm 1.0cm 0.0cm},clip]{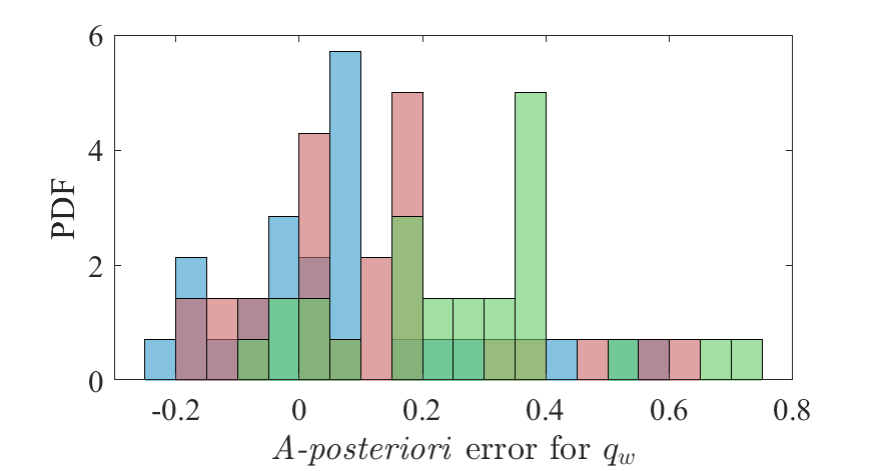}
    \put(-185,113){$(b)$}
    \put(-135,0){\colorbox{white}{\emph{A-posteriori} error for $q_w$}}
    \put(-190,50){\rotatebox{90}{{\colorbox{white}{{PDF}}}}}
    \caption{\emph{A-posteriori} evaluation of EQWM-$k_s$ in WMLES of
      compressible turbulent channels. PDFs of relative error for
      $(a)$ wall shear stress and $(b)$ wall heat flux. Results are
      shown at three different grid resolutions: $\Delta/\delta =
      1/20$ (blue), $1/10$ (red), and $1/5$ (green).
\label{fig:post_comp_EQWMks}}
\end{center}
\end{figure}

For comparison, WMLES using EQWM-$k_s$ are also performed, and the
results are presented in figure~\ref{fig:post_comp_EQWMks}. The errors
in both wall shear stress and wall heat flux are larger than those
obtained with BFWM-rough-v2, with the mean relative errors for
$\tau_w$ and $q_w$ are -7.60\% and 19.30\%, respectively. In
particular, cases at high $M_b$ exhibit significantly higher errors,
with the wall heat flux results from EQWM-$k_s$ reaching up to
80\%. This indicates that the performance of EQWM-$k_s$ deteriorates
in supersonic flow regimes, where compressibility effects are strong,
whereas BFWM-rough-v2 demonstrates superior predictive capability
under these conditions.

\begin{figure}
\centering
    \includegraphics[width=65mm,trim={0.5cm 0.0cm 1.0cm 0.0cm},clip]{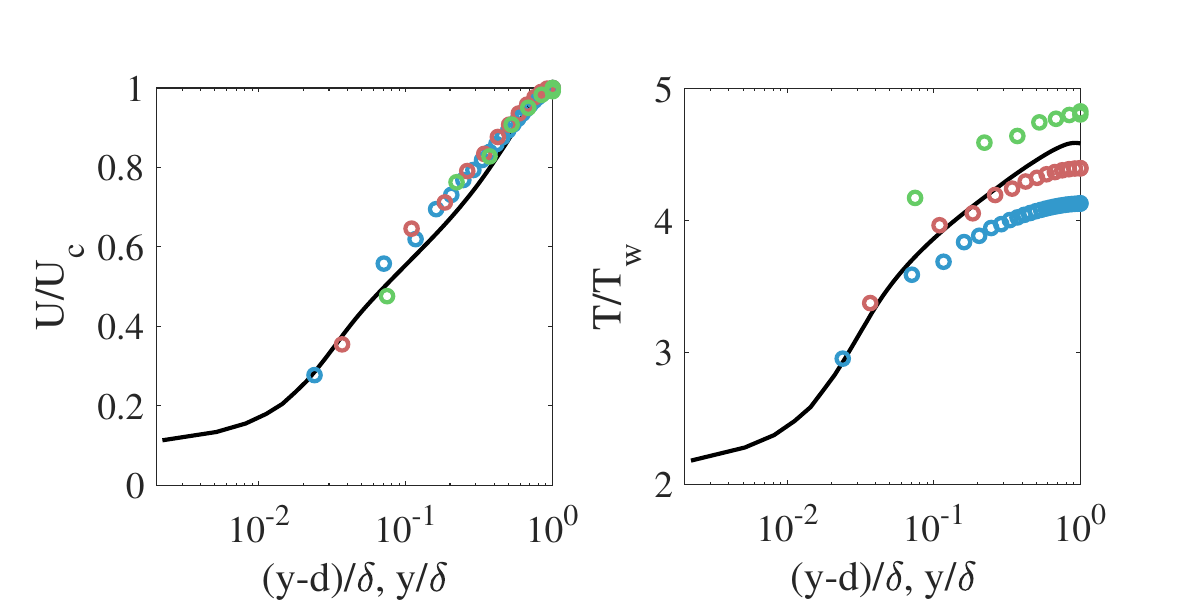}
    \put(-180,96){$(a)$}
    \put(-190,40){\rotatebox{90}{{\colorbox{white}{{$U/U_c$}}}}}
    \put(-150,-3){\color{white}\rule{40pt}{10pt}}
    \put(-142,-3){\colorbox{white}{$y/\delta$}}
    \put(-97,40){\rotatebox{90}{{\colorbox{white}{{$T/T_w$}}}}}
    \put(-60,-3){\color{white}\rule{40pt}{10pt}}
    \put(-52,-3){\colorbox{white}{$y/\delta$}}
    \includegraphics[width=72mm,trim={0.5cm 0.2cm 0cm 0.5cm},clip]{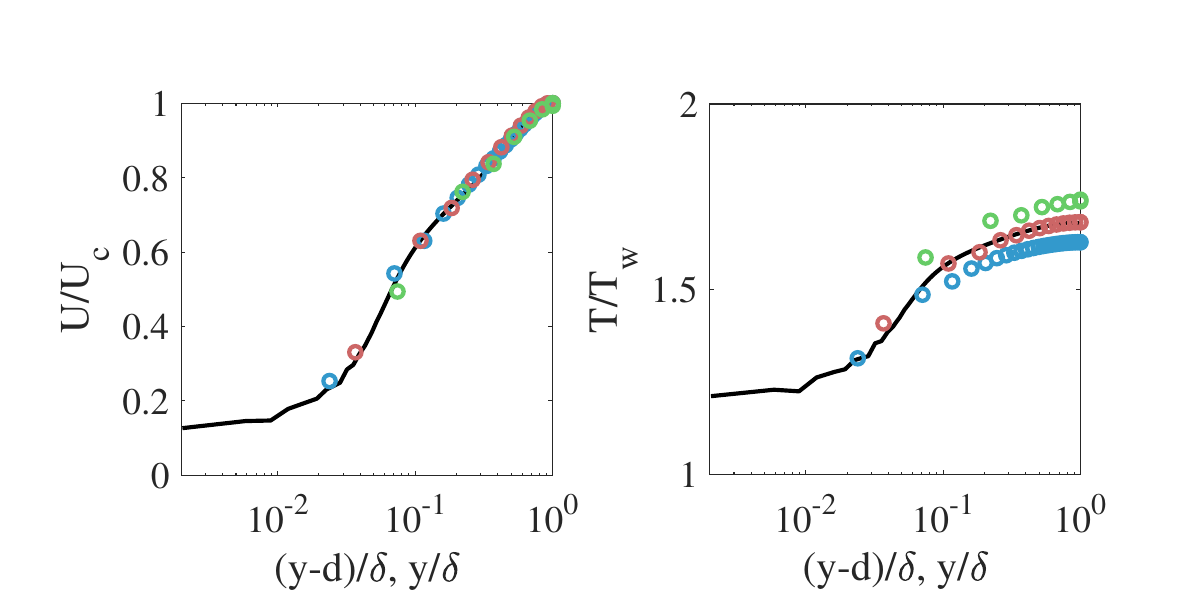}
    \put(-190,96){$(b)$}
    \put(-205,40){\rotatebox{90}{{\colorbox{white}{{$U/U_c$}}}}}
    \put(-167,-3){\color{white}\rule{40pt}{10pt}}
    \put(-156,-3){\colorbox{white}{$y/\delta$}}
    \put(-112,40){\rotatebox{90}{{\colorbox{white}{{$T/T_w$}}}}}
    \put(-75,-3){\color{white}\rule{40pt}{10pt}}
    \put(-64,-3){\colorbox{white}{$y/\delta$}}
    \hspace{5mm}
    \includegraphics[width=65mm,trim={0.5cm 0.0cm 1.0cm 0.0cm},clip]{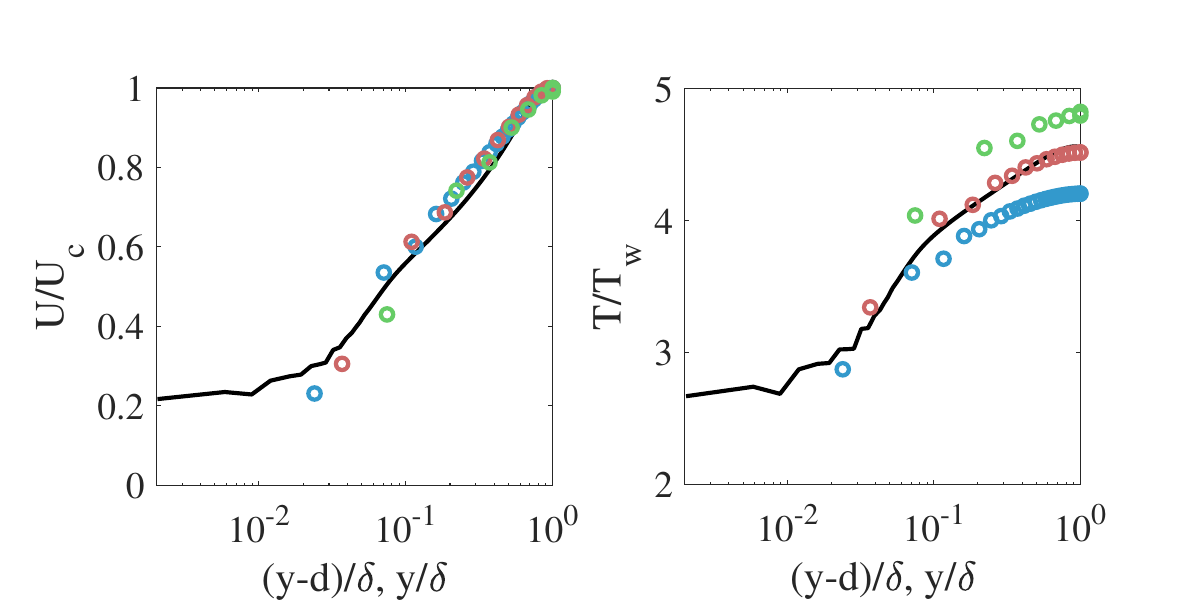}
    \put(-180,96){$(c)$}
    \put(-190,40){\rotatebox{90}{{\colorbox{white}{{$U/U_c$}}}}}
    \put(-150,-3){\color{white}\rule{40pt}{10pt}}
    \put(-142,-3){\colorbox{white}{$y/\delta$}}
    \put(-97,40){\rotatebox{90}{{\colorbox{white}{{$T/T_w$}}}}}
    \put(-60,-3){\color{white}\rule{40pt}{10pt}}
    \put(-52,-3){\colorbox{white}{$y/\delta$}}
    \includegraphics[width=68.5mm,trim={0.5cm 0.2cm 1.2cm 0.5cm},clip]{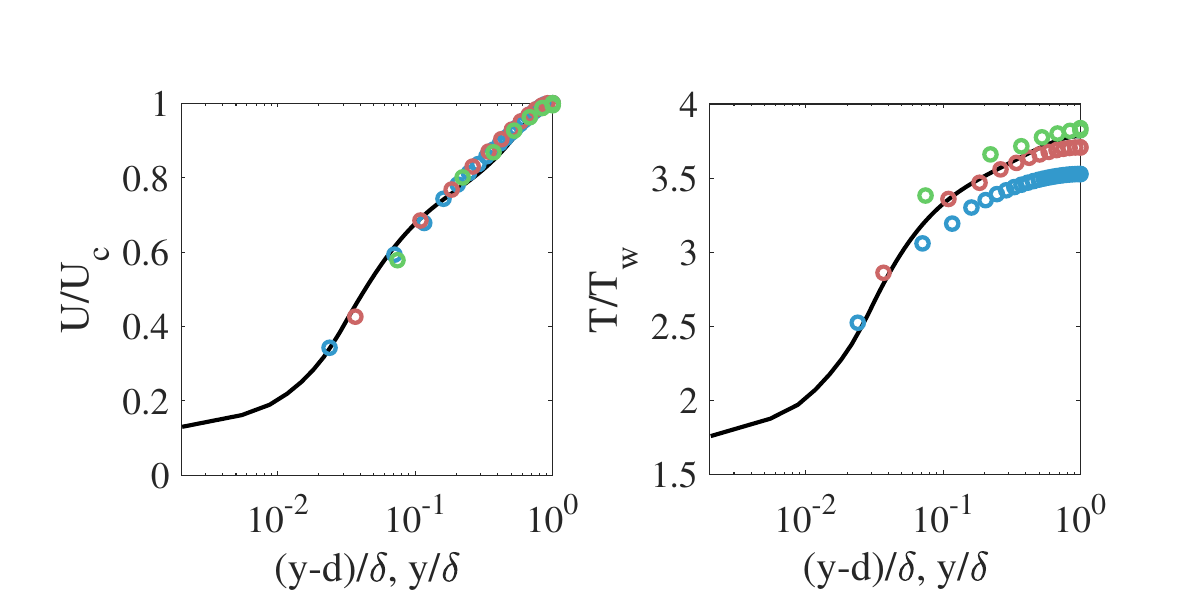}
    \put(-180,96){$(d)$}
    \put(-195,40){\rotatebox{90}{{\colorbox{white}{{$U/U_c$}}}}}
    \put(-152,-3){\color{white}\rule{40pt}{10pt}}
    \put(-146,-3){\colorbox{white}{$y/\delta$}}
    \put(-101,40){\rotatebox{90}{{\colorbox{white}{{$T/T_w$}}}}}
    \put(-60,-3){\color{white}\rule{40pt}{10pt}}
    \put(-52,-3){\colorbox{white}{$y/\delta$}}
    \caption{\emph{A-posteriori} evaluation of BFWM-rough-v2 in WMLES of
      compressible turbulent channels. Mean velocity and mean
      temperature profiles for DNS (line) and WMLES (symbols). $(a)$
      GS1 at $M_b=3.3$ and $Re_b$=15500; $(b)$ GS12 at $M_b=1.7$ and
      $Re_b$=15500; $(c)$ GS13 at $M_b=3.3$ and $Re_b$=15500; $(d)$
      GS15 at $M_b=3.3$ and $Re_b$=10000. Results are shown at three
      different grid resolutions: $\Delta/\delta = 1/20$ (blue),
      $1/10$ (red), and $1/5$ (green).
\label{fig:umean_comp}}
\end{figure}
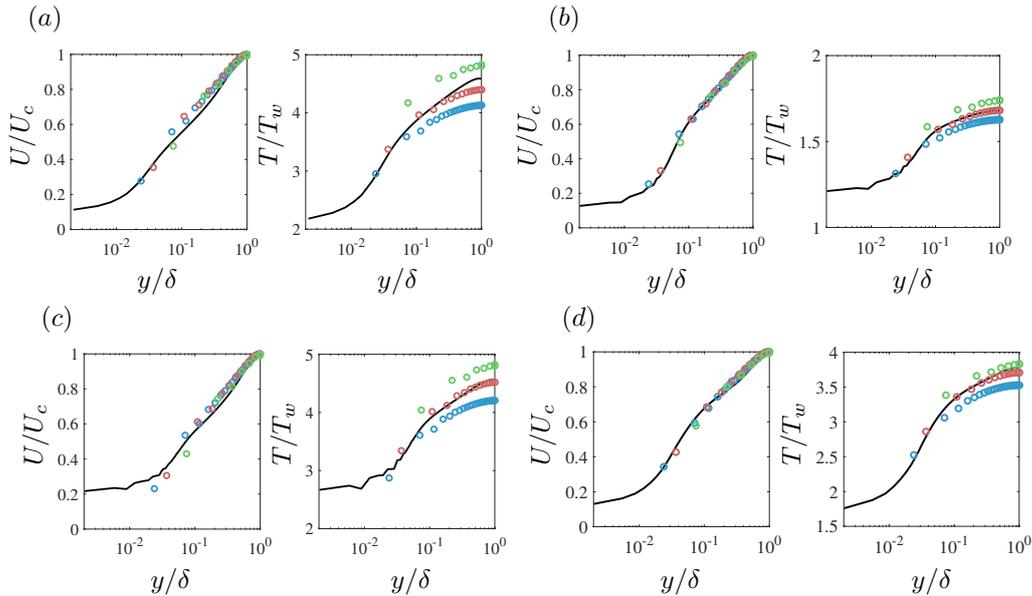

The mean velocity and mean temperature profiles for a selection of
test cases are shown in figure~\ref{fig:umean_comp}. The mean velocity
profiles exhibit good agreement with DNS across different grid
resolutions. In contrast, larger deviations are observed in the mean
temperature profiles.  Given that the \textit{a-priori} results
demonstrated stronger performance of BFWM-rough-v2, this discrepancy
suggests that the observed errors in temperature are primarily due to
external errors from the SGS model. 

\subsection{HPT blade with roughness}

The high-pressure turbine (HPT) blade with a rough surface represents
a challenging validation scenario due to its complex geometry and the
presence of multiple flow phenomena, including strong pressure
gradient effects, laminar–turbulent transition, and shock waves. Here,
we use as reference case the DNS by \cite{nardini2024direct},
conducted under engine-relevant conditions with an exit Reynolds
number of $Re_{exit} = 590{,}000$ based on the axial chord $C_{ax}$,
and an exit Mach number of $M_{exit} = 0.9$. The rough surface
considered on the blade corresponds to the rough case $k_3^s$ in
\citet{nardini2024direct}. It consists of a Gaussian roughness
representative of the surface texture associated with in-service
degradation.

\begin{figure}
\begin{center}
    \includegraphics[width=1\textwidth,trim={0.0cm 0.0cm 0.0cm 0.0cm},clip]{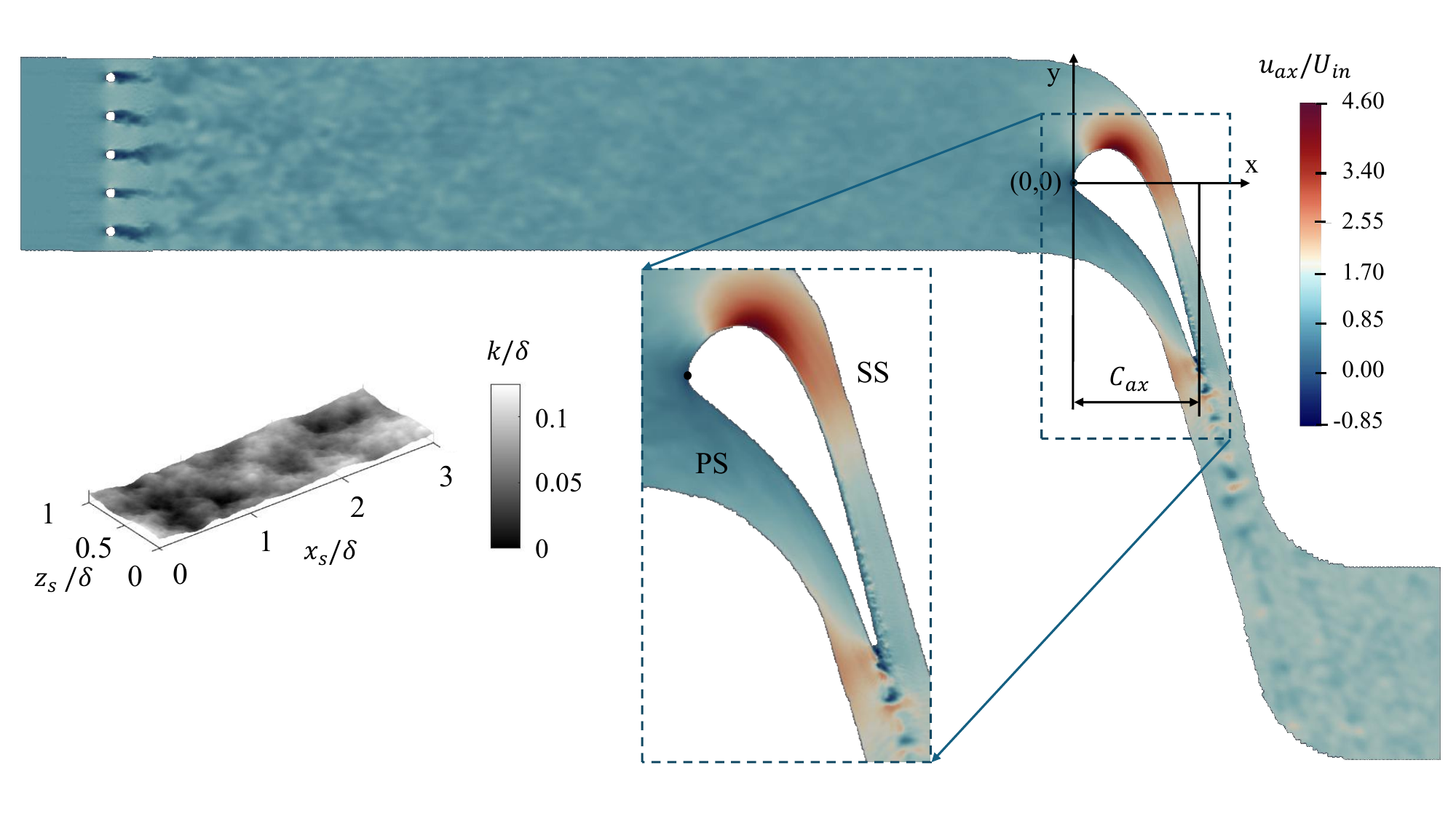} 
    \put(-450,250){$(a)$}
    \put(-450,125){$(b)$}
    \put(-306,143){\colorbox{white}{$k/\delta$}}
    \put(-365,80){\colorbox{white}{$x_s/\delta$}}
    \put(-447,70){\colorbox{white}{$z_s/\delta$}}
    \put(-112,135){\colorbox{white}{\scriptsize{$C_{ax}$}}}
    \put(-66,233){\colorbox{white}{$u_{ax}/U_{in}$}}
    \caption{\emph{A-posteriori} evaluation of BFWM-rough-v2 in WMLES of
      an HPT blade. $(a)$ Visualization of the computational setup and
      instantaneous streamwise velocity field from the WMLES of HPT
      blade.  The suction side of the blade is denoted by SS, while
      the pressure side is denoted by PS; $(b)$ Rough surface tile
      corresponding to case $k_3^s$ in \cite{nardini2024direct}.
      \label{fig:HPT_blade}}
\end{center}
\end{figure}
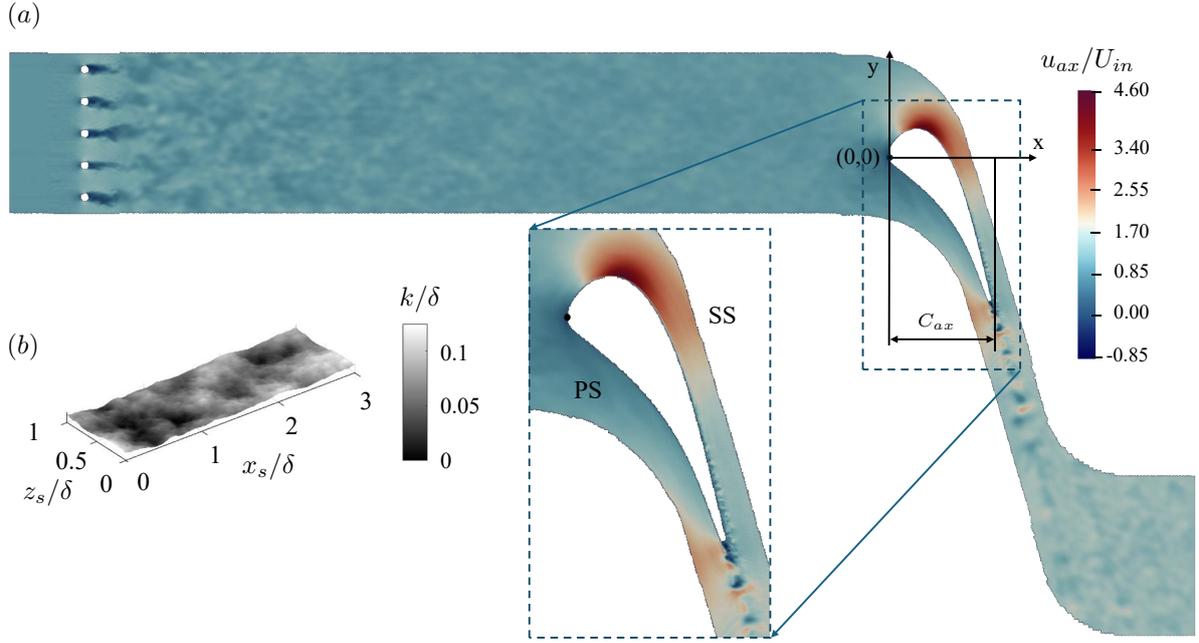

We carried out WMLES with BFWM-rough-v2 and EQWM-$k_s$. An
illustration of the simulation setup and flow field is shown in
figure~\ref{fig:HPT_blade}$(a)$.  The spanwise domain is $0.4C_{ax}$
to ensure that the largest structures develop correctly. The
turbulence intensity is set to $8\%$ by adjusting the distance of the
bars upstream of the blade. The Voronoi grids are generated with a
background grid size of $\Delta_b = 0.0311C_{ax}$.  The grids near the
upstream bars are refined to $\Delta_b/2$ with 20 layers, and the flow
near the bars is wall-resolved. The grids near the blade are refined
by 4 levels, with each level reducing the grid size by $50\%$ and
adding 30 layers.  With this mesh, the blade boundary layer is
resolved by approximately $0$ to $30$ control volumes, with near-zero
values occurring at the leading edge. The roughness sample tile is
visualized in figure~\ref{fig:HPT_blade}$(b)$, and the key roughness
parameters in Table~\ref{tab:blade_roughness} are extracted from the
sample tile and used as input information for BFWM-rough-v2.
%
\begin{table}
\begin{center}
\def~{\hphantom{0}}
\renewcommand{\arraystretch}{2}
    \begin{tabular}{ccccccc}
    \hline
     Case & $k_{rms}/C_{ax}$ & $R_{a}/C_{ax}$ & $S_k$ & $K_u$ & $ES$  \\
     $k_3^s$ &  $0.60 \times 10^{-3}$ & $0.48 \times 10^{-3}$ & 0.0 & 3.0 & 0.18  &\\
     \hline
    \end{tabular}
    \caption{\label{tab:blade_roughness} The key geometrical roughness
      parameters of the blade surface roughness $k_3^s$ from
      \cite{nardini2024direct}.}
    \end{center}
\end{table}

\begin{figure}
\centering
    \includegraphics[width=70mm]{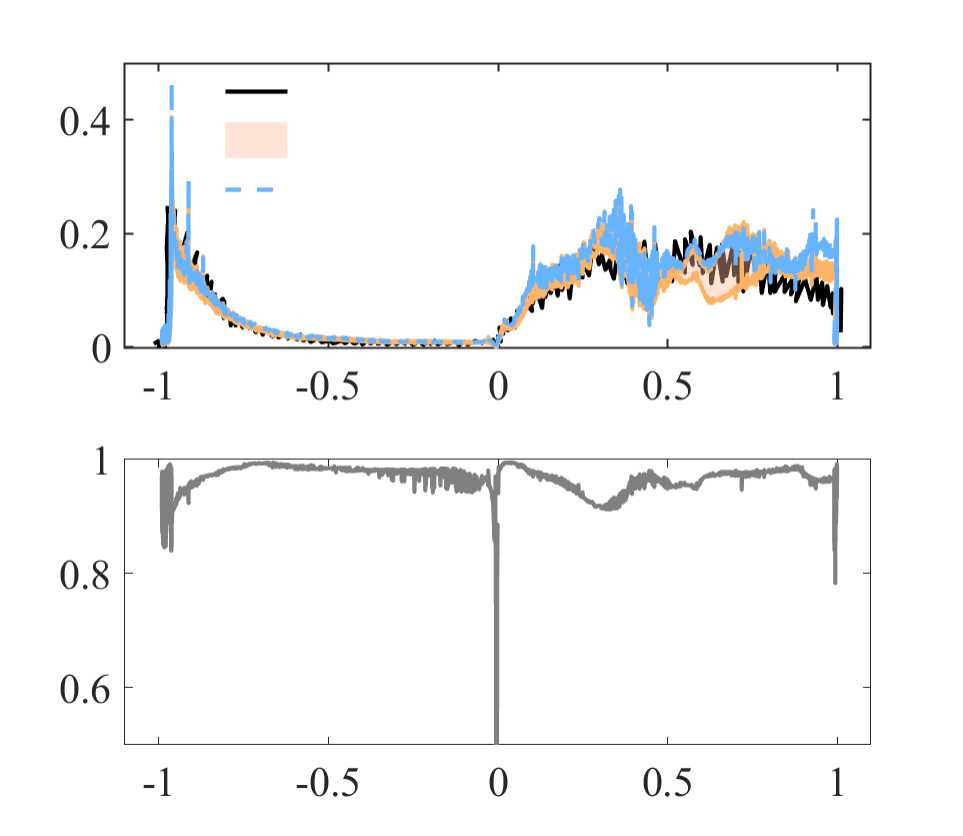}
    \put(-200,167){$(a)$}
    \put(-135,151){\scriptsize{DNS, $k_3^s$, Nardini et al.}}
    \put(-135,141){\scriptsize{WMLES, EQWM-$k_s$}}
    \put(-135,131){\scriptsize{WMLES, BFWM-rough-v2}}
    \put(-202,125){$C_f$}
    \put(-202,45){$C_{\tau_w}$}
    \put(-113,-3){$x/C_{ax}$}
    \includegraphics[width=70mm]{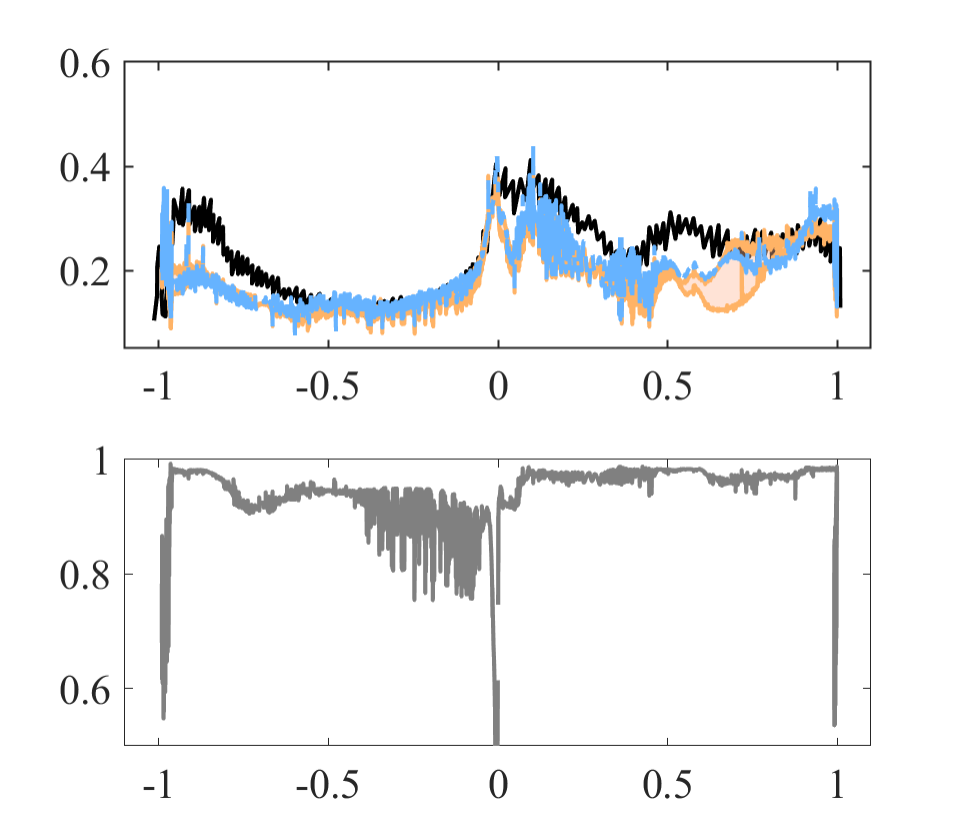}
    \put(-200,167){$(b)$}
    \put(-202,125){$Q$}
    \put(-202,45){$C_{q_w}$}
    \put(-113,-3){$x/C_{ax}$}
    \caption{\emph{A-posteriori} evaluation of BFWM-rough-v2 and
      EQWM-$k_s$ in WMLES of an HPT blade. $(a)$ Time and
      spanwise-averaged skin friction coefficient $C_f$ and $(b)$ time
      and spanwise-averaged wall heat flux $Q$ along the axial
      position of the blade normalized by axial chord length
      $x/C_{ax}$. The time and spanwise-averaged confidence score for
      the prediction of $C_f$ and $Q$ is presented at the bottom
      line. The DNS results are from \cite{nardini2024direct}. The
      shaded area denotes the WMLES with EQWM-$k_s$, where $k_s=A
      k_{rms}$, and $A=3\sim7$. Note that $x/C_{ax} > 0$ and $x/C_{ax}
      < 0$ correspond to the SS and PS, respectively, with $x/C_{ax} =
      0$ locate at the leading edge of the blade.
\label{fig:blade_C}}
\end{figure}
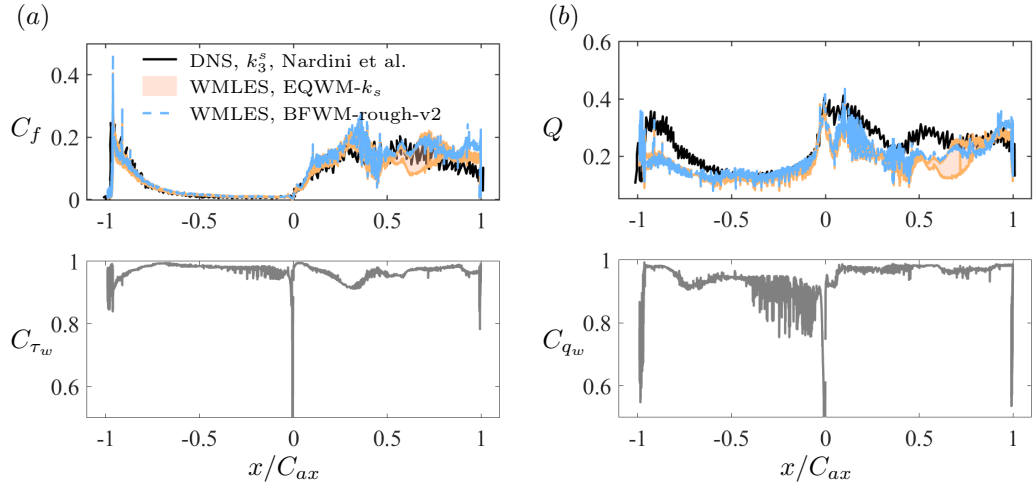

Figures~\ref{fig:blade_C}$(a)$ and $(b)$ show the skin friction
coefficient $C_f$ and the wall heat flux $Q$ along the axial position
of the blade for WMLES using BFWM-rough-v2 and EQWM-$k_s$.  Following
\cite{nardini2024direct}, the skin friction coefficient is defined as
$C_f = 2\tau_w / (\rho_{in} U_{in})$, and the wall heat flux is
defined as $Q = q_w T_{in} / U_{in}^2$, where subscript $in$ denotes
quantities at the inlet. For EQWM-$k_s$, the equivalent sand-grain
roughness height is expressed as $k_s = A k_{\mathrm{rms}}$, where $A$
is a constant scaling factor.  Although \cite{hama1954boundary}
originally proposed $A = 5$ for Gaussian roughness surfaces, this
value is not universally applicable, as other roughness
characteristics, such as $ES$, can also influence the value of $A$. In
our DNS database, the value of $A$ varies considerably, ranging from
3.5 to 6.5 for Gaussian roughness surfaces. This variability
introduces uncertainty in the estimation of $k_s$. To account for
this, the EQWM-$k_s$ model is evaluated over an extended range of $A =
3$--$7$.

The WMLES results are compared to the DNS from
\cite{nardini2024direct}. Both BFWM-rough-v2 and EQWM-$k_s$ capture
the overall trends of $C_f$ and $Q$ on both the suction side (SS) and
pressure side (PS) of the blade. However, the uncertainty in $k_s$
introduces a pronounced deviation in EQWM-$k_s$ predictions of $C_f$
and $Q$ within $0.5 < x/C_{ax} < 0.8$, where BFWM-rough-v2
demonstrates better predictive accuracy. In this region,
roughness-induced normal shock waves give rise to complex interactions
among shock-induced vortices, separation bubbles, and the boundary
layer. On the PS, both wall models underpredict $Q$ near the trailing
edge, while on the SS, the underprediction extends over $0 < x/C_{ax}
< 0.8$, where transition occurs.

In terms of confidence score, the predictions are generally associated
with high values (typically above 0.9). Some exceptions occur near the
stagnation point at the leading edge, where the first few points
exhibit near-zero confidence. The former could be due to lack of grid
points at the thin leading edge boundary layer whereas the latter
could be associated to the strong curvature effects. While the
confidence levels for $Q$ are comparable to those for $C_f$, the
prediction errors in $Q$ are slightly larger in some regions. This
discrepancy may reflect that, even when the model encounters inputs
similar to those in the training set (and therefore assigns high
confidence), the current building-block flows do not fully capture the
complex physics of the HPT blade. In addition, the observed errors may
be affected by limitations of the SGS model, or by a combination of
both factors.

\subsection{Hypersonic compression ramp with surface roughness}

\begin{figure}
\begin{center}
    \includegraphics[width=130mm,trim={0.0cm 5.0cm 0.0cm 5.0cm},clip]{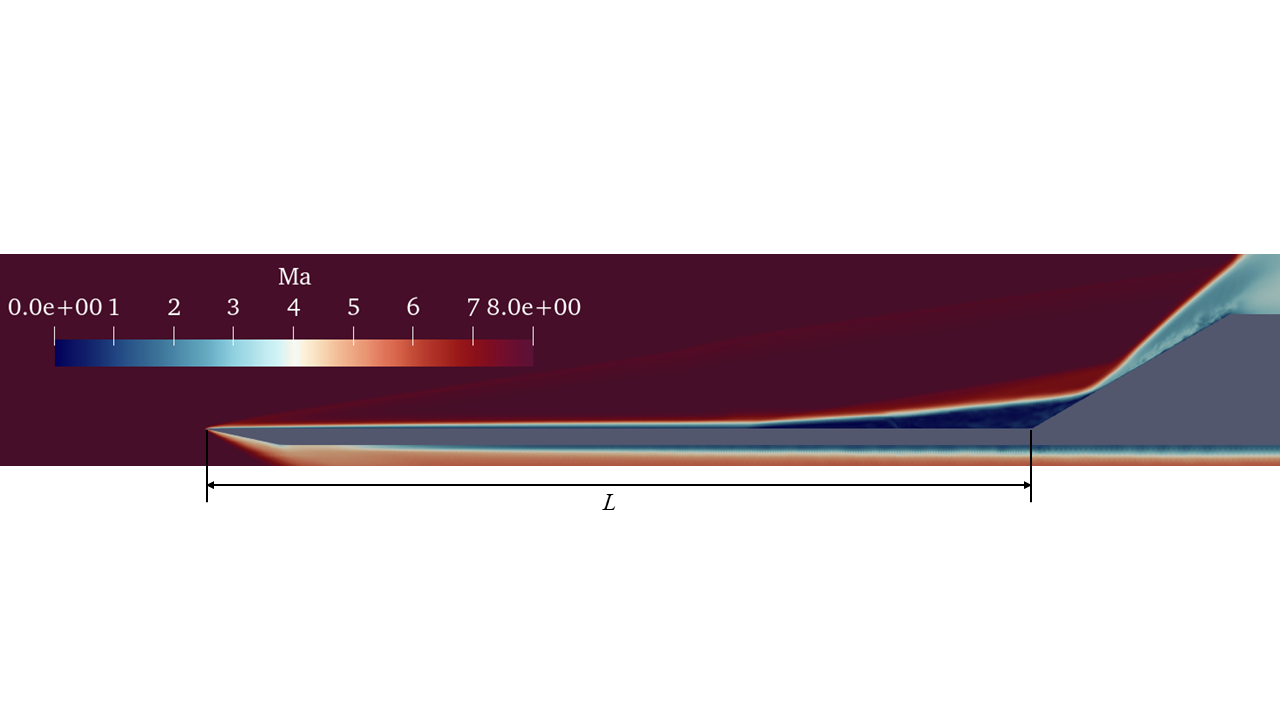}
    \put(-370,85){$(a)$}
    \hspace{3mm}
    \includegraphics[width=40mm]{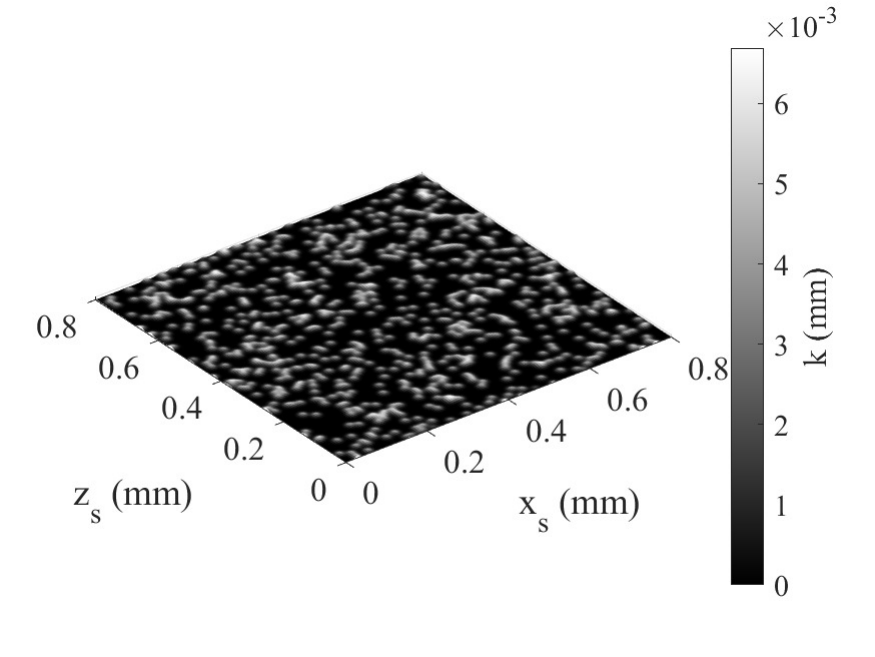}
    \includegraphics[width=40mm]{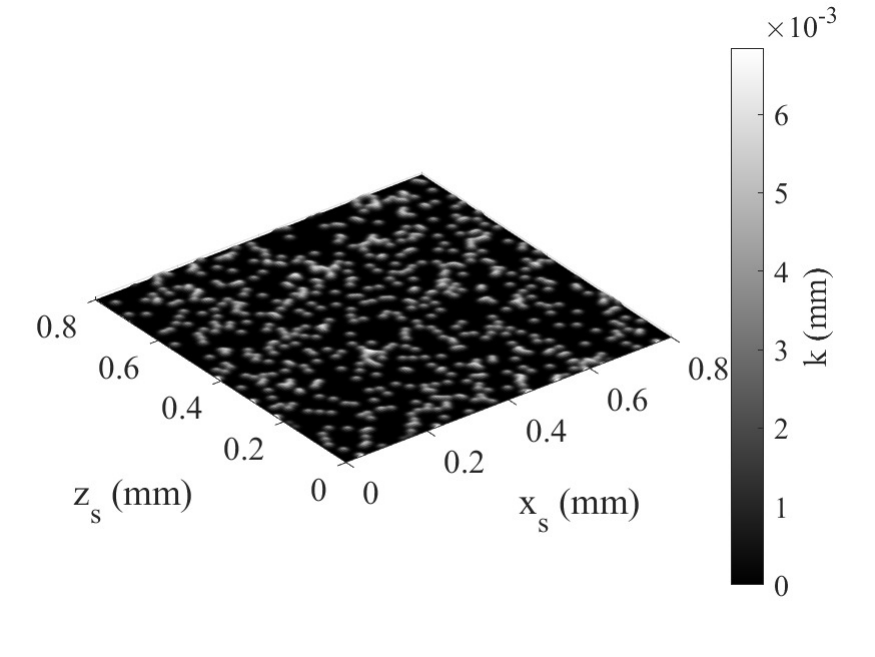}
    \includegraphics[width=40mm]{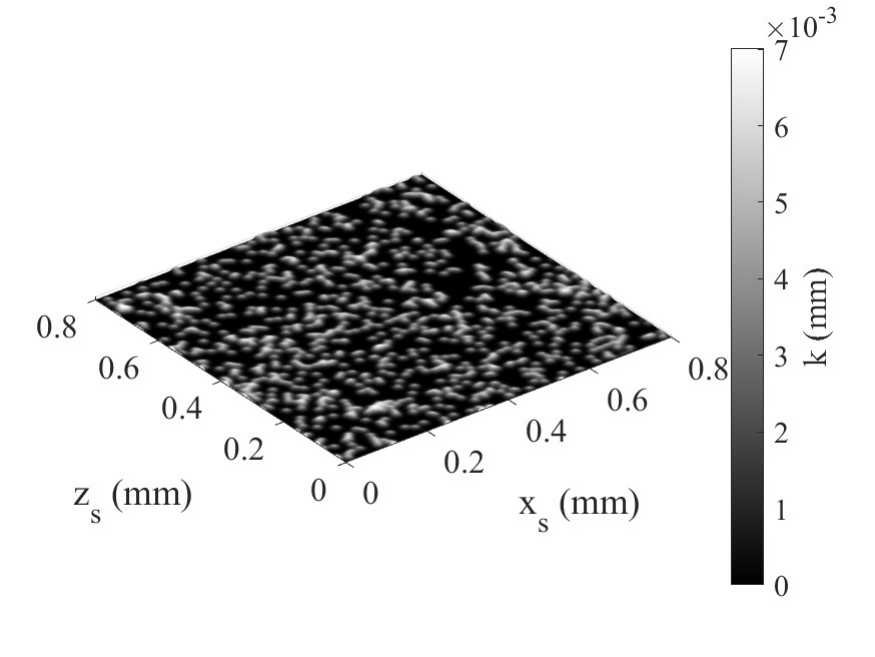}
    \put(-358,82){$(b)$}
    \caption{\emph{A-posteriori} evaluation of BFWM-rough-v2 in WMLES of
      a hypersonic compression ramp. $(a)$ Flow field over the
      compression ramp visualized by the contour plot of instantaneous
      Mach number. $(b)$ The roughness tiles of sand-grain roughness,
      SG1 (left), SG2 (middle), and SG3 (right) corresponding to the
      roughness in \cite{prince2005}. The contour plot shows the
      surface elevation of the roughness, with units in millimeters.
\label{fig:ramp_roughness}}
\end{center}
\end{figure}

We evaluate the model in a hypersonic compression ramp with surface
roughness. This configuration is relevant to high-speed entry
vehicles, particularly for understanding the aerothermal loads on
control surfaces at high flap angles, including surface heating and
pressure forces during atmospheric entry. The setup, illustrated in
Figure~\ref{fig:ramp_roughness}(a), follows the experimental
configuration of \citet{prince2005}. The ramp angle is $30^\circ$, and
the roughness was found to significantly influence peak heating and
pressure at high flap angles ($\alpha_f \ge 30^\circ$). The
free-stream Mach number is $\textit{M}_\infty = 8.2$ and the flow
occurs under low-enthalpy conditions, with no chemical reactions
involved.  The Reynolds number is $\textit{Re} = 1.44 \times 10^6$,
based on the distance from the leading edge to the flap hinge and
free-stream conditions, denoted by $L$. Surface roughness is
introduced by applying sandpaper to both the plate and flap surfaces.

For WMLES, a uniform velocity profile is imposed at the inflow
boundary, while a sponge layer is applied at the outlet. Periodic
boundary conditions are used in the spanwise direction. The farfield
grid spacing is uniform with $\Delta = 0.1L$. Near the solid surfaces,
mesh refinement is applied using seven successive levels, each
reducing the grid spacing by 50\%, resulting in approximately 30 grid
points across the boundary layer at the flap hinge.  Additionally, a
refinement zone with constant spacing $\Delta \approx 30 \times
10^{-4} L$ is applied above both the plate and the ramp to accurately
resolve the separation shock wave, reattachment shock wave, and the
shock wave originating from the leading edge. In the simulations, a
synthetic sand-grain rough surface labeled `SG1' is generated to match
the experimentally reported equivalent sand-grain roughness height of
$k_s = 0.3\,\text{mm}$. To evaluate the sensitivity of wall heat flux
to variations in roughness, two additional surfaces (`SG2' and `SG3')
are constructed by adjusting the number of sand-grain elements to
achieve $k_s = 0.25\,\text{mm}$ and $k_s = 0.35\,\text{mm}$,
respectively.  A summary of the roughness characteristics is provided
in Table~\ref{tab:ramp_roughness}, and a visualization of the flow
field and rough surfaces is shown in figure~\ref{fig:ramp_roughness}.
\begin{table}
    \centering
    \begin{tabular}{lcccccc}
        Case   & No. of elements & $k_{rms}$ (mm)  & $R_a$ (mm) & $k_{rms}/R_a$ & ES & $k_s$ (mm)\\
        \hline
        SG1    & 800 & 0.022  & 0.0186 & 1.183 & 0.12 & 0.30 \\
        SG2    & 600 & 0.020  & 0.0175 & 1.143 & 0.11 & 0.25 \\
        SG3    & 1000 & 0.024  & 0.0203 & 1.182 & 0.13 & 0.35 \\
        \hline
    \end{tabular}
    \caption{The key roughness parameters of sand-grain roughness SG1,
      SG2, and SG3.}
    \label{tab:ramp_roughness}
\end{table}

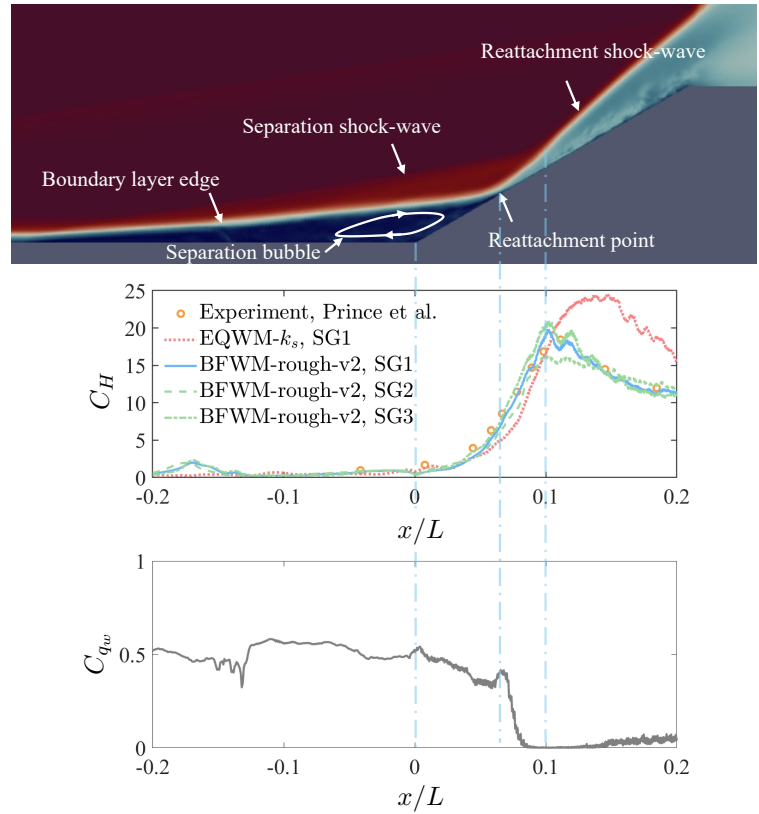
\begin{figure}
\centering
    \includegraphics[width=170mm,trim={3.0cm 0cm 1.0cm 0.0cm},clip]{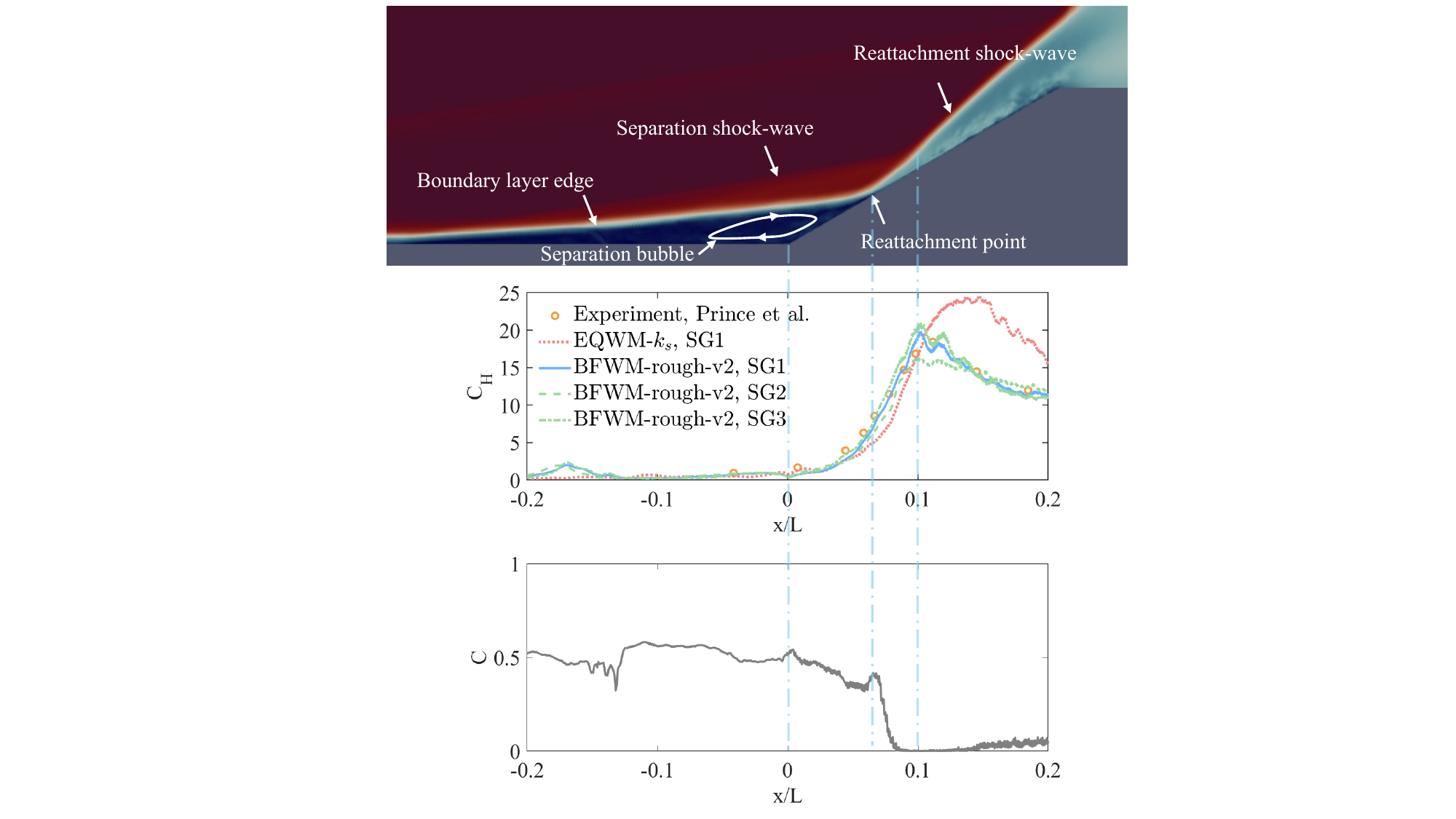}
    \put(-362,50){\rotatebox{90}{{\colorbox{white}{{$C_{q_w}$}}}}}
    \put(-360,150){\rotatebox{90}{{\colorbox{white}{{$C_H$}}}}}
    \put(-245,4){{\colorbox{white}{{$x/L$}}}}
    \put(-245,105){{\colorbox{white}{{$x/L$}}}}
    \caption{\emph{A-posteriori} evaluation of BFWM-rough-v2 and
      EQWM-$k_s$ in WMLES of a hypersonic compression ramp. The top
      panel illustrates the shock–boundary-layer interaction at the
      compression corner. The middle panel shows the wall heat
      transfer coefficient along the surface, comparing WMLES
      predictions using BFWM-rough-v2 (lines) and EQWM-$k_s$ (dots)
      with experimental data from \cite{prince2005} (symbols). The
      bottom panel displays the corresponding time and
      spanwise-averaged confidence score. The origin corresponds to
      the start of the ramp flap.}
\label{fig:ramp_h}
\end{figure}

Figure~\ref{fig:ramp_h} presents the flow dynamics of the
shock--boundary layer interaction at the compression corner, together
with the heat transfer coefficient and the confidence score along the
streamwise direction. The heat transfer coefficient is defined as $C_H
= q_w/(\rho_\infty U_\infty C_p (T_w - T_r))$, where $T_r$ is the
recovery temperature. Upstream of the ramp, the heat transfer levels
remain low, and both BFWM-rough-v2 and EQWM-$k_s$ predict these values
with reasonable accuracy. In this region, however, the model exhibits
only moderate confidence, with scores around 0.5. This reduced
confidence likely reflects the departure of the flow from a
quasi-equilibrium boundary-layer state due to strong adverse pressure
gradients, which drive the formation of a separated shear layer and a
recirculation bubble. Such separation-induced dynamics lie outside the
equilibrium assumptions of BFWM-rough-v2, reducing the reliability of
the predictions despite the agreement in heat transfer levels.

At the reattachment point, the separated shear layer impinges on the
wall and produces a reattachment shock that compresses the flow. The
shear layer impingement and the reattachment shock inject a large
amount of momentum and energy into the near-wall region, resulting in
excessive energy transfer to the wall and giving rise to the heat flux
overshoot phenomenon, where the wall heat flux spikes and exceeds the
theoretical inviscid level. The WMLES results show that the
BFWM-rough-v2 model predictions align well with the experimental data
from \cite{prince2005}, capturing the sharp rise in heat flux at the
ramp induced by surface roughness. The results also indicate that
roughness density variation has minimal impact on the overall
distribution of $C_H$, affecting only the peak values. In contrast,
EQWM-$k_s$ overpredicts the heating on the ramp by approximately 60\%
and incorrectly predicts the peak further downstream.

Finally, we observe that the confidence drops almost to zero at the
location where the heat flux overshoot occurs. This behavior is
attributed to the fact that the local flow inputs in this region lie
outside the range represented in the training data. As shown in
figure~\ref{fig:ramp_roughness}, the boundary layer downstream of the
reattachment point becomes supersonic, with a peak Mach number at the
matching location of $M_m \approx 2.6$, whereas the maximum value in
the training dataset is $M_m \approx 0.7$.  It is important to note
that low confidence does not necessarily imply inaccurate
predictions. Rather, it indicates that the model is operating in an
extrapolative regime. Accurate predictions may still be obtained under
extrapolation if the model captures the correct dimensionless scaling
of the problem, as is the case here with $M_m$. Nonetheless, the
likelihood of model underperformance increases in low-confidence
regions, and such areas should be carefully inspected to identify
potential issues.

\subsection{Hypersonic blunt body with surface roughness}

The last case considered is a blunt body with surface roughness. This
case is relevant to EDL vehicles with ablated surfaces. The
experimental data are obtained from NASA Langley Research Center’s
20-Inch Mach 6 Air Tunnel~\citep{hollis2014distributed}. The test
model features a hemispherical nose with sand-grain roughness, and the
physical radius of the body is $R_B = 0.0762~\text{m}$. Validation is
performed against the ``20-Mesh'' roughness case reported in
\cite{hollis2014distributed}, where the roughness height in viscous
units, $k^+$, ranges from 50 to 200. The sand-grain roughness on the
model is created by applying an adhesive layer to a bed of uniformly
manufactured spherical glass particles. For this problem, the flow
over the hemisphere with a smooth surface remains laminar, whereas the
presence of roughness triggers early transition and the development of
a turbulent boundary layer on the hemisphere surface. Specifically,
transition is initiated near the stagnation point.  Three cases at
different Reynolds numbers (per unit length) are considered, and the
corresponding flow conditions are summarized in
Table~\ref{tab:EDL_condition}. The flow takes place under low-enthalpy
conditions, with no chemical reactions involved.
\begin{table}
    \centering
    \begin{tabular}{ccccccc}
        $Re_{\infty} (\text{1/ft})$ & $M_{\infty}$ & $T_{\infty}$ (K) & $P_{\infty}$ (Pa) & $\rho_{\infty}$ ($\text{kg/$\text{m}^3$}$) & $U_{\infty}$ (m/s) & $h_{FR}$ ($\text{kg/$\text{m}^2$/s}$) \\
        \hline
        $3.03 \times 10^6$     & 5.99         & 62.5             & 844.1             & $4.708 \times 10^{-2}$ & 948.7  &  $2.163 \times 10^{-1}$\\
        $5.04 \times 10^6$     & 6.02         & 63.2             & 1421.0            & $7.843 \times 10^{-2}$ & 957.5 & $2.213 \times 10^{-1}$  \\
        $8.34 \times 10^6$     & 6.03         & 58.6             & 2090.8            & $1.249 \times 10^{-1}$ & 918.1 & $3.407 \times 10^{-1}$ \\
        \hline
    \end{tabular}
    \caption{Test conditions for the hemisphere model, corresponding
      to the experiment conditions for LaRC 20-Inch Mach 6 air tunnel
      Test 6975 in \citet{hollis2014distributed}.}
    \label{tab:EDL_condition}
\end{table}

\begin{figure}
\begin{center}
    \includegraphics[width=70mm,trim={8.0cm 2.0cm 8.0cm 1.0cm},clip]{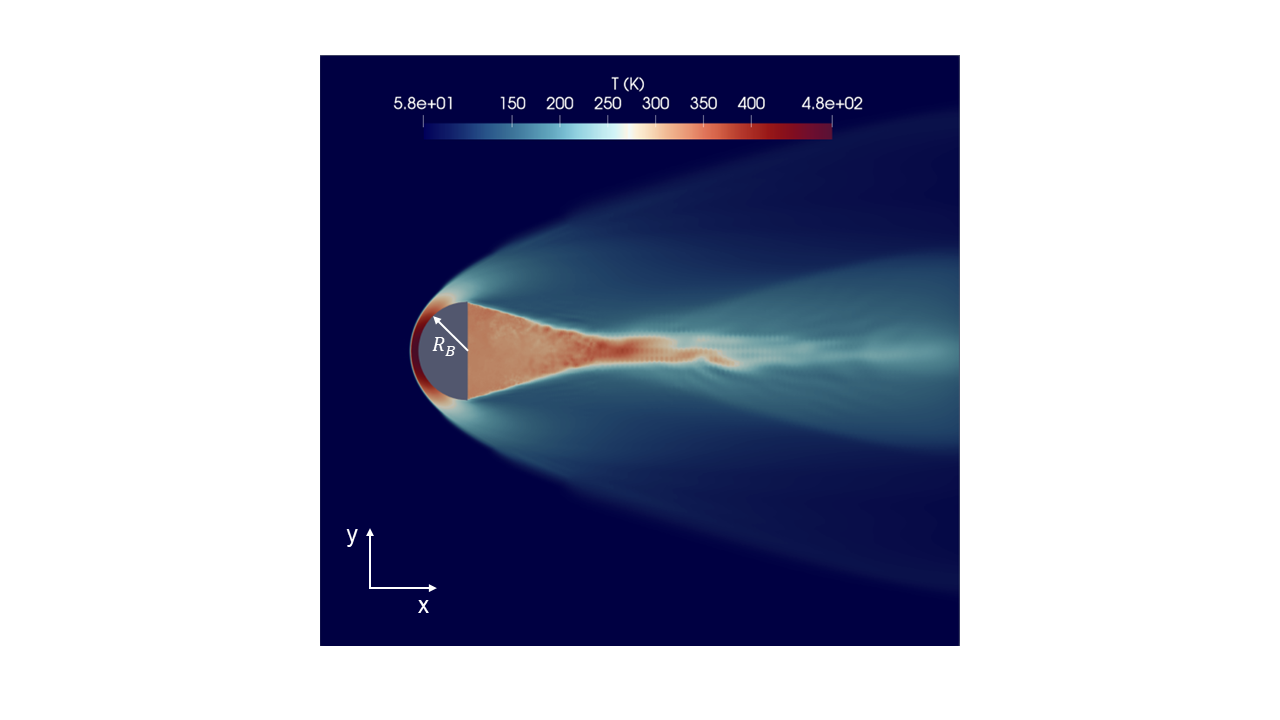}
    \includegraphics[width=60mm]{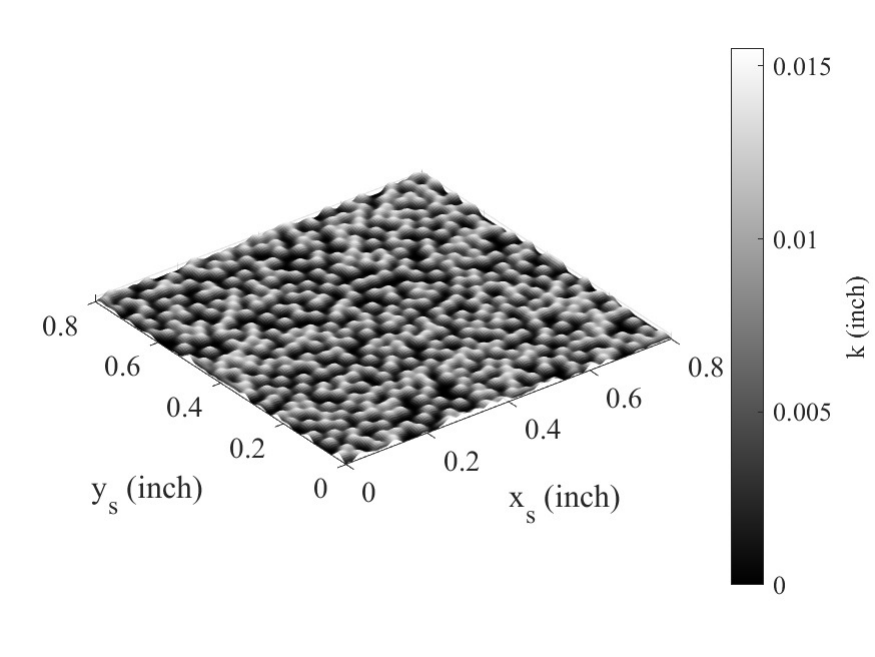}
    \put(-370,180){$(a)$}
    \put(-165,106){$(b)$}
    \caption{\emph{A-posteriori} evaluation of BFWM-rough-v2 in WMLES of
      a hypersonic blunt body. $(a)$ Instantaneous flow field for
      WMLES of hemisphere with 20-Mesh sand-grain roughness at Mach 6
      across a free stream Reynolds number $Re_{\infty}=8.34 \times
      10^6/$ft. The contour shows the temperature variation at the
      plane $z=0$. $(b)$ A representative tile of randomly distributed
      sand-grain roughness, corresponding to the 20-Mesh roughness
      layout provided in \cite{hollis2014distributed}. The contour
      plot illustrates the surface elevation of the roughness, with
      units in inches.
\label{fig:EDL_roughness}}
\end{center}
\end{figure}

For WMLES, the freestream velocity, pressure, and temperature are
prescribed at the inflow and lateral boundaries ($y$ and $z$
directions). To suppress spurious acoustic reflections, a
characteristic boundary condition with a damping (sponge) region is
imposed near the outflow. The computational grid employs a Voronoi
mesh with a base grid size of $\Delta = 0.6 R_B$. Near the
hemispherical surface, the grid is locally refined over 10 levels,
halving the grid size at each level, to achieve approximately six
control volumes across the boundary layer at $x/R_B =
0.4$. Additionally, a 5-level refinement region is applied around the
shock wave to improve resolution in those
areas. Figure~\ref{fig:EDL_roughness}$(a)$ displays the configuration
setup and the temperature field obtained from the WMLES for one case.

To obtain the roughness parameters required as input for
BFWM-rough-v2, a synthetic roughness tile representative of the
20-Mesh case from \cite{hollis2014distributed} is generated by
randomly placing hemispherical elements on a flat substrate. The
diameter of each hemispherical element is set to $D_s =
0.0336~\text{in}$, consistent with the experimental setup. The number
of elements is adjusted to replicate both the roughness density
observed in the experimental roughness tile and the equivalent
sand-grain roughness height of $k_s = 0.0168$ inches. An example of
the resulting surface is shown in
Figure~\ref{fig:EDL_roughness}$(b)$. The key roughness parameters are
summarized in Table~\ref{tab:EDL_roughness}.
\begin{table}
    \centering
    \begin{tabular}{lcccccc}
    \hline
        Case                  & Element & Diam. (mm)  & Diam. (inch) & $k_s$ (inch) & $k_{rms}/R_a$ & ES \\
        20-Mesh               & hemispherical & 0.8530  & 0.0336 & 0.0168 & 1.156 & 0.330 \\
        \hline
    \end{tabular}
    \caption{The key roughness parameters of 20-Mesh sand-grain
      roughness.}
    \label{tab:EDL_roughness}
\end{table}

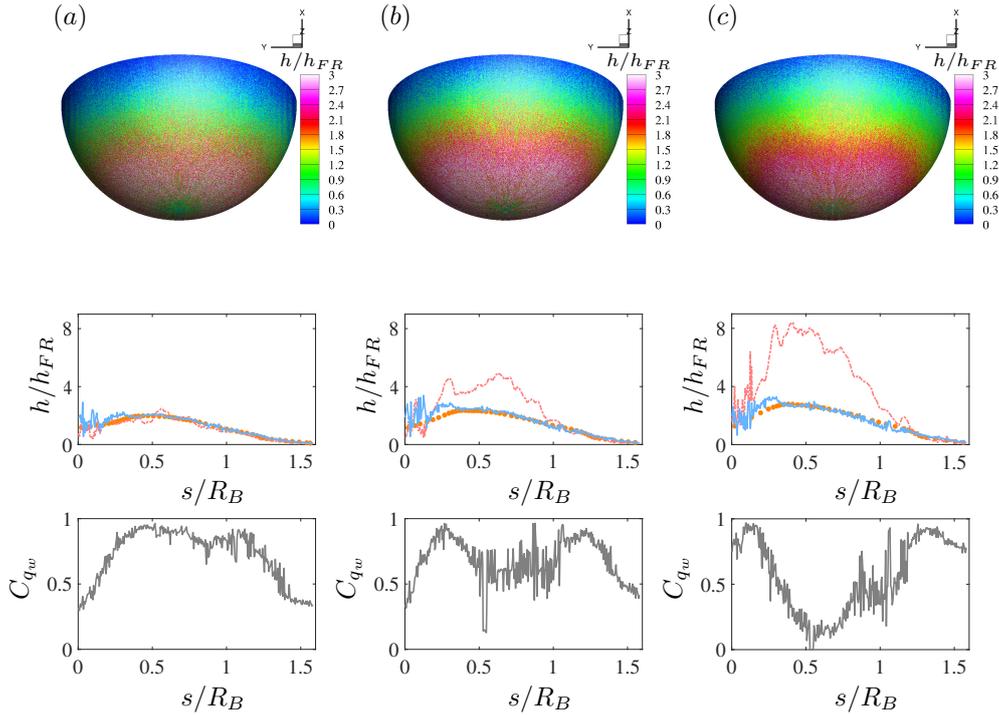
\begin{figure}
\begin{center}
    \includegraphics[width=43mm,trim={1.0cm 2.0cm 0.5cm 0.2cm},clip]{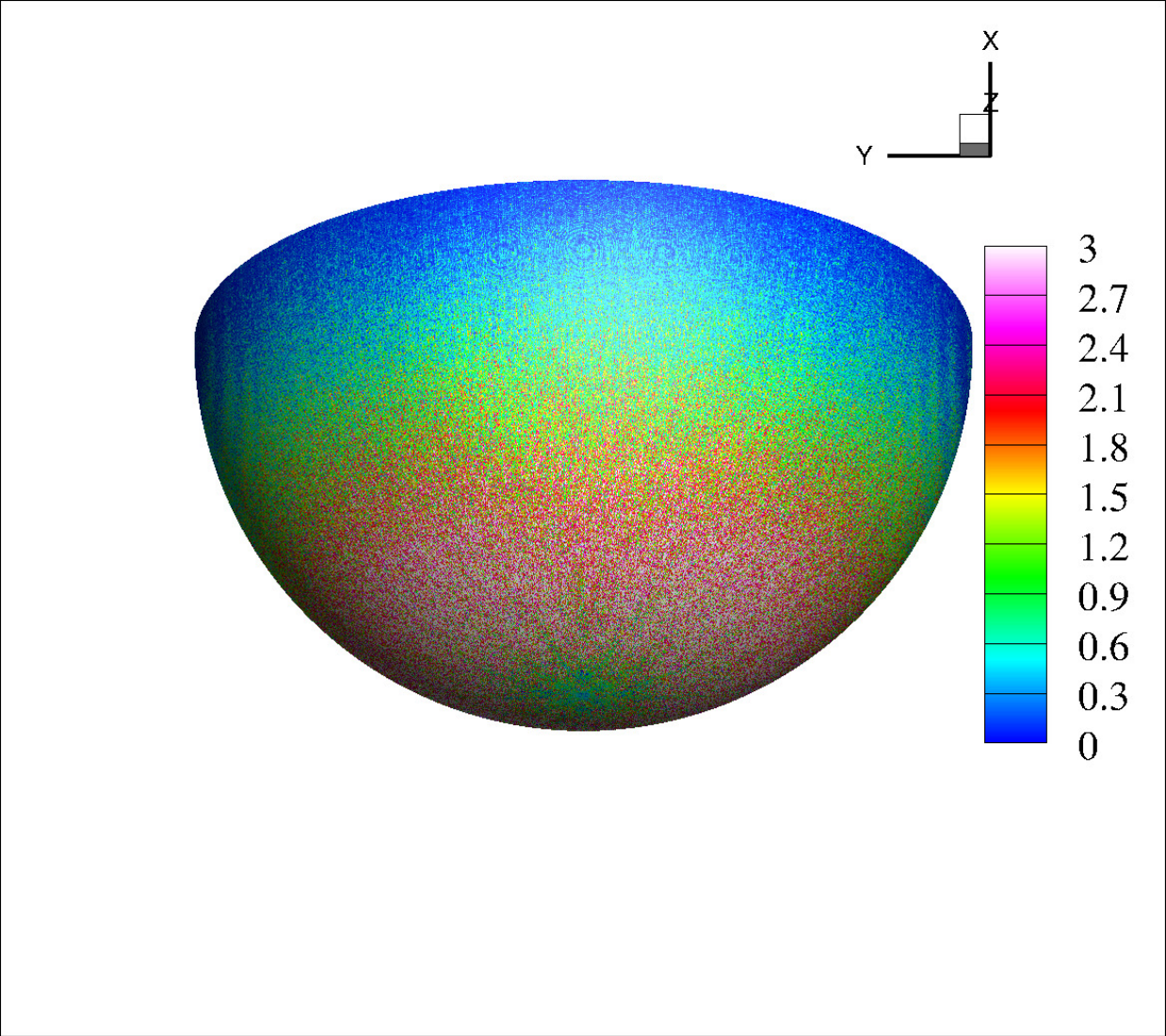}
    \put(-110,96){$(a)$}
    \put(-25,80){\scriptsize{$h/h_{FR}$}}
    \includegraphics[width=43mm,trim={1.0cm 2.0cm 0.5cm 0.2cm},clip]{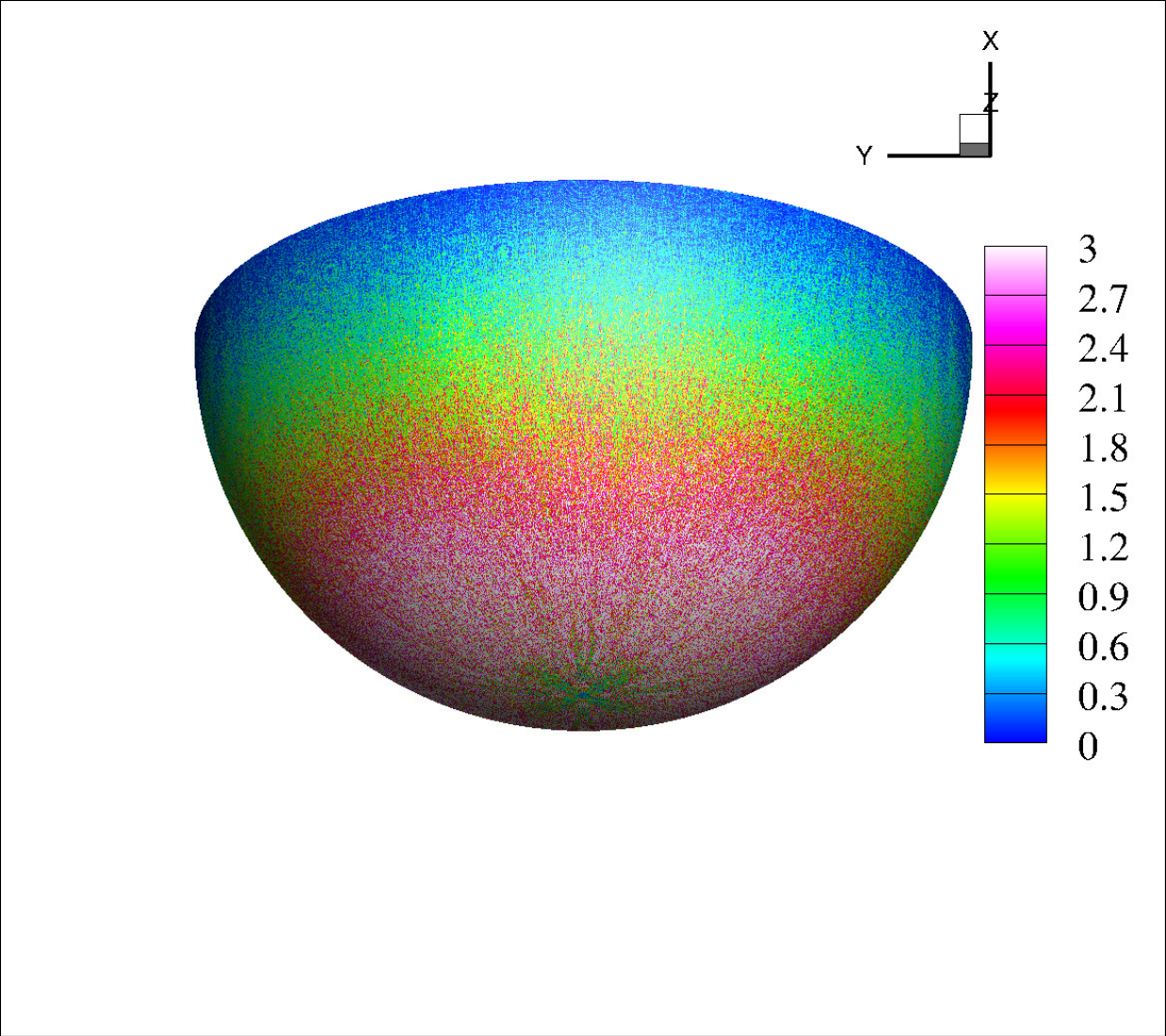}
    \put(-110,96){$(b)$}
    \put(-25,80){\scriptsize{$h/h_{FR}$}}
    \includegraphics[width=43mm,trim={1.0cm 2.0cm 0.5cm 0.2cm},clip]{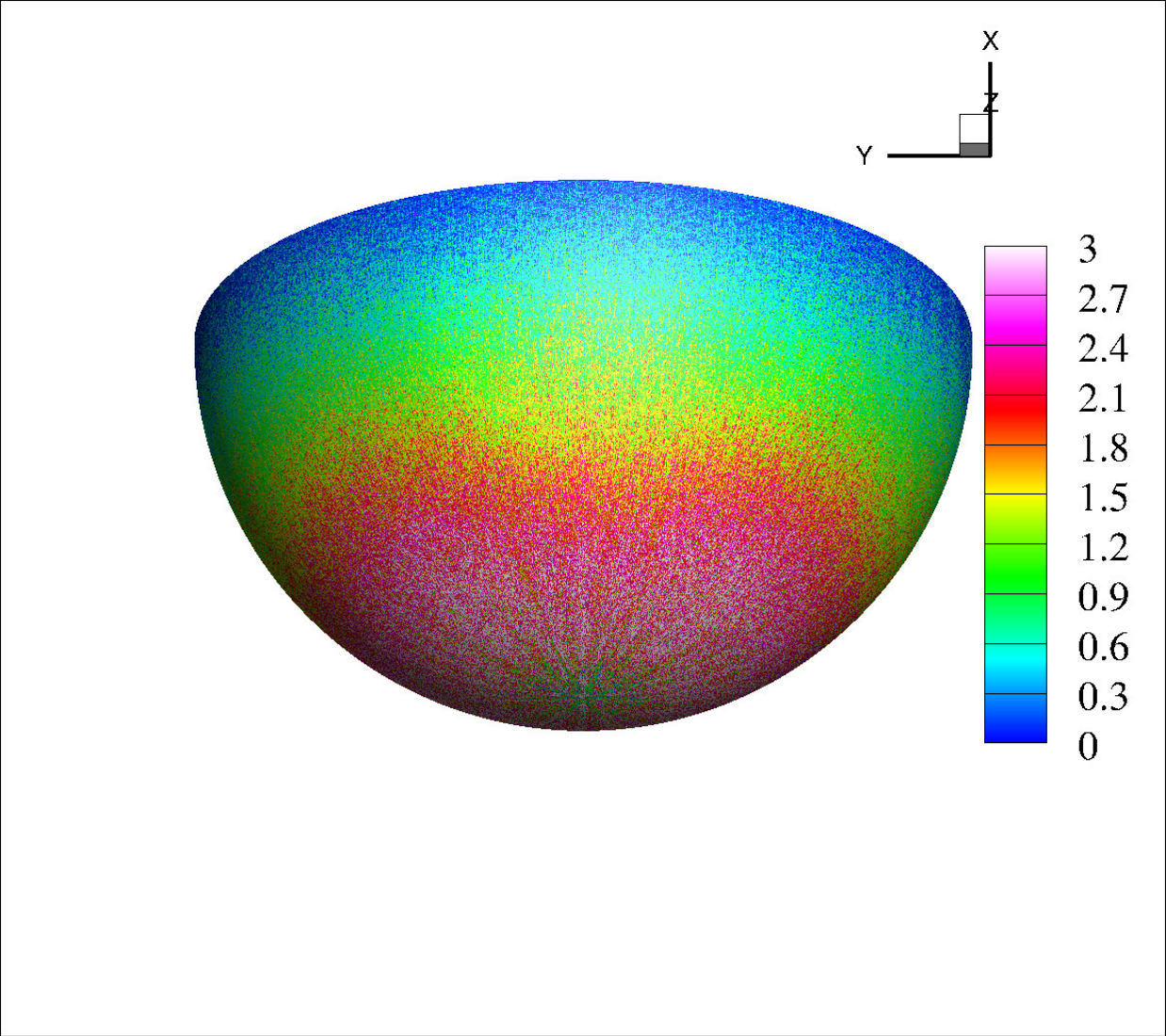}
    \put(-110,96){$(c)$}
    \put(-25,80){\scriptsize{$h/h_{FR}$}}
    \hspace{3mm} 
    \includegraphics[width=43mm]{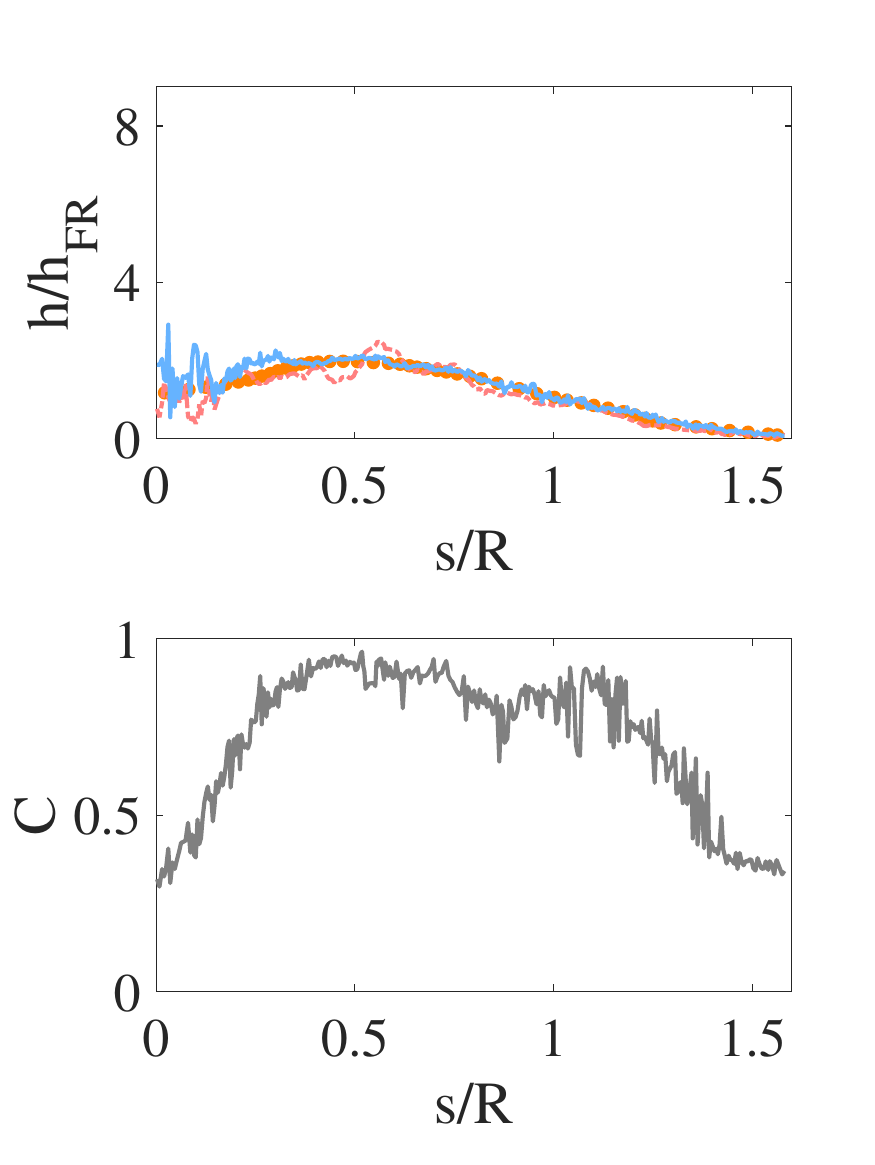}
    \put(-122,110){\rotatebox{90}{{\colorbox{white}{{$h/h_{FR}$}}}}}
    \put(-129,40){\rotatebox{90}{{\colorbox{white}{{$C_{q_w}$}}}}}
    \put(-65,3){\colorbox{white}{$s/R_B$}}
    \put(-65,81){\colorbox{white}{$s/R_B$}}
    \includegraphics[width=43mm]{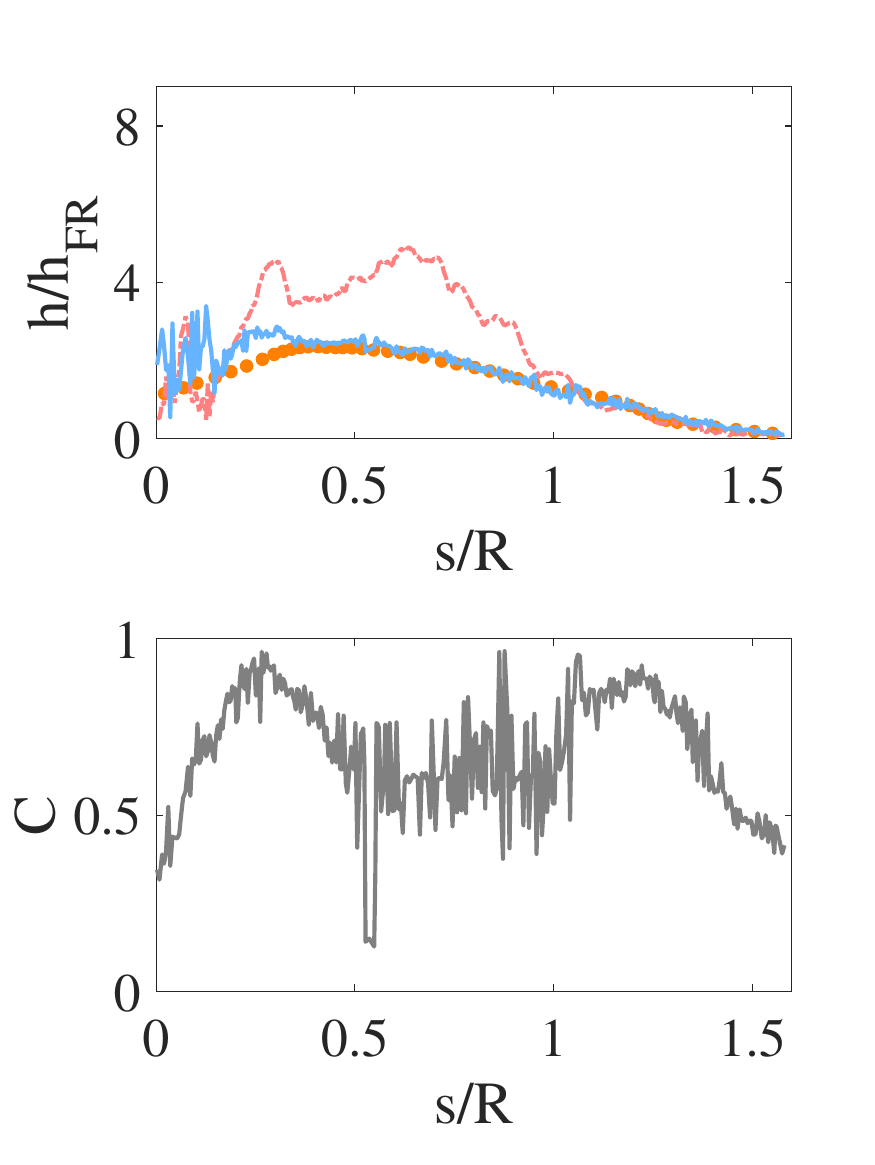}
    \put(-122,110){\rotatebox{90}{{\colorbox{white}{{$h/h_{FR}$}}}}}
    \put(-129,40){\rotatebox{90}{{\colorbox{white}{{$C_{q_w}$}}}}}
    \put(-65,3){\colorbox{white}{$s/R_B$}}
    \put(-65,81){\colorbox{white}{$s/R_B$}}
    \includegraphics[width=43mm]{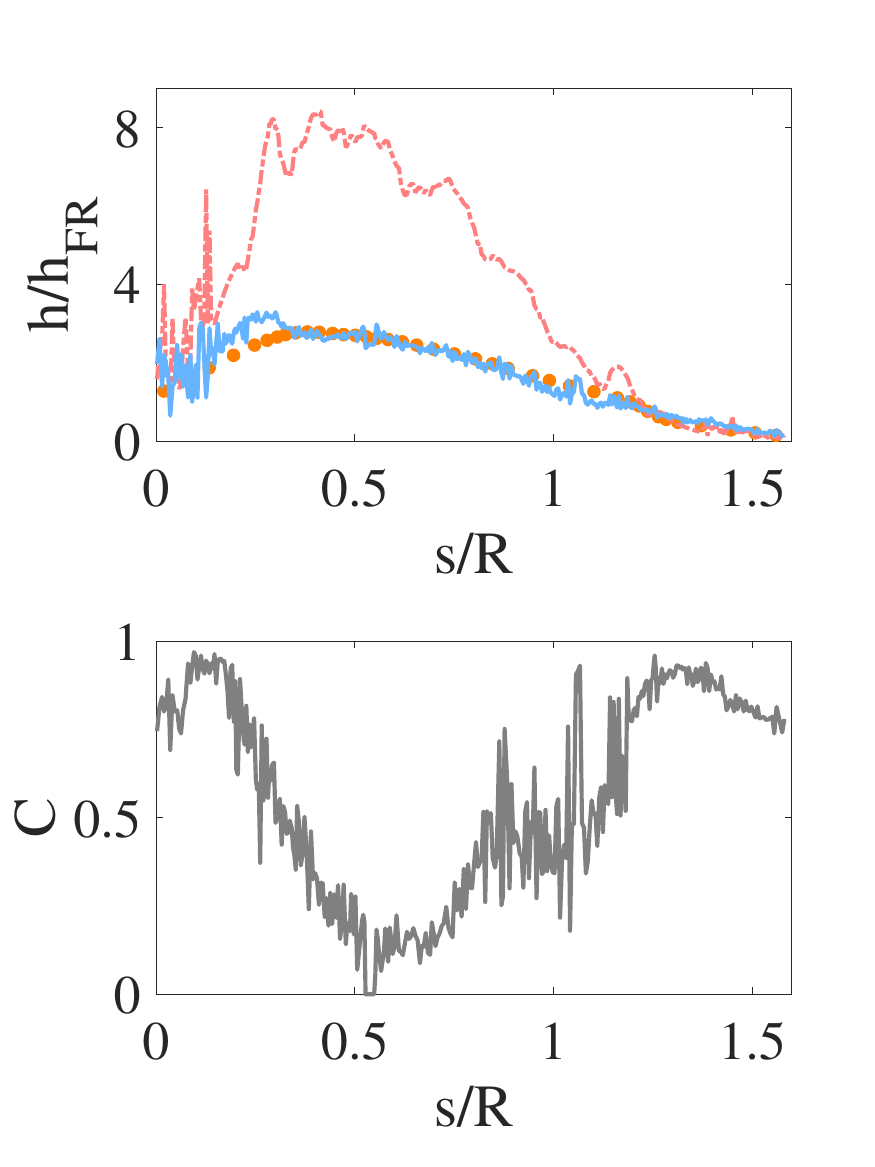}
    \put(-122,110){\rotatebox{90}{{\colorbox{white}{{$h/h_{FR}$}}}}}
    \put(-129,40){\rotatebox{90}{{\colorbox{white}{{$C_{q_w}$}}}}}
    \put(-65,3){\colorbox{white}{$s/R_B$}}
    \put(-65,81){\colorbox{white}{$s/R_B$}}
    \caption{\emph{A-posteriori} evaluation of BFWM-rough-v2 and
      EQWM-$k_s$ in WMLES of a hypersonic blunt body. WMLES results
      using BFWM-rough-v2 for the 20-Mesh case (blue solid) are
      compared against experimental data from
      \cite{hollis2014distributed} (orange dot) and WMLES with
      EQWM-$k_s$ (red dash dot) at three different Reynolds numbers
      (per unit length): $(a)$ $Re_{\infty}=3.0 \times 10^6 /ft$;
      $(b)$ $Re_{\infty}=5.0 \times 10^6 /ft$; $(c)$ $Re_{\infty}=8.3
      \times 10^6 /ft$. Top panels: instantaneous heat transfer
      distribution on the hemisphere surface in terms of the
      normalized film coefficient, $h/h_{FR}$, where $h$ is the
      heat-transfer film coefficient and $h_{FR}$ is the theoretical
      prediction based on Fay–Riddell theory
      \citep{fay1958theory}. Middle and bottom panels: time- and
      azimuthally-averaged wall heat transfer and confidence score
      along the surface of the hemisphere from the front stagnation
      point to the trailing edge, covering the arc length ($s$) from
      $s=0$ to one-quarter of the circumference.  }
\label{fig:EDL_H}
\end{center}
\end{figure}

The instantaneous and time–azimuthally-averaged heat transfer
predicted by WMLES with BFWM-rough-v2 and EQWM-$k_s$ are shown in
Figure~\ref{fig:EDL_H}, along with experimental measurements for a
20-Mesh rough surface under three test conditions. The heat transfer
coefficient $h$ is normalized by the theoretical reference value
$h_{FR}$.  The theoretical value $h_{FR}$ is computed
from the equilibrium stagnation-point boundary layer theory, as
described by the correlation formula in \citet{fay1958theory}, with
further details provided in Appendix~\ref{appE}. The specific values
of $h_{FR}$ for each test condition are reported in
\citet{hollis2017experimental}.

The results show that BFWM-rough-v2 captures the enhanced level of
heat transfer due to surface roughness across different test
conditions, closely matching the experimental data. The oscillations
and overprediction for $s/R_B < 0.4$, where $s$ is the arc length from
the stagnation point, may be linked to the under-resolution of the
thin boundary layer in WMLES. We observe relatively low confidence
near the front stagnation region ($s/R_B < 0.3$) and the trailing edge
($s/R_B > 1.2$), possibly because strong pressure gradient and
curvature effects reduce the validity of the model in these
areas. Interestingly, the confidence in the mid-arc region ($0.3 <
s/R_B < 1.2$) decreases with increasing $Re_{\infty}$, indicating that
higher $Re_{\infty}$ pushes the local inputs farther from what the
model learned during training. We find that this is caused by the
Reynolds number at the matching location $Re_m$ exceeding the range
represented in the training dataset at high $Re_{\infty}$.

In contrast, the WMLES with EQWM-$k_s$ yields accurate predictions for
$Re_{\infty} = 3.0 \times 10^6/\mathrm{ft}$, but fails to do so for
the two higher Reynolds numbers. In those cases, it overpredicts the
heating augmentation by approximately 60\% at $Re_{\infty} = 5.0
\times 10^6/\mathrm{ft}$ and 120\% at $Re_{\infty} = 8.3 \times
10^6/\mathrm{ft}$ in the region $0.2 < s / R_B < 1$.  This
overprediction of wall heating may arise from the application of $k_s$
correlations calibrated for low-speed turbulent boundary layers, which
may not reliably extend to the flow conditions near the hemispherical
surface. Additionally, the temperature log law used in EQWM-$k_s$ may
be insufficient to capture the temperature transport dynamics in this
complex flow scenario.

\section{Conclusions}
\label{conclusion}

We have developed a wall model for low- and high-speed turbulent flows
that accounts for irregular subgrid-scale roughness features. The
model is capable of predicting both wall shear stress and wall heat
flux across a broad range of flow regimes, along with a confidence
score in the predictions. The approach, implemented using artificial
neural networks, is based on the building-block flow
assumption~\citep{lozano2023machine} and constitutes the second
version of our earlier model for incompressible flows,
\emph{BFWM-rough}, introduced in \citet{ma2025machine}. The new model,
referred to as \emph{BFWM-rough-v2}, improves upon its predecessor by
incorporating a revised input formulation, a new approach for
confidence score estimation, a richer training database, and expanded
applicability for WMLES of rough-wall flows.

To train the model, we constructed a DNS roughness database comprising
both incompressible and compressible turbulent channel flows over
irregular rough surfaces. The database includes 192 incompressible
rough cases and 180 compressible rough cases spanning subsonic to
supersonic flow regimes.  The mean flow state, obtained via time and
spatial averaging in the wall normal range $0.02 < y/\delta < 0.15$,
is extracted from the DNS and used as input to the wall model. These
mean-flow quantities, combined with statistical roughness parameters,
are used to construct dimensionless input variables for the model. To
identify the most predictive set of inputs, we apply dimensionless
learning based on information-theoretic principles from
\citet{yuan2025dimensionless}.  The resulting optimized wall model
achieves $L_2$ prediction errors of 4.07\% for wall shear stress and
2.89\% for wall heat flux.

The model also incorporates a confidence score to assess the
reliability of predictions under unseen flow conditions. This is
implemented using a Spectral-normalized Neural Gaussian Process, which
combines spectral-normalized hidden representations with a
random-Fourier-feature Gaussian process layer. This approach yields
calibrated predictive distributions that capture epistemic
uncertainty, enabling the evaluation of the confidence in the
predictions of the wall model.

We have validated BFWM-rough-v2 extensively across a wide range of
flow conditions. The first validation is performed in canonical
turbulent channel flows. WMLES of unseen incompressible and
compressible turbulent channel flows are conducted, comprising a total
of 78 incompressible and 84 compressible cases across various grid
resolutions. The predicted frictional drag exhibits errors primarily
within 10\%, while heat flux predictions are mostly within 15\%. The
PDF of confidence scores indicates high model confidence ($0.8 \le C
\le 1$) for the majority of test cases.  The results also reveal that
wall heat flux predictions are generally more sensitive to grid
resolution and prone to larger errors compared to wall shear
stress. For comparison, WMLES using an equilibrium-based wall model
with prescribed $k_s$ (EQWM-$k_s$) was also performed. While
EQWM-$k_s$ achieves wall shear stress predictions within 5\% in
incompressible channel flows, its accuracy significantly deteriorates
in compressible flows, where errors in wall shear and wall heat flux
can reach up to 30\% and 80\%, respectively, substantially higher than
those observed with BFWM-rough-v2.

We have also validated BFWM-rough-v2 in more practical flow scenarios.
The first such case is a transonic HPT blade with Gaussian
roughness. BFWM-rough-v2 demonstrates good predictive accuracy for the
skin friction coefficient and outperforms EQWM-$k_s$ in this
regard. The model also captures the overall variation of wall heat
flux along both the suction and pressure sides of the blade. However,
it underpredicts the heat flux near the trailing edge on the pressure
side, as well as in regions where normal shock waves are induced by
the surface roughness.  The confidence score of the model for the HPT
blade is generally high ($C > 0.9$), suggesting that the local flow
conditions are similar to those represented in the training dataset.

The next validation case involved a high-speed turbulent compression
ramp with sandpaper roughness. BFWM-rough-v2 predictions show
agreement within 5\% with experimental measurements of wall heat
transfer, while EQWM-$k_s$ overpredicts the peak heating by
approximately 60\% and shifts the location of the peak further
downstream.  Although BFWM-rough-v2 successfully captures the
enhancement in wall heat transfer, it exhibits moderate confidence ($C
\approx 0.5$) upstream of the reattachment location and nearly zero
confidence downstream. The reduced confidence upstream is likely due
to flow separation at the ramp corner. The near-zero confidence
downstream of reattachment is caused by to the local near-wall flow
becoming supersonic, which deviates substantially from the conditions
represented in the training dataset.

Finally, we have performed WMLES of a blunt body with sand-grain
roughness at $M_{\infty} = 6$ and three different $Re_{\infty}$. The
BFWM-rough-v2 model accurately predicts the surface heating
augmentation for $s / R_B \ge 0.4$, but tends to overpredict it for $s
/ R_B < 0.4$.  This overprediction is attributed to limited resolution
in the thin boundary layer near the front of the hemisphere.  In
contrast, EQWM-$k_s$ increasingly overpredicts the heating
augmentation as $Re_{\infty}$ increases. Reduced confidence scores are
observed near both the front and the trailing edge of the body, likely
due to strong pressure-gradient and curvature effects. Furthermore, in
the mid-arc region, the confidence of the model decreases with
increasing $Re_{\infty}$, suggesting that higher Reynolds numbers
amplify wall-heating gradients and shift the local input distribution
further away from the training data.

Finally, we discuss the limitations of the model. In general,
departures from the physical assumptions outlined in
\S\ref{sec:assumption} lead to degraded model performance. Below, we
summarize some of these key limitations.  First, the model is trained
primarily on canonical channel flows, which exhibit statistically
stationary and fully developed turbulence. As a result, its ability to
generalize to flows featuring strong pressure gradients, separation,
reattachment, or shock--boundary-layer interactions is not guaranteed,
as these regimes are not represented in the training dataset.  Second,
the training roughness, which includes Gaussian and Weibull surfaces,
spans only a subset of possible roughness statistics, as they are
isotropic and irregular. Engineering surfaces may exhibit anisotropy
or possess roughness distributions that deviate substantially from the
Gaussian or Weibull forms. Such conditions may lie outside the learned
physics of the model, potentially leading to degraded or unreliable
predictions.  Third, the training data cover a limited range of Mach
numbers (predominantly subsonic and supersonic). At higher Mach
numbers, phenomena such as shock-induced heating and stronger
temperature and density gradients may introduce physics not
encountered during training, reducing reliability in hypersonic
regimes. Last, when the model is deployed in WMLES, its performance
depends on the choice of SGS model, since deficiencies in the latter
can propagate into the wall-model inputs and lead to inaccurate
predictions.

Our future work will focus on addressing the limitations discussed
above and expanding the training database to encompass a broader range
of flow conditions and roughness geometries. These efforts aim to
further improve the overall performance and reliability of WMLES in
complex, high-speed flow environments.

\appendix

\section{Roughness parameter definition}
\label{appA}

The statistical roughness parameters are defined in table
\ref{tab:roughness_para}.
\begin{table}
\begin{center}
\def~{\hphantom{0}} \renewcommand{\arraystretch}{2}
    \begin{tabular}{c|c}
     \hline
     Mean height & $k_{avg} = \frac{1}{A_t}\int_{x,z} k(x,z) dA$ \\
     Crest height & $k_c = \max\{k(x,z)\} -\min\{k(x,z)\}$ \\
     Mean peak-to-valley height & $k_t = \text{mean}\{\max|_{\delta \times \delta}\{k(x,z)\} - \min|_{\delta \times \delta}\{k(x,z)\}\}$ \\
     Root-mean-square height & $k_{rms} = \sqrt{\frac{1}{A_t}\int_{x,z} (k(x,z)-k_{avg})^2 dA}$ \\
     First-order moment of height fluctuations & $R_a = \frac{1}{A_t}\int_{x,z} |k(x,z)-k_{avg}| dA$ \\
     Skewness & $S_k = \frac{1}{A_tk_{rms}^3}\int_{x,z} (k(x,z)-k_{avg})^3 dA$ \\
     Kurtosis & $K_u = \frac{1}{A_tk_{rms}^4}\int_{x,z} (k(x,z)-k_{avg})^4 dA$ \\
     Effective slope & $ES = \frac{1}{A_t}\int_{x,z} |\frac{\partial k(x,z)}{\partial x}| dA$ \\
     Surface porosity & $P_o = \frac{1}{A_tk_c}\int_{0}^{k_c} A_f(y) dy$ \\
     Frontal solidity & $\lambda_f = \frac{A_p}{A_t}$ \\
     Correlation length &  $L_{cor}=\min_{\delta x }\{R_h(\delta x,0) \le 0.2 \}$\\
     \hline
    \end{tabular}
    \caption{\label{tab:roughness_para} Definitions of roughness
      geometrical parameters. $k(x,z)$ is the roughness height
      function, and the reference $k=0$ for the rough surface is
      defined at the minimum roughness height. $A_f(y)$ is the fluid
      area at the $y$ location, $A_p$ is the frontal projected area of
      the roughness elements, and $A_t$ is the total plan area. The
      correlation lengths are computed as the horizontal separation at
      which the roughness height autocorrelation function $R_h(\delta
      x,\delta z)=\frac{1}{k_{rms}^2}\langle k(x+\delta x, z+\delta
      z)k(x,z) \rangle_{xz}$ drops below 0.2, where $\langle \cdot
      \rangle_{xz}$ denotes average over $x$ and $z$. Given that the
      rough surfaces considered are isotropic, the parameters $ES$ and $L_{cor}$ are equivalent along any wall-parallel
      direction.  }
    \end{center}
\end{table}

\section{Validation of DNS setup}
\label{appB}

\begin{figure}
\centering
\includegraphics[width=130mm]{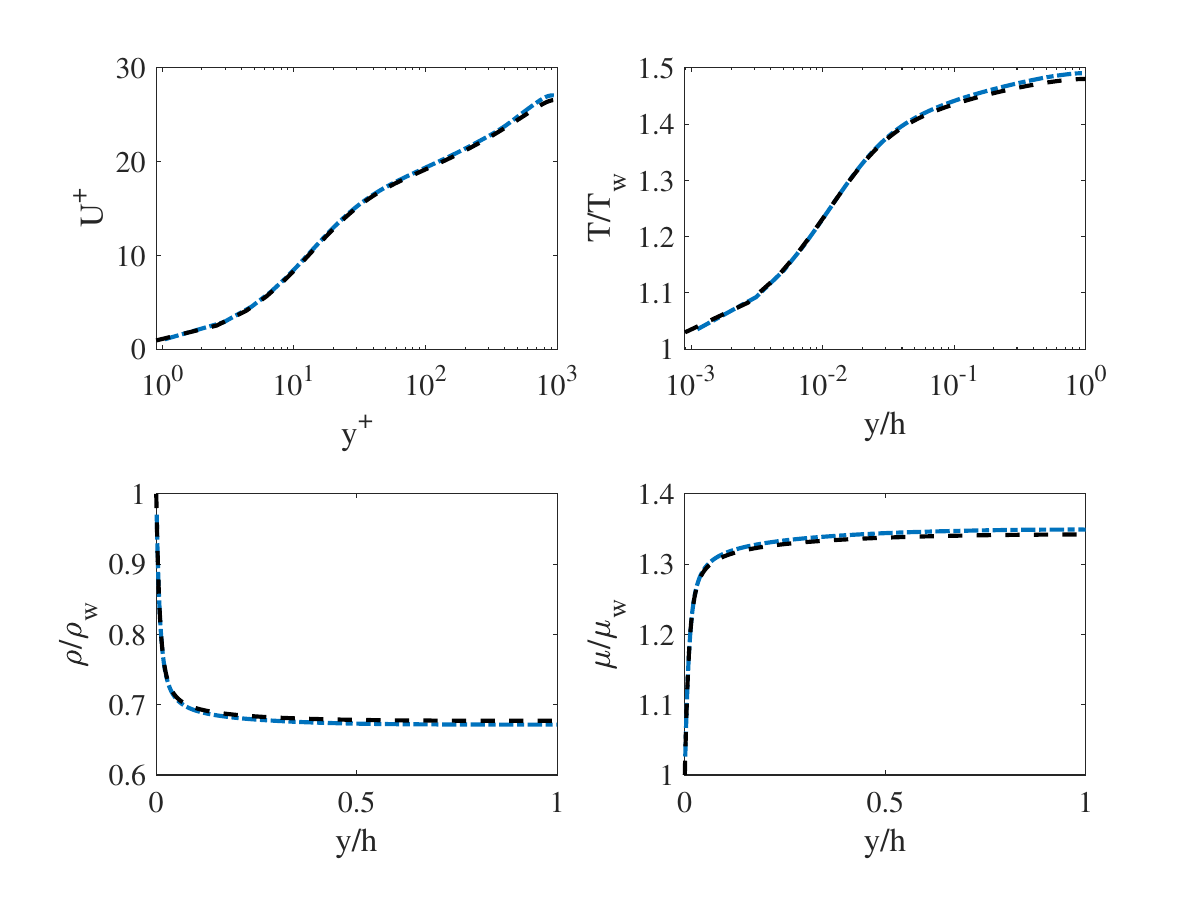}
\put(-270,143){\colorbox{white}{$y^+$}}
\put(-270,15){\colorbox{white}{$y/\delta$}}
\put(-106,143){\colorbox{white}{$y/\delta$}}
\put(-106,15){\colorbox{white}{$y/\delta$}}
\caption{Validation for DNS of turbulent channel flow against Case
  M1.7R600 in \cite{trettel2016mean}: present (blue dot dash),
  \cite{trettel2016mean} (black dash).}
\label{fig:M1_7R600}
\end{figure}
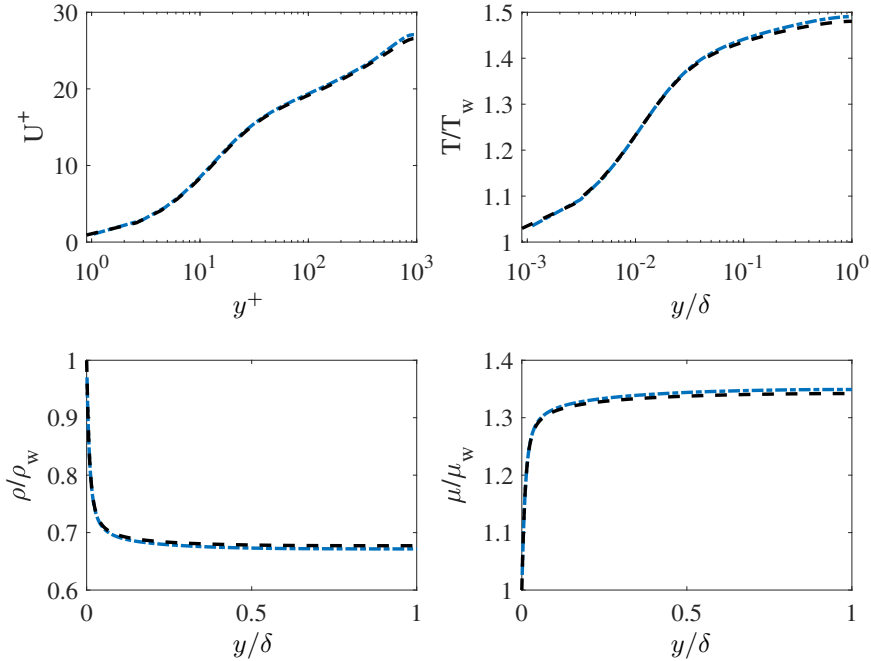

The DNS performed in the present work has been validated by comparison
with the reference DNS of
\citet{trettel2016mean}. Table~\ref{validation_DNS} summarizes the
simulation details of both cases. Specifically, case M1.7R600 in
\citet{trettel2016mean} corresponds to our case M1.7-R15500.  A grid
convergence study was carried out for this case using three grid
resolutions: `coarse', `medium', and `fine'. The results demonstrate
that the `medium' grid is sufficiently resolved, as it closely matches
the `fine' grid. For simplicity, we only present the results from the
`medium' in Figure~\ref{fig:M1_7R600}. Good agreement is observed
between our DNS and the reference, despite the fact that
\citet{trettel2016mean} employed a significantly larger domain size.
\begin{table}
\begin{center}
\vskip -0.1in
\begin{tabular}{l c c c c c c c c c l}
Case & $L_x\times L_y\times L_z$ & $M_b$ & $Re_b$ & $Re_{\tau}$ & $Re_{\tau}^*$ & $-B_q$  & $\Delta x^+$ & $\Delta y_{min}^+$ & $\Delta y_{max}^+$ & $\Delta z^+$  \\
\hline
\cite{trettel2016mean} & $10\delta \times 2\delta \times 3\delta$ & 1.7 & 15500 & 972 & 596 & 0.05 & 10.85 & 0.87 & 7.97 & 6.07\\
M1.7-R15500 & $3\delta \times 2\delta \times \delta$  & 1.7 & 15500 & 970 & 600 & 0.05 & 4.85 & 0.15 & 8.73 & 4.04 \\
GS2-M1.7-R7500-minimal & $3\delta \times 2\delta \times \delta$ & 1.7 & 7500 & 706 & 430 & 0.042 & 5.30 & 0.10 & 5.96 & 4.41 \\
GS2-M1.7-R7500-full & $6\delta \times 2\delta \times 3\delta$ & 1.7 & 7500 & 710 & 420 & 0.046 & 5.33 & 0.10 & 5.99 & 4.43 \\
\hline
\end{tabular}
\caption{Simulation details of the DNS validation cases for smooth-
  and rough-wall turbulent channel flows. \label{validation_DNS}}
\end{center}
\end{table}

For rough-wall cases, a dedicated validation study was conducted to
confirm the suitability of the minimal-span approach for wall model
development. The selected roughness surface, GS2, features the largest
roughness wavelength among all surfaces in the DNS database. As such,
it provides the sparsest sampling of geometrical elements at this
scale and represents the most challenging case for minimal-span
modeling. A comparison is performed between a full-span channel
($6\delta \times 2\delta \times 3\delta$) and a minimal-span channel,
both simulated under the same bulk Mach number $M_b$ and bulk Reynolds
number $Re_b$ in the supersonic regime. The corresponding simulation
parameters are provided in Table~\ref{validation_DNS}.

\begin{figure}
\centering
\includegraphics[width=130mm]{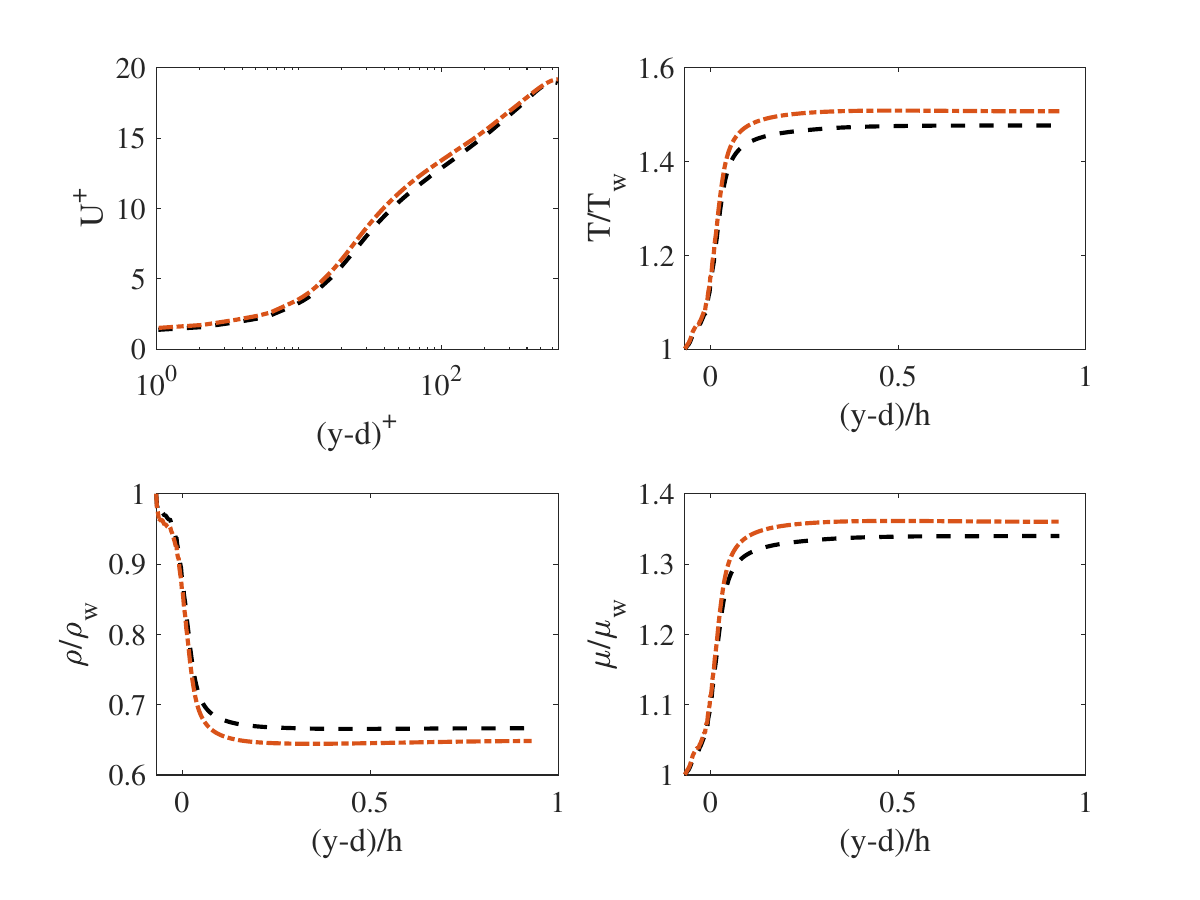}
\put(-275,138){\color{white}\rule{50pt}{12pt}}
\put(-270,143){\colorbox{white}{$y^+$}}
\put(-275,10){\color{white}\rule{50pt}{12pt}}
\put(-270,15){\colorbox{white}{$y/\delta$}}
\put(-111,143){\color{white}\rule{50pt}{12pt}}
\put(-106,143){\colorbox{white}{$y/\delta$}}
\put(-111,10){\color{white}\rule{50pt}{12pt}}
\put(-106,15){\colorbox{white}{$y/\delta$}}
\caption{Comparison between DNS of rough-wall turbulent channel flow
  with a minimal-span (black dash) and a full-span (red dot dash)
  domain for Case GS2-M1.7-R7500.}
\label{fig:min_span}
\end{figure}
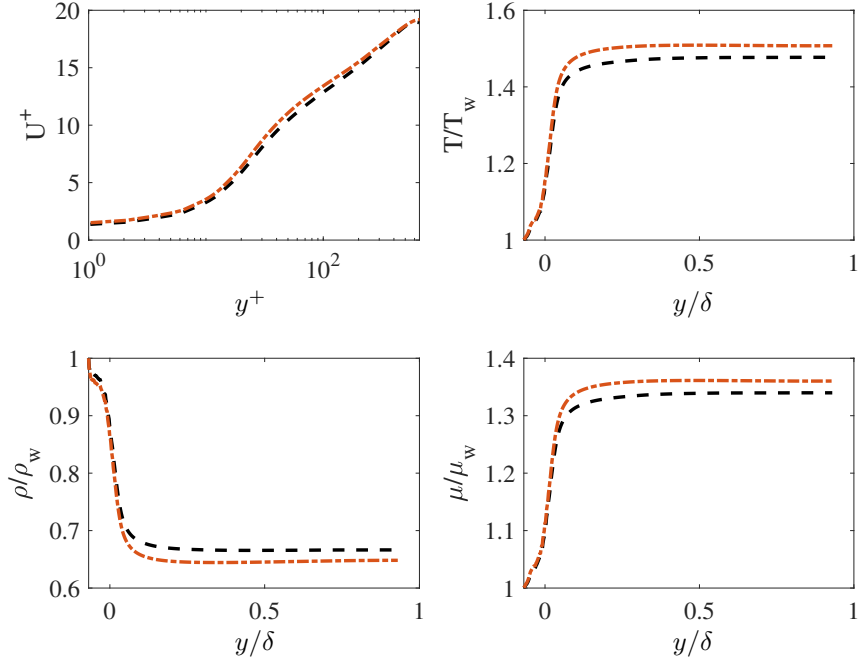

A comparison of mean velocity profiles is shown in
Figure~\ref{fig:min_span}. Good agreement is observed in both
$Re_\tau$ and $-B_q$, confirming the similarity between the two
simulations. The mean velocity profiles are nearly identical, while
the mean temperature and density profiles agree closely below $(y -
d)/\delta = 0.05$, with deviations remaining within 3\% beyond this
region. Based on these results, the DNS mean data for $y/\delta <
0.15$ are considered reliable for wall-model training, with an
estimated error below 3\%.
%
%

\section{Mathematical formulation for a random-feature GP}
\label{appC}

The SNGP model is implemented as a residual-based deep neural network
(DNN) consisting of two fully connected residual blocks, each
containing 64 hidden units and employing ReLU activation
functions. Each residual block is defined as
$
\mathbf{h}_l(\boldsymbol{\Pi}) = \boldsymbol{\Pi} + g_l(\boldsymbol{\Pi})$,
where $\boldsymbol{\Pi} \in \mathbb{R}^d$ denotes the input vector to the $l$-th block and
$\mathbf{h}_l(\boldsymbol{\Pi}) \in \mathbb{R}^d$ is the corresponding output vector.
The residual mapping $g_l(\boldsymbol{\Pi})$ is given by
$
g_l(\boldsymbol{\Pi}) = a\!\left( W_l \boldsymbol{\Pi} + \mathbf{b}_l \right),
$
where $W_l \in \mathbb{R}^{d \times d}$ and $\mathbf{b}_l \in \mathbb{R}^d$ denote the weight
matrix and bias vector, respectively, and $a(\cdot)$ denotes the ReLU activation function
applied element-wise.
A final random-Fourier-feature Gaussian
process layer is appended to the output to yield calibrated predictive
distributions.

The SNGP is formulated by incorporating two key components into this
residual DNN:
\begin{itemize}
    \item[(i)] \emph{Distance-preserving} hidden mapping via spectral
      normalization: \\
    To ensure that the hidden mapping $\mathbf{h}$
is distance preserving, it is sufficient to require each nonlinear residual
block $g_l$ to be $\alpha$-Lipschitz continuous,
that is,
\begin{equation}
\bigl\| g_l(\mathbf{\Pi}) - g_l(\mathbf{\Pi}') \bigr\|_{\mathcal{H}}
\;\le\;
\alpha \, d_s(\mathbf{\Pi}, \mathbf{\Pi}'),
\label{eqn:gl}
\end{equation}
where $\mathbf{\Pi}, \mathbf{\Pi}' \in \mathbb{R}^d$ denote input vectors and
$\|\cdot\|_{\mathcal{H}}$ denotes the norm in the hidden space.

Under this condition, the overall mapping $\mathbf{h}$ of an $L$-layer residual
network satisfies
\begin{equation}
L_1 \, d_s(\mathbf{\Pi}, \mathbf{\Pi}')
\;\le\;
\bigl\| \mathbf{h}(\mathbf{\Pi}) - \mathbf{h}(\mathbf{\Pi}') \bigr\|
\;\le\;
L_2 \, d_s(\mathbf{\Pi}, \mathbf{\Pi}'),
\end{equation}
where
\[
L_1 = (1 - \alpha)^{L-1}
\quad \text{and} \quad
L_2 = (1 + \alpha)^{L-1}
\]
represent the lower (contractivity) and upper (expansivity) bounds,
respectively, for an $L$-layer residual network. The condition in Equation \eqref{eqn:gl} can be satisfied by ensuring that the
spectral norm of each weight matrix ${W}_l$ is upper bounded.
Specifically, since the residual mapping is given by
$g_l(\mathbf{\Pi}) = a({W}_l \mathbf{\Pi} + \mathbf{b}_l)$, its
Lipschitz constant satisfies
$
\| g_l \|_{\mathrm{Lip}}
\;\le\;
\| {W}_l \mathbf{\Pi} + \mathbf{b}_l \|_{\mathrm{Lip}}
\;\le\;
\| {W}_l \|_2
$
\citep{behrmann2019invertible}.
In this work, we
enforce
\begin{equation}
W_l \;=\;
\begin{cases}
c \, W_l / \hat{\sigma}_l, & \text{if } \hat{\sigma}_l > c, \\
W_l, & \text{otherwise},
\end{cases}
\label{eq:sn}
\end{equation}
where $\hat{\sigma}_l \approx \| W_l \|_2$ is estimated via power
iteration and $c > 0$ is a tunable hyperparameter that is determined
based on the validation data to specify the desired upper bound on the
spectral norm.
    \item[(ii)] \emph{Distance awareness} through a random-feature GP
      layer: \\ The random Fourier feature method is employed to
      project the final hidden representation into a high-dimensional
      feature space. This method provides an efficient approximation
      to the kernel function and allows the subsequent probabilistic
      model to operate on a feature space that encodes non-linearities
      without requiring explicit kernel computations. 
\end{itemize}

To enforce distance awareness, we adopt a Gaussian Process (GP) with a
radial basis function (RBF) kernel, which is the classical model that
satisfies this property \citep{rasmussen2006gaussian}. Although the
RBF kernel is used here for simplicity, the framework can be readily
extended to other kernels by adjusting the activation function and
random feature distribution.

Given a training dataset $\mathcal{D} = \{(\boldsymbol{\Pi}, {\Pi_{o}})\}$ with $N$ data points, the hidden representation of the DNN for each training point is denoted as  $h_i = h(\Pi_i)$ where $i=1,2,..., N$, and the GP conditioned on the hidden 
representations is denoted as $\mathbf{g}_{N\times1} = [g(h_1), \dots, g(h_N)]^\top$. The prior distribution of a GP model equipped with an RBF kernel is a multivariate normal distribution:
\begin{equation}
\mathbf{g}_{N\times1} \sim \mathcal{N}(\mathbf{0}_{N\times1}, \, m^2 \mathbf{K}_{N \times N}),
\label{eqn:GP_prior}
\end{equation}
where $m^2$ is the kernel amplitude, and the RBF kernel is defined as:
\begin{equation}
\mathbf{K}_{ij} = \exp\!\left(-\frac{\|h_i - h_j\|_2^2}{2}\right).
\end{equation}
However, computing the GP prior in Equation (\ref{eqn:GP_prior}) is computationally expensive, as it requires the inversion of an \( N \times N \) kernel matrix \(\mathbf{K}\), which scales cubically with the number of samples, i.e., \( \mathcal{O}(N^3) \). 
To deal with this challenge, the GP prior in Equation (\ref{eqn:GP_prior}) is approximated by constructing a low-rank approximation of the kernel matrix \( K = \Phi \Phi^\top \) using random Fourier features \citep{rahimi2007random}.  
This yields a random-feature Gaussian process:
\[
g_{N\times1} \sim \mathcal{N}\!\left(\mathbf{0}_{N\times1},\, \Phi \Phi^\top_{N \times N}\right),
\quad \text{where} \quad
\Phi_{i, D_{L}\times1} = \sqrt{\frac{2 m^2}{D_L}} \cos\!\left(-\mathbf{W}_L h_i + \mathbf{b}_L\right).
\tag{10}
\]
Here, \( D_{L} \) is the number of random features. We tested $D_L=512,1024$ and $2048$ and find that the model performance is not very sensitive to different values of $D_L$, and $D_L=512$ is sufficient in our work. \( \Phi_{i,D_{L}\times1} \) is the final layer feature, where  
the random weights \( \mathbf{W}_L \in \mathbb{R}^{D_L \times D_{L-1}} \) are drawn independently from \( \mathcal{N}(0, I) \), and the bias term \( \mathbf{b}_L \in \mathbb{R}^{D_L} \) is sampled independently from \( \text{Uniform}(0, 2\pi) \).  

Consequently, we replace the standard deterministic output layer with a Bayesian linear regression head. The GP prior can be expressed as a neural network layer with fixed hidden weights \( \mathbf{W}_L \) and learnable output weights \( \beta \) with a prior variance $\sigma_p^2$:
\begin{equation}
g(h_i) = \Phi_{i, D_{L}\times1}^\top \beta,
\quad \text{with prior} \quad
\beta_{D_{L}\times1} \sim \mathcal{N}(0, \sigma_p^2 I_{{D_L}\times{D_L}}).
\end{equation}
By Bayes’ rule, the posterior distribution of $\beta$ is proportional to the likelihood times the prior:
\begin{equation}
p(g \mid \mathcal{D}) \propto p(\mathcal{D} \mid g) \, p(g),
\label{eqn:GP_posterior}
\end{equation}
where \( p(g) \) denotes the GP prior (not to be confused with
pressure).  For a regression task, the data likelihood \(
p(\mathcal{D} \mid g) \) with an observation noise variance
$\sigma_n^2$ can be modeled using the exponentiated squared loss:
\begin{equation}
p(\mathcal{D} \mid g) = \exp\!\left( - \frac{1}{2\sigma_n^2}\|\Pi_o - g\|_2^2 \right).
\end{equation}
Note that the observation noise variance $\sigma_n^2$ represents the variance of the residuals (i.e., the mismatch between the model predictions and the observed targets) in the training data. It is typically estimated once during training and then treated as a constant hyperparameter throughout the GP inference and prediction steps.

The negative log-posterior of the Bayesian linear regression head is given by
\begin{equation}
- \log p(g \mid \mathcal{D})
= \frac{1}{2\sigma_n^2}\|\Pi_o - \Phi\beta\|_2^2 + \frac{1}{2\sigma_p^2}\|\beta\|_2^2 + \text{const.}
\end{equation}  
In this equation, it can be observed that the prior variance $\sigma_p^2$ controls the strength of regularization and defines the initial uncertainty before seeing the data. A common default value $\sigma_p=2$ is used in our work. By rearranging terms, the posterior distribution over the regression weights $\beta$ can be expressed in closed form as a multivariate Gaussian:
\begin{equation}
p(\beta \mid \mathcal{D}) = 
\mathcal{N}\!\left(\mu_\beta,\, \Sigma_\beta\right),
\quad
\mu_\beta = \frac{1}{\sigma_n^2}\Sigma_\beta \Phi^\top \Pi_o,
\quad
\Sigma_\beta = \left(\frac{1}{\sigma_n^2}\Phi^\top\Phi + \frac{1}{\sigma_p^2}I\right)^{-1}.
\end{equation}

For a new input $\boldsymbol{\Pi}_*$ with feature representation $\boldsymbol{\phi}_* = \boldsymbol{\phi}(\boldsymbol{h}_*)$, the predictive distribution is obtained by marginalizing over the posterior of $\beta$:
\begin{equation}
p(\Pi_{o*} \mid \boldsymbol{\Pi}_*, \mathcal{D})
= \mathcal{N}\!\left(\boldsymbol{\phi}_*^\top \mu_\beta,\,
\boldsymbol{\phi}_*^\top \Sigma_\beta \boldsymbol{\phi}_* + \sigma_n^2\right).
\end{equation}
The predictive variance consists of two components:
\[
\underbrace{\boldsymbol{\phi}_*^\top \Sigma_\beta \boldsymbol{\phi}_*}_{\text{epistemic uncertainty}}
\;+\;
\underbrace{\sigma_n^2}_{\text{aleatoric uncertainty}}.
\]
The first term, $\sigma_e^2=\boldsymbol{\phi}_*^\top \Sigma_\beta \boldsymbol{\phi}_*$, represents the \textit{epistemic uncertainty}. It increases in regions far from the training data, reflecting the model’s lack of knowledge and assessing the reliability of predictions in unseen or extrapolative regimes.  
The second term, $\sigma_n^2$, corresponds to the \textit{aleatoric uncertainty} caused by inherent observation noise and remains constant regardless of the data size. The epistemic uncertainty obtained from the SNGP model is therefore used to quantify the model uncertainty in this work.

\section{EQWM for rough walls with prescribed $k_s$}
\label{appD}

The EQWM-$k_s$ provides an algebraic formulation based on the
logarithmic law of the wall for both velocity and temperature,
extended to account for surface roughness effects. It assumes a fully
developed, equilibrium turbulent boundary layer characterized by a
logarithmic profile.

For smooth walls, the log law is expressed as
\begin{equation}
U^{+} = \frac{1}{\kappa_v} \ln(y^{+}) + 5,
\end{equation}
where $\kappa_v$ is the von Kármán constant.  For rough walls, this
relationship is modified to include the velocity deficit as
\begin{equation}
U^{+} = \frac{1}{\kappa_v} \ln\!\left(\frac{y}{k_s}\right) + 8.5,
\end{equation}
where $k_s$ is the equivalent sand-grain roughness height.
Consequently, the wall shear stress is obtained algebraically from the
local velocity $u_m$ and density $\rho_m$ at the matching location
$y_m$ as
\begin{equation}
\tau_w = \rho_m 
\left[
  \frac{u_m}{
    \frac{1}{\kappa_v}\ln\!\left(\frac{y_m}{k_s}\right) +8.5
  } \right]^{2}.\label{eq:app_eqwm_tauw}
\end{equation}

The temperature law of the wall provides an empirical relationship
between the nondimensional temperature and wall-normal distance in
turbulent boundary layers, analogous to the velocity log-law for
momentum transfer~\citep{kader1981temperature,
  bradshaw2013introduction}. The correction function $T^*$ ensures a
smooth transition from the linear sublayer to the logarithmic region
and is defined as:
\begin{equation}
    T^* = 
    \begin{cases}
        0.5\,Re_m, & Re_m \le y^* \\[6pt]
        \dfrac{0.5\,y^{*2}}{Re_m} + C_t\!\left(1 - \dfrac{y^*}{Re_m}\right) + \dfrac{\ln(Re_m/y^*)}{\kappa_t}, & Re_m > y^*
    \end{cases}
\end{equation}
where the Reynolds number $Re_m = \tfrac{\rho_m u_\tau y_m}{\mu_m}$,
$\kappa_t = \tfrac{\kappa_v}{Pr_t}$, $Pr_t = 0.85$, $C_t = 3.9$, and
$y^* = 8.28$ are empirical constants.

The algebraic wall model for wall heat flux is then derived by
relating the mean temperature gradient at the matching location $y_m$
to the wall heat flux through the total thermal conductivity
$\lambda_{\text{tot}}$ and the temperature law correction:
\begin{equation}
    q_w = \frac{\lambda_{\text{tot}} (T_m - T_w)}{y_m \,[1 + \kappa_t T^*]}.
    \label{eqn:C4_reordered}
\end{equation}
The term $[1 + \kappa_t T^*]$ acts as a turbulent diffusion damping
factor that modulates the temperature gradient near the wall.  The
total thermal conductivity $\lambda_{\text{tot}}$ includes both
molecular and turbulent contributions,
\begin{equation}
    \lambda_{\text{tot}} = \lambda + \lambda_t,
    \qquad
    \lambda_t = \frac{\mu_t c_p}{Pr_t},
\end{equation}
where the eddy viscosity $\mu_t$ is obtained using a mixing-length
model with near-wall damping \citep{kawai2010dynamic} with $A^{+} =
17$:
\begin{equation}
    \mu_t = \kappa_v \, \rho \, y \, \sqrt{\frac{\tau_w}{\rho}} \, D,
    \qquad D = 1 -
    \exp\!\left[-\left(\frac{y^{+}}{A^{+}}\right)^{2}\right],
\end{equation}
where $\tau_w$ is obtained from Equation~(\ref{eq:app_eqwm_tauw}).

\section{Fay--Riddell stagnation-point heat-transfer theory}
\label{appE}

We follow previous works and use the Fay--Riddell stagnation-point
heat-transfer theory~\citep{fay1958theory} to obtain a reference heat
flux for the blunt body~\citep{hollis2014distributed,
  cheatwood2024mars}. The theory provides a semi-empirical expression
for the convective heat-transfer rate to a blunt body at the
stagnation point under high-enthalpy, equilibrium-flow conditions. The
model assumes a steady, axisymmetric, laminar boundary layer in
chemical and thermal equilibrium between the post-shock flow and the
wall.  For an equilibrium laminar boundary layer with a Lewis number
(ratio of thermal to mass diffusivity) of $Le=0.71$, the
stagnation-point convective heat flux is given by
\begin{equation}
 q_q
= 0.76\,Pr^{-0.6}\,(\rho_w\mu_w)^{0.1}\,(\rho_e\mu_e)^{0.4}\,
\Big[1+\big( Le^{0.52}-1\big)\,(h_D/h_e)\Big]\,
(h_e-h_w)\,
\sqrt{\left(\frac{du_e}{dx}\right)_s},
\label{eq:fay_riddell}
\end{equation}
where $Pr$ is the Prandtl number, $\rho$ and $\mu$ are the
density and dynamic viscosity (subscripts $w$ and $e$ denote wall and
boundary-layer-edge conditions), $h_e$ and $h_w$ are the static
enthalpies at the boundary-layer edge and at the wall, $h_D$ is the
dissociation (diffusion) enthalpy scale, and $(du_e/dx)_s$ is the
stagnation-point edge velocity gradient (often approximated using the
nose radius $R_n$, edge pressure $p_e$, freestream pressure
$p_\infty$, and edge density $\rho_e$).


\section*{Acknowledgements}
We gratefully acknowledge Y. Yuan for
  the helpful discussion on dimensionless learning based on
  information. We also gratefully acknowledge M. Whitmore,
  A. Elnahhas, R. Agrawal, and G. Arranz for their contributions to
  the DNS compressible flow database. We thank C. Pederson for the
  insightful discussion on this work.

\section*{Funding}
This work was supported by the National Science
  Foundation under Award Number 2317254 and by an Early Career Faculty
  grant from NASA’s Space Technology Research Grants Program (Grant
  Number 80NSSC23K1498). The authors acknowledge ACCESS allocation and
  the Oak Ridge Leadership Computing Facility, which is a DOE Office
  of Science User Facility supported under Contract DE-AC05-00OR22725,
  for providing HPC resources that have contributed to the research
  results reported within this work.

\section*{Declaration of interests}
The authors report no conflict of interest.



\bibliographystyle{jfm}
\bibliography{jfm-instructions}


\end{document}